\documentclass[11pt,a4paper]{article}
\pdfoutput=1
\usepackage{jheppub}
\usepackage[sort&compress]{natbib}
\usepackage{url}
\usepackage{hyperref}
\usepackage{amsfonts}
\usepackage{epsfig}\usepackage{dcolumn}
\usepackage{bm}
\usepackage{slashed}
\graphicspath{{./figuras/}}
\usepackage{subfigure}

\newcommand{\gev}{\; \hbox{GeV}}
\newcommand{\ifb}{\; \hbox{fb}^{-1}}
\newcommand{\zp}{Z^{\prime}}
\newcommand{\gzp}{g_{Z'}}
\newcommand{\wzp}{\Gamma_{Z'}}

\newcommand{\mzp}{M_{Z'}}
\newcommand{\mchi}{M_{\chi}}

\newcommand{\gx}{g_{\chi}}

\title{ {\color{blue} \mbox{Dirac-Fermionic Dark Matter in $U(1)_X$ Models}}}

\author{Alexandre Alves$^a$,}
\author{Asher Berlin$^{b,c}$,}
\author{Stefano Profumo$^d$,}
\author{Farinaldo S. Queiroz$^{d,e,f}$}

\affiliation{$^a$Departamento de Ci\^encias Exatas e da Terra,
Universidade Federal de S\~ao Paulo, Diadema-SP, 09972-270, Brazil}
\affiliation{$^b$Enrico Fermi Institute, University of Chicago, Chicago, IL 60637}
\affiliation{$^c$Kavli Institute for Cosmological Physics, University of Chicago, Chicago, IL 60637}
\affiliation{$^d$Department of Physics and Santa Cruz Institute for Particle Physics
University of California, Santa Cruz, CA 95064, USA}
\affiliation{$^e$Department of Physics and Astronomy, Mitchell Institute for Fundamental Physics and Astronomy,Texas A \& M University, College Station, TX 77843-4242}
\affiliation{$^f$Max-Planck-Institut f\"ur Kernphysik, Saupfercheckweg 1, 69117 Heidelberg, Germany\\}

\emailAdd{aalves@unifesp.br}
\emailAdd{berlin@uchicago.edu}
\emailAdd{profumo@ucsc.edu}
\emailAdd{queiroz@mpi-hd.mpg.de}

\abstract{We study a number of $U(1)_X$ models featuring a Dirac fermion dark matter particle. We perform a comprehensive analysis which includes the study of corrections to the muon magnetic moment, dilepton searches with LHC data, as well as direct and indirect dark matter detection constraints. We consider four different coupling structures, namely $U(1)_{B-L}, U(1)_{d-u}, U(1)_{universal}$, and $U(1)_{10+\bar{5}}$, all motivated by compelling extensions to the standard model. We outline the viable and excluded regions of parameter space using a large set of probes. Our key findings are that (i) the combination of direct detection and collider constraints rule out dark matter particle masses lighter than $\sim 1$~TeV, unless rather suppressed $\zp$-fermion couplings exist, and that (ii) for several of the models under consideration, collider constraints rule out $\zp$ masses up to $\sim 3$ TeV. Lastly, we show that we can accommodate the recent Diboson excess reported by ATLAS collaboration within the $U(1)_{d-u}$ model.}
    
\begin{document} 
\maketitle
\flushbottom
    
\section{Introduction}

The existence of dark matter (DM) has been ascertained by numerous lines of evidence, originating from different scales and epochs in the history of the universe. However, the fundamental particle nature of DM is yet to be unveiled. From a particle physics standpoint, thermally produced DM  can have have mass in the keV-TeV range. Within this mass range, WIMPs (Weakly Interacting Massive Particles) are some of the most compelling dark matter candidates (see for example the reviews in Ref. \cite{Baer:2014eja,Strigari:2013iaa,Bertone:2004pz,Dutta:2009uf,
Bringmann:2012ez,Bergstrom:2009ib,Jungman:1995df}).

WIMPs have been intensively searched for in experiments that use a broad variety of techniques, customarily grouped in the three categories of (i) direct detection, (ii) indirect detection, and (iii) collider searches \cite{Cushman:2013zza}.  

(i) Direct detection  relies on the measurement of the energy recoil deposited by WIMPs on nuclei in underground experiments typically operating at very cold temperatures. With a good control over the background, current experiments seek to observe excesses of scattering events, which are potentially translated to  measurements of the WIMP-nucleon scattering cross section and mass, for a given velocity distribution. The difficulty in precisely assessing the number of background events has in the past misled several direct detection experiments to claim the observation of excesses which could be plausibly explained by $7-30$~GeV WIMPs \cite{Kelso:2011gd,DelNobile:2014sja}. However, null results from other experiments, including CDMSlite, XENON, and LUX \cite{Akerib:2013tjd,Agnese:2013jaa} are in blatant tension with this possibility, despite theoretical efforts to accommodate a WIMP interpretation by means of isospin violation \cite{Feng:2013vod,Kumar:2011dr,Cirigliano:2013zta} or scattering with impurities in the detectors \cite{Profumo:2014mpa}. At present, no conclusive signal from WIMP dark matter has been reported in direct detection experiments, and increasingly strong limits on the nucleon-WIMP cross section probe theories beyond the Standard Model with unprecedented accuracy.

(ii) Indirect detection refers to the observation of DM annihilation products, such as cosmic- and gamma-rays (see Ref.~\cite{Profumo:2013yn} for a pedagogical introduction). This is an exciting and promising avenue, which makes use of a variety of different {\em messenger} particles and of observational devices, and which relies on direct or indirect information about the DM density distribution in the Galaxy and beyond. Ref.~\cite{Cirelli:2013igx} provides an overview of recent results. Since potential discoveries in indirect detection generically suffer both from large and vastly unknown background and systematic uncertainties, we focus here on bounds derived using the relatively robust limits currently set by Fermi-LAT observations of dwarfs spheroidal galaxies \cite{Ackermann:2015zua}.

(iii) Collider searches for dark matter are often based on the search for events with large missing energy, associated with pair-produced WIMPs escaping the detector \cite{Goodman:2010ku,Rajaraman:2011wf,Bauer:2013ihz}. In models where the mediator between the visible and WIMP sector has sizable couplings to SM fermions, either dijet or dilepton limits are typically most stringent \cite{Patra:2015bga,Alves:2013tqa,Alves:2015pea,Arcadi:2014lta,Harris:2014hga,Buchmueller:2014yoa,
Fairbairn:2014aqa,Cline:2014dwa,Busoni:2014haa,Alves:2014yha,Lebedev:2014bba,
deSimone:2014pda,Cogollo:2014jia,Arcadi:2013qia,Profumo:2013sca,Dudas:2013sia,
Dudas:2009uq,Primulando:2015lfa,Chala:2015ama,Feldman:2011ms,FileviezPerez:2012mj,
Arnold:2012fm,Duerr:2014wra,Han:2013mra,An:2012ue,An:2012va,Kapukchyan:2013hfa,
Freitas:2004hq,Carena:2003aj,Langacker:2008yv}, whereas mono-jet ones are complementary, and often competitive in the low WIMP mass regime 
\cite{Jacques:2015zha,Khachatryan:2014rra,Agrawal:2013hya,Chatrchyan:2012me,
Racco:2015dxa,Frandsen:2012rk}. As explained below in Sec.~\ref{sec:colliders}, the dilepton searches provide the most stringent collider bounds for the models that we consider in this study.

Of additional relevance for the class of models under discussion here are limits from contributions to  the muon magnetic moment, especially in the regime where the DM couplings to the mediator are suppressed, i.e. in a regime where direct detection and indirect detection limits weaken. Here, we use an adapted version of the public code \cite{Queiroz:2014zfa} to assess such limits.

In view of the current bounds, several DM portals have been studied with the purpose of outlining the remaining viable parameter space after combining existing limits. Some of the most interesting and studied candidates are Dirac fermions. In this context, $Z^{\prime}$ gauge bosons are natural mediators since they appear in a large multitude of gauge extensions of the SM. Most often, this $Z^{\prime}$ portal is studied in simplified models such as Ref.~\cite{Alves:2015pea,Arcadi:2014lta,Alves:2013tqa}. The simplified model approach is valid and interesting. However, the conclusions drawn often cannot be directly applied to particular models, due to the large variations in the structure and magnitude of the vector and axial-vector couplings. Therefore, it is worthwhile to study particular gauge structures, which appear in several extensions of the SM. These types of extensions often address intriguing open questions such as neutrino masses, number of family generations, and matter and anti-matter asymmetry. 


An additional motivation to focus on $Z^{\prime}$ models originates from the recent excess reported by the ATLAS collaboration in the resonant diboson to hadronic final states channel at an energy of around 2 TeV \cite{Aad:2015owa}; such excess, currently featuring a local statistical significance of 3.4$\sigma$, 2.6$\sigma$ and 2.9$\sigma$ respectively in the $WZ$, $WW$ and $ZZ$ channels \cite{Aad:2015owa}, has been ascribed to an extra 2 TeV bosonic particle \cite{Fukano:2015hga, Hisano:2015gna, Franzosi:2015zra}. In particular, Ref.~\cite{Hisano:2015gna} constructs an explicit leptophobic $U(1)^\prime$ extension inspired by an $E_6$ supersymmetric GUT. While we do not consider this specific model here, our results provide a direct connection between the ATLAS diboson excess and models with a dark matter particle candidate, and a roadmap for further studies in this direction.

In summary, the current experimental landscape allows us to probe models of new physics with a WIMP dark matter candidate from several independent directions. Here, for the first time, we perform a comprehensive analysis taking advantage of collider, $(g-2)_{\mu}$, direct, and indirect WIMP dark matter detection limits for four theoretically simple and well-motivated $U(1)$ gauge extensions. In particular, we outline the remaining viable region of parameter space and that which is now experimentally excluded, for a broad range of WIMP masses ranging from 8 GeV up to 5 TeV. 

The remainder of this study is as follows: in the following section we describe and define the four $U(1)$ gauge extensions that we consider throughout our study; in Sec.~\ref{sec:dmpheno} we outline the dark matter phenomenology (relic density, direct, and indirect searches) in these scenarios; Sec.~\ref{sec:gmu} explores the impact of contributions to the muon anomalous magnetic moment, while Sec.~\ref{sec:colliders} estimates the impact of collider searches. Section \ref{sec:results} summarizes and compiles all of our results, while Sec.~\ref{sec:conclusions} concludes.

\section{$U(1)$ Extensions}
\label{section: exotic}

One of the most minimal ways of extending the SM gauge structure is to include an additional spontaneously broken $U(1)_X$ gauge group. As a result, a new neutral, massive spin-1 gauge boson ($Z^\prime$) arises. In the scenarios of interest here, the DM is a new Dirac fermion charged under the $U(1)_X$ group. As far as the setup's phenomenology is concerned, the relevant parameters of the theory are the $\zp$ mass ($\mzp$), its coupling to fermions (including the DM), its width ($\wzp$), as well as the DM mass ($\mchi$). \footnote{Some recent model-independent attempts to study the DM phenomenology in the context of $U(1)_X$ theories include Ref.~\cite{Alves:2015pea, Alves:2013tqa,Hooper:2014fda,Bell:2014tta,Arcadi:2014lta,Arcadi:2013qia,Chu:2013jja, Dudas:2009uq}.} In this work we focus on extensions of the SM which (i) avoid flavor changing neutral currents at tree level, and (ii) allow for cancellation of triangle anomalies by the introduction of vector-like fermions. It has been shown that these two requirements greatly restrict the list of possible coupling structures. We will focus on four interesting cases \cite{Appelquist:2002mw,Carena:2004xs}, namely:

(i) $U(1)_{B-L}$: baryon and lepton numbers are accidental anomalous global symmetries of the SM, whereas B-L is not. Thus an interesting extension of the SM consists of gauging the B-L symmetry \cite{Appelquist:2002mw}; 

(ii) $U(1)_{d-u}$: this structure arises after spontaneous symmetry breaking in some $E_6$ grand unified theories \cite{PDG};

(iii) $U(1)_{10+\bar{5}}$: this structure arises after the spontaneous symmetry breaking of flipped SU(5) grand unified theories \cite{Carena:2004xs}. 

(iv) $U(1)_{universal}$: an extensively explored extension which inherits the universal coupling structure of the SM $Z$ \cite{Appelquist:2002mw}.

As mentioned above, we seek suppression of flavor-changing neutral current processes at tree level, thus we focus on models where the $Z^\prime$ couplings to the fermions are generation-independent \cite{Carena:2004xs}. The relevant charges of the left-handed quark doublet ($Q_L$), right-handed quarks ($u_R, d_R$), lepton doublet $l_L$, and right-handed leptons $l_R$ are shown in Table~\ref{table: charge}. In what follows, we do not assume any specific scalar structure inducing spontaneous symmetry breaking of the $U(1)_X$ gauge group and that the SM Higgs is neutral under $U(1)_X$ so that there is no $Z-\zp$ mixing at tree level, which relaxes otherwise strong electroweak precision constraints. Finally, since the DM particle is postulated to be charged under $U(1)_X$, it will affect the anomaly cancellation requirement; however, in principle additional charged fermions might be postulated, with no impact on the DM phenomenology (see Ref.~\cite{Duerr:2014wra} for an explicit example). Therefore, with the exception of the $\zp$ and of the DM particle, we assume any additional fields beyond the SM is effectively decoupled, and will not discuss them further in association with the models' phenomenology.

\begin{table}[t]
\centering
\begin{tabular}{| c || c | c | c | c |}
\hline
Field & $U(1)_\text{universal}$ & $U(1)_{B-xL}$ & $U(1)_{10+x\bar{5}}$ & $U(1)_{d-xu}$\\ \hline \hline
$Q_L$ & $x$ &  $\frac{1}{3}$ &  $\frac{1}{3}$ & $0$ \\ \hline
$u_R$  & $x$ &  $\frac{1}{3}$ & $-\frac{1}{3}$ & $-\frac{x}{3}$ \\ \hline
$d_R$  & $x$ &  $\frac{1}{3}$ &  $-\frac{x}{3}$ & $\frac{1}{3}$ \\ \hline
$l_L$   & $x$ &  $-x$ &   $\frac{x}{3}$ & $\frac{-1+x}{3}$\\ \hline
$e_R$  & $x$ &  $-x$ & $-\frac{1}{3}$ & $\frac{x}{3}$\\ \hline
\end{tabular}
\caption{The generation-independent exotic $\zp$ $U(1)$ charge assignments explored in this paper. In all cases there is a free continuous parameter $x$ which, for simplicity, we take equal to 1 throughout our analysis.}
\label{table: charge}
\end{table}

The four $U(1)_X$ charge assignments shown in Table~\ref{table: charge} depend on a continuous free parameter denoted by ``$x$". Although the successful cancellation of triangle anomalies is guaranteed for any value of $x$,  for simplicity, we  fix $x=1$. We will also restrict the new $U(1)$ $\zp$ gauge coupling ($g_{\zp}$) to two numerical values: $\gzp = $ 1 or 0.5. With $x$ and $g_{\zp}$ specified, we can then determine the numerical values of the $\zp$-SM fermion couplings as discussed in Sec.~\ref{sec:dmpheno} below.

\section{Dark Matter Phenomenology}\label{sec:dmpheno}

A new massive and neutral $\zp$ is an interesting feature of the low-energy limit of many extensions to the SM possessing a viable DM candidate \cite{Profumo:2013sca,Cogollo:2014jia,Kopp:2014tsa,Das:2013jca,Agrawal:2014ufa,Chang:2014tea,
Hektor:2014kga,Bell:2014tta,Freitas:2014jla,Cao:2014cda}. As discussed above, the number of possible simple $U(1)$ charge extensions is greatly reduced after considering anomaly cancellations and electroweak precision measurements. In particular, we focus on the the $U(1)$ charge assignments in Table~\ref{table: charge}. We will parametrize the simplified Lagrangian responsible for a $\zp$ interacting with a Dirac fermion DM ($\chi$) and SM fermion ($f$) as

\begin{equation}
\label{equation: lagrangian}
\mathcal{L} \supset Z^\prime_\mu~\big[ \bar{\chi} \gamma^\mu \left( g_{\chi \text{v}} + g_{\chi a} \gamma^5 \right) \chi +  \sum_{f \in \text{SM} } \bar{f} \gamma^\mu \left( g_{f \text{v}} + g_{f a} \gamma^5 \right) f \big]
~,
\end{equation}
where the sum is over all quarks and leptons of the SM (including neutrinos when relevant). The $\zp$-SM couplings above can be specified in terms of the $U(1)_X$ charges and gauge coupling ($g_{\zp}$) since

\begin{align}
\label{equation: SMcouplings}
g_{u \text{v}} &= \frac{1}{2} g_{\zp} \left( z_{u_R} + z_{Q_L} \right) ~,~  g_{u a} = \frac{1}{2} g_{\zp} \left( z_{u_R} - z_{Q_L} \right)
\nonumber \\
g_{d \text{v}} &= \frac{1}{2} g_{\zp} \left( z_{d_R} + z_{Q_L} \right) ~,~  g_{d a} = \frac{1}{2} g_{\zp} \left( z_{d_R} - z_{Q_L} \right)
\nonumber \\
g_{l \text{v}} &= \frac{1}{2} g_{\zp} \left( z_{e_R} + z_{l_L} \right) ~~,~~ g_{l a} = \frac{1}{2} g_{\zp} \left( z_{e_R} - z_{l_L} \right)
\nonumber \\
g_{\nu \text{v}} &= \frac{1}{2} g_{\zp} z_{l_L} ~~~~~~~~~~~~~,~~  g_{\nu a} = -\frac{1}{2} g_{\zp} z_{l_L}
~,
\end{align}
where the $z$'s are the $U(1)_X$ charges of the the gauge eigenstates of Table~\ref{table: charge} and $g_{\zp}$ is the new $U(1)$ gauge coupling constant. 


  
We note that our restriction to Dirac DM allows for more general interactions with the $\zp$ than Majorana DM, since Dirac DM allows for vector interactions with the $\zp$. Therefore, restricting to Majorana DM instead would affect mostly the scattering behavior of DM with quarks inside nuclei since vector interactions are coherent over the entire nucleus as will be discussed in Sec.~\ref{section: directdetection}. We also assume that the $\chi$'s vector and axial interactions are of the same magnitude and sign, i.e. $g_{\chi \text{v}} = g_{\chi a}$. Hence, we define the coupling $g_\chi$ as $g_\chi \equiv g_{\chi \text{v}} = g_{\chi a}$. The case where the vector and axial couplings have opposite signs, $g_{\chi \text{v}} = - g_{\chi a}$, would not change any qualitative aspect of our results (see Ref.~\cite{Arcadi:2014lta} for studies which assume $g_{\chi v} \neq g_{\chi a}$).

\begin{figure}[!t]
\centering
\subfigure[\label{fig01}]{\includegraphics[scale=0.5]{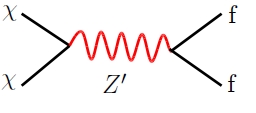}}
\subfigure[\label{fig02}]{\includegraphics[scale=0.5]{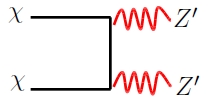}}
\subfigure[\label{fig03}]{\includegraphics[scale=0.5]{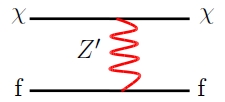}}
\subfigure[\label{fig04}]{\includegraphics[scale=0.5]{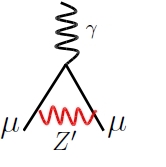}}
\caption{Relevant tree-level processes for: (a,b) pair-annihilation; (c) elastic scattering off of SM fermions; (d) muon anomalous magnetic moment.}
\label{fig:feynmandiagrams}
\end{figure}

\subsection{Direct Detection}
\label{section: directdetection}

The general parametrization of the $\zp$ interactions in Eq.~({\ref{equation: lagrangian}}) induces both spin-independent (SI) and spin-dependent (SD) scattering with nuclei. In particular, elastic scattering occurs through the $t$-channel exchange of a $\zp$ (see Fig.\ref{fig:feynmandiagrams}, c). As mentioned previously, in the case where both $\chi$ and the valence quarks of nucleons possess vector interactions with $\zp$, large coherent spin-independent scattering may occur, and this process is severely constrained by current bounds from direct detection experiments \cite{Akerib:2013tjd,Aprile:2013doa}. An approximate form for the SI cross section at low-momentum transfer is
\begin{align}
\label{equation: SI}
\sigma^\text{SI (per nucleon)} &\approx \frac{\mu^2_{\chi n}}{\pi} \Big[ \frac{Z f_\text{prot} + (A-Z) f_\text{neut}}{A} \Big]^2
\nonumber \\
f_\text{prot} &\equiv \frac{g_{\chi \text{v}}}{\mzp^2} \left( 2g_{u \text{v}} + g_{d \text{v}} \right)
\nonumber \\
f_\text{neut} &\equiv \frac{g_{\chi \text{v}}}{\mzp^2} \left( g_{u \text{v}} + 2 g_{d \text{v}} \right),
\end{align}

\noindent where $\mu_{\chi n}$ is the WIMP-nucleon reduced mass, and where $Z$ and $A$ are the atomic number and atomic mass of the target nucleus, respectively. In scanning through the parameter space of different models, we demand that the scattering cross section of Eq.~(\ref{equation: SI}) be below the most stringent current upper bound, given at present by the LUX experiment~\cite{Akerib:2013tjd}. 

Similarly, if the nucleon valence quarks and $\chi$ both possess axial-vector interactions with the $\zp$, one additionally has spin-dependent scattering, which, again in the low momentum transfer limit, has a cross section of the form

\begin{align}
\label{equation: SD}
\sigma^\text{SD (per neutron)} &\approx \frac{3 \mu^2_{\chi \text{neut}}}{\pi} \frac{g_{\chi a}^2}{\mzp^4} \Big[ g_{u a}  \Delta_u^\text{neut} +  g_{d a} \left( \Delta_d^\text{neut} + \Delta_s^\text{neut}  \right) \Big]^2
~,
\end{align}
where $\Delta_q^\text{neut}$ are the quark spin fractions of the neutron. We will take these to be $\Delta_u^\text{neut} = -0.42$, $\Delta_d^\text{neut} = 0.85$, $\Delta_s^\text{neut} = -0.08$~\cite{Cheng:2012qr}. We then require that the spin-dependent rate to be below current published limits from XENON100~\cite{Aprile:2013doa}. For spin-dependent bounds, we  focus on scattering with neutrons since these limits are at present the most stringent.

\subsection{Indirect Detection}
\label{section: indirect}
From the Lagrangian of Eq.~(\ref{equation: lagrangian}), if $\chi$ possesses non-negligible interactions with $\zp$, then, during the early universe, $\chi$ was in thermal equilibrium with SM particles. 
Residual pair annihilation can still occur in the late universe, producing SM quark and lepton pairs. In turn, this leads to high energy gamma-rays via neutral pion production and final state radiation, detectable with gamma-ray telescopes such as the Fermi Large Area Telescope \cite{Ackermann:2011wa}. In particular, constraints can be derived from the non-observation of signals in DM-dominated targets such as dwarf spheroidal galaxies \cite{Ackermann:2015zua,Ackermann:2013yva,Abdo:2010ex}.
%

From the Lagrangian interaction terms indicated in Eq.~(\ref{equation: lagrangian}), $\chi$ pairs can annihilate through an s-channel $\zp$ exchange process to a pair of SM fermions, as long as $\mchi > m_f$ (see (a) of Fig.~\ref{fig:feynmandiagrams}). The non-relativistic form for this annihilation cross section is
\begin{align}
\label{equation: annihilation}
\sigma v \left( \chi \bar{\chi} \to f \bar{f} \right) &\approx \frac{n_c \sqrt{1-\frac{m_f^2}{\mchi^2}}}{2 \pi \mzp^4 \left( 4 \mchi^2 - \mzp^2 \right)^2} \bigg\{ g_{fa}^2 \Big[ 2 g_{\chi \text{v}}^2 \mzp^4 \left(\mchi^2 - m_f^2 \right) + g_{\chi a}^2 m_f^2 \left( 4 \mchi^2 - \mzp^2\right)^2 \Big]
\nonumber \\
& + g_{\chi \text{v}}^2 g_{f \text{v}}^2 \mzp^4 \left( 2 \mchi^2 + m_f^2 \right) \bigg\}
~,
\end{align}
where $v$ is the relative velocity of the annihilating DM pair and $n_c$ is the number of colors of the final state SM fermion. Sufficiently near resonance, the width of $\zp$, $\Gamma_{\zp}$, should be included in Eq.~(\ref{equation: annihilation}). Using again the interaction terms of Eq.~(\ref{equation: lagrangian}), the $\zp$ width takes the simple form
\begin{align}
\Gamma_{\zp} &= \sum_{f \in \text{SM}}\theta \left( \mzp - 2 m_f \right) \frac{n_c \mzp}{12 \pi} \sqrt{1 - \frac{4 m_f^2}{\mzp^2}}~\left[ g_{f \text{v}}^2 \left(1 + \frac{2 m_f^2}{\mzp^2}\right) + g_{f a}^2 \left( 1 - \frac{4 m_f^2}{\mzp^2}\right) \right]
\nonumber \\
&+ \theta \left( \mzp - 2 \mchi \right) \frac{\mzp}{12 \pi} \sqrt{1 - \frac{4 \mchi^2}{\mzp^2}}~\left[ g_{\chi \text{v}}^2 \left(1 + \frac{2 \mchi^2}{\mzp^2}\right) + g_{\chi a}^2 \left( 1 - \frac{4 \mchi^2}{\mzp^2}\right) \right]
~,
\end{align}
where $\theta$ is the unit step function.

Furthermore, if $\mchi > \mzp$, then $\chi$ may also annihilate directly into pairs of on-shell $\zp$ bosons, which subsequently decay to SM fermions. The non-relativistic form for this annihilation channel 
is
\begin{align}
\label{equation: cascade}
\sigma v \left( \chi \bar{\chi} \to \zp \zp \right) &\approx \frac{1}{16 \pi \mchi^2 \mzp^2} \left(1 - \frac{\mzp^2}{\mchi^2}\right)^{3/2} \left(1 - \frac{\mzp^2}{2 \mchi^2}\right)^{-2}
\nonumber \\
&\times \Big[ 8 g_{\chi \text{v}}^2 g_{\chi a}^2 \mchi^2 + \left( g_{\chi \text{v}}^4 + g_{\chi a}^4 - 6 g_{\chi \text{v}}^2 g_{\chi a}^2\right) \mzp^2 \Big]
~.
\end{align}
To calculate the gamma-ray yield for this annihilation mode, one needs to account for the $\zp$ decay channels.

Whenever appropriate, we utilize limits on the gamma-ray flux from Fermi observations. For a given annihilation final state, we thus need to calculate the differential or integrated gamma-ray yield.
To do this, we utilize the {\tt PPPC4DMID} code, which calculates the spectrum of gamma-rays for direct annihilations to SM fermions~\cite{Cirelli:2010xx}. Compared to the direct annihilation case, annihilations to pairs of $\zp$ bosons lead to a spread in the gamma-ray energy due to the Lorentz boost between the $\chi$ and $\zp$ rest frames. In this case, the spectra from {\tt PPPC4DMID}, which corresponds to direct annihilations to SM fermions, are convolved over a finite energy range, corresponding to the kinematics of the boosted final state fermions~\cite{Mardon:2009rc}.

Astrophysical observations place upper bounds on the quantity $\frac{1}{\mchi^2}  \sum_f \frac{1}{2} \langle \sigma v \rangle_f N_{\gamma, f}$, where $\langle \sigma v \rangle_f$ and $N_{\gamma,f}$ are the annihilation rate and number of photons, within a given energy-bin, produced in a single annihilation to some final state $f\bar f$, and the factor of $\frac{1}{2}$ takes into account that Dirac DM is not self-conjugate. There are several possible annihilation channels whose relative importance depends on the dark matter mass and the $U(1)_X$ model. In order to derive the indirect detection limits in a consistent way, we implement the recent Dwarf PASS8 results from FERMI-LAT \cite{Ackermann:2015zua}. In particular, we utilize the upper limits that FERMI Dwarf observations place on direct annihilations to $b \bar{b}$ and rescale accordingly to account for the total photon yield in the $\zp$ model under consideration. We point out that competitive but less robust indirect detection limits from gamma-ray observations of the center of the Milky Way could be used~\cite{Hooper:2012sr}. For other complementary limits applicable to this model see Refs.~\cite{Aguilar:2013qda,Cholis:2013psa,Lin:2015taa,Hamaguchi:2015wga,Jin:2015sqa,
Bringmann:2014lpa}

It is important to note that when $\mzp \ll \mchi$ non-perturbative effects such as Sommerfeld enhancement might become relevant, especially after freeze-out, and would thus significantly strengthen the bounds from indirect detection~\cite{Hisano:2004ds,Hisano:2003ec}. Since most of the viable parameter space in our models corresponds to larger $\zp$ masses, we do not incorporate this effect into our analysis. We also point out that if had included couplings between the $\zp$ and gauge bosons, the shape of the abundance curve would change and mild $\gamma$-ray line bounds would be applicable \cite{Duerr:2015wfa}. 


\section{The Muon Anomalous Magnetic Moment}\label{sec:gmu}

Significant contributions to the muon anomalous magnetic moment $a_\mu$ are expected in $U(1)_X$ models where the neutral $\zp$ gauge boson  has sizable couplings to leptons.
The E821 experiment at Brookhaven National Laboratory, which measured the precession of muons and anti-muons in a constant external magnetic field as they circulated in a confining storage ring, reported the value $a_{\mu}^{E821}= (116592080 \pm 63) \times 10^{-11}$ \cite{Bennett:2004pv,Bennett:2006fi}. Thus, 

\begin{equation}
\Delta a_{\mu} ({\rm E821} -{\rm SM}) = (295 \pm 81) \times 10^{-11},
\label{deltaa}
\end{equation}
i.e. an excess of about $3.6~\sigma$ compared to the SM expectation. This excess might or not be associated with new physics,  but the result stands in any case as a general constraint to models producing significant contributions to $a_\mu$.

It is important to notice that due to large uncertainties in the SM hadronic contributions to $a_\mu$, caution should be used in taking this excess at face value.  The contribution to $a_\mu$ stemming from $Z^{\prime}$ gauge bosons is of the form \cite{Queiroz:2014zfa}, 
\begin{eqnarray}
&&
\Delta a_{\mu} (Z^{\prime}) = \frac{m_\mu^2}{8\pi^2 M_{Z^\prime}^2} \int_0^1 dx \frac{g^2_{\mu \text{v}} P_{\text{v}}(x)+ g^2_{\mu a} P_{a}(x) }{(1-x)(1-\lambda^2 x) +\lambda^2 x},
\label{vectormuon1}
\end{eqnarray} where $g_{\mu \text{v}}$ and $g_{\mu \text{a}}$ are the vector and axial couplings to muons, respectively, $\lambda = m_{\mu}/M_{Z^{\prime}}$ and,
\begin{eqnarray}
P_{\text{v}}(x) & = & 2 x^2 (1-x) \nonumber\\
P_{a}(x) & = & 2 x(1-x)\cdot (x-4)- 4\lambda^2 \cdot x^3.
\label{vectormuon2} 
\end{eqnarray} 
In the limit $M_{Z^{\prime}} \gg m_{\mu}$, we find \cite{Kelso:2014qka,Kelso:2013zfa},
\begin{equation}
\Delta a_{\mu}(Z^{\prime}) = \frac{m_{\mu}^2}{4 \pi^2 M_{Z^\prime}^2}\left(\frac{1}{3}g^2_{\mu \text{v}} - \frac{5}{3}g^2_{\mu a}\right).
\label{vectormuon3}
\end{equation}
It is worth pointing out that the overall $\zp$ correction to the muon magnetic moment can take positive or negative values depending on the relative magnitude of the vector and axial couplings. Rather light $\zp$ gauge bosons are required to explain the excess, and as we shall see further those masses are already excluded by collider searches. Nevertheless, one can still use the measurement to place $1\sigma$ limits by demanding the contributions to lie within the error bar. This is precisely what we do here. In Table 3, we summarize our limits for the four models discussed in our analysis. Those exclusion bounds are also shown as horizontal gray lines in Figs.~4-27. We also exhibit in thick gray regions the $Z^{\prime}$ mass region that accommodates the muon magnetic moment excess.

\begin{table}[t]
\centering 
{\color{blue}Bounds from Muon Magnetic Moment}
\begin{tabular}{|c|c|c|}
\hline
\hline 
Model & Overall Sign & Bound\\ 
\hline
$U(1)_{\rm universal}$& $\Delta a_{\mu} > 0$ & $\gzp=1$, $\mzp > 323$ GeV\\
& $\Delta a_{\mu} > 0$ & $\gzp=0.5$, $\mzp > 162.5$ GeV \\
\hline 
$U(1)_{\rm B-L}$ & $\Delta a_{\mu} > 0$ &  $\gzp=1$, $\mzp > 323$ GeV\\
& $\Delta a_{\mu} > 0$&  $\gzp=0.5$, $\mzp > 162.5$ GeV \\
\hline 
$U(1)_{10+\bar{5}}$& $\Delta a_{\mu} < 0$ & $\gzp=1$, $\mzp > 240$ GeV\\
& $\Delta a_{\mu} < 0$& $\gzp=0.5$, $\mzp > 120$ GeV \\
\hline 
$U(1)_{\rm d-u}$ & $\Delta a_{\mu} < 0$ & $\gzp=1$, $\mzp > 107$ GeV\\
& $\Delta a_{\mu} < 0$ & $\gzp=0.5$, $\mzp > 53.5$ GeV \\
\hline
\end{tabular}
\caption{Bounds on the $\zp$ mass rising from the muon magnetic moment for each of the models. Those limits are also indicated in Figs.~4-27 as horizontal gray lines.}
\label{g2bounds}
\end{table}


\section{Collider Bounds: search for new resonances in dilepton events}\label{sec:colliders}

The new $\zp$ boson associated with the spontaneously broken $U(1)_X$ gauge group generically decays to charged leptons, light jets, top quarks, and, invisibly, to dark matter and/or neutrinos. Those decays enable $\zp$ searches with the Large Hadron Collider (LHC) in a variety of channels, including dileptons, dijets and mono-X signatures. As pointed out in Ref.~\cite{Alves:2015pea}, the constraints from dijet and monojet searches are competitive only in the regime of very weak couplings to charged leptons -- e.g. in the ``leptophobic'' scenario studied in~\cite{Alves:2013tqa}. Incidentally, we find that the features of dilepton and dijet events are very similar to those presented in~\cite{Alves:2015pea}. 

%
\begin{figure}[!t]
\centering
\includegraphics[scale=0.5]{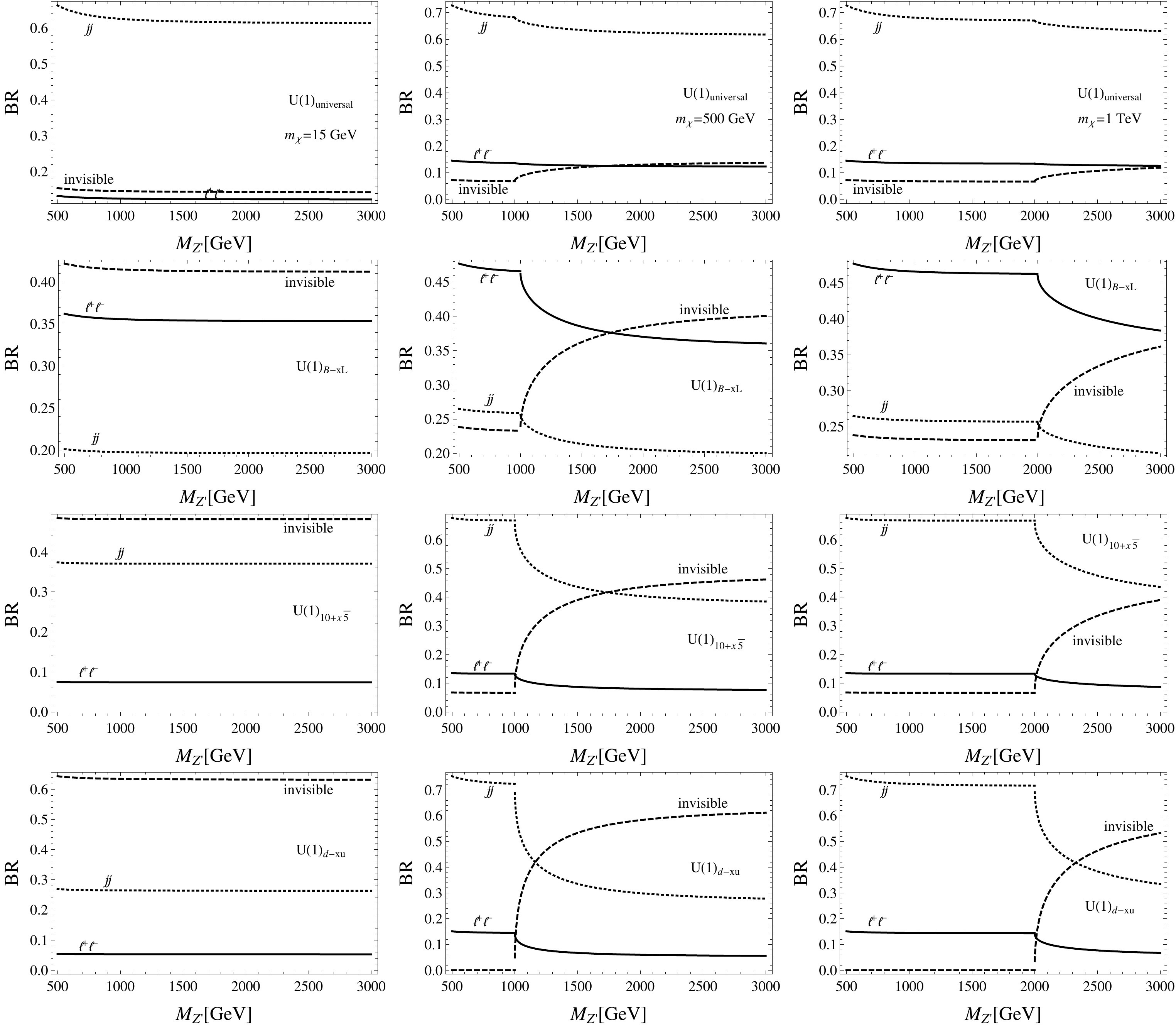}
\caption{Branching ratios of $\zp$ into jets (dotted lines), leptons (solid lines) and an invisible (neutrinos plus DM) mode (dashed lines) as a function of the new gauge boson mass, for DM masses $\mchi=15$ GeV (left column), 500 GeV (middle column) and 1 TeV (right column). The first row shows the branching ratios of the $U(1)_{universal}$ model, the second row the $U(1)_{B-L}$, the third row the $U(1)_{10+\bar{5}}$, and the fourth row the $U(1)_{d-u}$ model. We have fixed $g_{\zp}=\gx=1$ in all plots. In all cases there is also a small branching fraction to top quarks, which is not shown.}
\label{figbr}
\end{figure}
%
Fig.~\ref{figbr} illustrates the decay branching ratios for the $\zp$ into jets (dotted lines), leptons (solid lines) and invisible channels (dashed lines). The left-most column assumes $\mchi=15$ GeV, the middle 500 GeV, and the right-most 1 TeV, while the four rows correspond to the four charge structures under consideration: $U(1)_{\rm universal}$ (first row), $U(1)_{B-L}$ (second row), $U(1)_{10+x\bar 5}$ (third row) and $U(1)_{d-xu}$ (fourth row).

Depending on the $U(1)_X$ charge assignment structure, stronger couplings between charged leptons and the $\zp$ produce larger branching ratios to charged leptons. This is the case e.g. for $B-L$ models, as we can see in the second row of Fig.~\ref{figbr}. While a $\sim 40$\% branching ratio into charged leptons (including tau leptons decays) is typical for these types of models, the other three models considered in this work present smaller branching ratios, of order $\sim 10$\%. On the other hand, stronger couplings to quarks lead to larger $\zp$ production rates as in the $U(1)_{\rm universal}$ models. The trade-off between production cross section and branching fraction to charged leptons determines the relative number of events for each model given a specific mass spectrum; the cut efficiencies for dilepton searches are expected to be the similar for these models once the $\zp$ mass is fixed. 

Fig.~\ref{figbr} also illustrates that a dilepton search for a $\zp$ is critically sensitive to the model details. First of all, for $\gx=1$, the branching ratio to jets is considerably smaller than those found in the leptophobic scenario analyzed in Ref.~\cite{Alves:2013tqa} over the entire $\zp$ mass range for $\mchi < 500$ GeV, but increases in the $\zp$ mass range where the decay to a DM pair is kinetically forbidden. However, constraints from dileptons are much tighter than those from dijets even in those cases. This can be understood in terms of the backgrounds associated with dileptons and dijets searches for new resonances. While the signal cross section for dileptons and dijets from $\zp$ production  have similar rates, the backgrounds for dijet resonances involve QCD production of jets, which demands much tighter cuts to clean up the SM events. On the other hand, the main background for dileptons searches is the Drell-Yan process, which is much easier to suppress.

As in~\cite{Alves:2015pea}, the branching ratio to an invisible final state is larger compared to a leptophobic scenario due  to potential decays to neutrinos. In all models but $U(1)_{\rm universal}$, the invisible branching ratio can reach the 60\% level. This is larger than the typical branching ratios of the leptophobic scenario~\cite{Alves:2013tqa}. However, this somewhat larger branching ratio to invisible does not make the monojet channel competitive in the case of these models with sizable lepton couplings.

In Ref.~\cite{Alves:2013tqa}, $\zp$ masses up to 2.1 TeV for light DM could be ruled out, based on dijet and monojet searches at the Tevatron and LHC for $g_\chi\le 1$. Dilepton searches are more efficient to exclude regions of parameter space for dark $\zp$ models which present sizable lepton couplings. This is the case for the models presented in this work and the model independent framework studied in Ref.~\cite{Alves:2015pea}.

\vskip0.5cm
\noindent{\underline{Dileptons at the 8 TeV LHC - ATLAS search}}
\vskip0.5cm

To evaluate the constraints from collider searches on the models under consideration, we used results from the 8 TeV LHC search for dilepton resonances after $20.3\ifb$ of integrated luminosity for the electron sample and $20.5\ifb$ for the muon sample~\cite{Aad:2014cka}. The following Drell-Yan type process was simulated
\begin{equation}
p\bar{p}\to \zp\to \ell^-\ell^+\; ,\; \ell=e,\mu
\label{ppxll}
\end{equation}
plus up to two extra jets using \texttt{Madgraph5}~\cite{Alwall:2011uj}--FeynRules~\cite{Alloul:2013fw}, clustering and hadronizing jets with \texttt{Pythia}~\cite{Sjostrand:2006za}, and simulating detector effects with \texttt{Delphes3}~\cite{deFavereau:2013fsa}. Soft and collinear jets from QCD radiation generated by \texttt{Pythia} were consistently merged with the hard radiation calculated from matrix elements in MLM scheme~\cite{Mangano:2006rw} at appropriate matching scales. We adopted the CTEQ6L parton distribution functions~\cite{Nadolsky:2008zw}, computed at the factorization/renormalization scales $\mu_F=\mu_R=\mzp$. 

Signal events were selected according with the same criteria adopted in Ref.~\cite{Aad:2014cka}:
\begin{eqnarray}
p_T(e_1) & > & 40\gev\; ,\; p_T(e_2) > 30\gev\; ,\; |\eta_e| < 2.47 \\
p_T(\mu_1) & > & 25\gev\; ,\; p_T(\mu_2) > 25\gev \; ,\; |\eta_\mu| < 2.47 \\
128 < & M_{\ell\ell} & < 4000 \gev 
\label{cuts}
\end{eqnarray}
Here $\ell_{1}(\ell_2)$ is the hardest (second hardest) lepton in the event, and $ M_{\ell\ell}$ the invariant mass of the lepton pair. The signal acceptance times efficiency found in our simulations are similar to those presented in~\cite{Aad:2014cka} and in our previous study~\cite{Alves:2015pea}.

All background simulations to $pp\to\zp\to\ell^-\ell^+$ were taken from Ref.~\cite{Aad:2014cka}. To constrain a $\zp$ model we calculated a $\chi^2$ statistic (at 95\% confidence level) based on $M_{\ell\ell}$ measured in Ref.~\cite{Aad:2014cka} in 6 invariant mass bins: $110-200$ GeV, $200-400$ GeV, $400-800$ GeV, $800-1200$ GeV, $1200-3000$ GeV, and $3000-4500$ GeV. We show in Fig.~\ref{mll} the number of events for the assumed luminosity in each $\ell^-\ell^+$ invariant mass bin for the total background and signal in the $U(1)_{\rm universal}$ model (with $g_{\zp}=\gx=1$ and $\mchi=100$ GeV) for various $\mzp$ values. The limits we obtain in this work do not take systematic uncertainties into account in the fitting procedure, thus our results might be somewhat overestimated.  
\begin{figure}[t]
\centering
\includegraphics[scale=0.55]{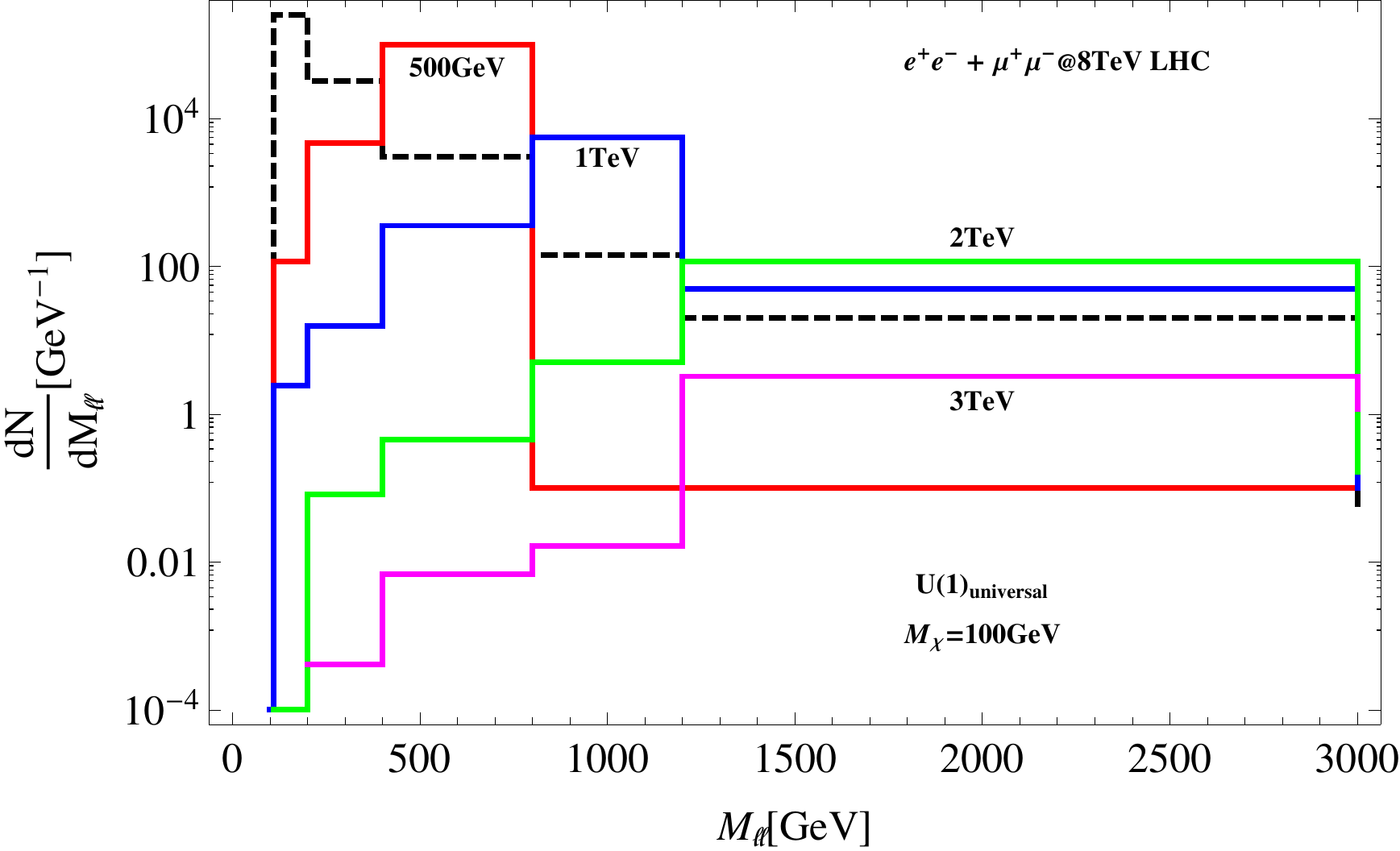}
\caption{The dilepton invariant mass distribution for the total background (dashed line) in six bins taken from Ref.~\cite{Aad:2014cka} and the signal for four different $\zp$ masses in the $U(1)_{\rm universal}$ couplings model: 500 GeV, 1, 2 and 3 TeV. We fixed $g_{\zp}=\gx=1$ and $\mchi=100$ GeV. The integrated luminosity assumed in this plot is the same as in the ATLAS study~\cite{Aad:2014cka}.}
\label{mll}
\end{figure}

The exclusion regions from collider bounds do not vary drastically from one model to another in the weak $\gx$ coupling regime, as can be seen in Figs.~4-27. This feature is the result of the relative contribution between the $\zp$ production cross section and the branching ratio to charged leptons. The branching ratio to light jets, as shown as the dotted lines in Fig.~\ref{figbr}, illustrates the relative size of the production cross section in each model. The $U(1)_{\rm universal}$ model is expected to produce the largest number of new gauge bosons, but precisely because of their stronger couplings to quarks, the branching ratio to $\ell^+\ell^-$ is suppressed, compensating for the larger production cross section compared to the other models. For the $d-u$ and $10+\bar{5}$ models, the branching ratios to charged leptons are larger compared to the universal model, but the couplings to quarks are smaller, again compensating each other and rendering the bounds very similar. The $B-L$ model presents a much larger branching fraction to leptons which translates into resonance searches probing $\zp$ masses roughly $\sim 300$ GeV larger than the other models under consideration.

The other prominent feature we observe in Figs.~4-27 is how the bounds become increasingly weaker as $\gx$ increases. As seen in Table~\ref{table: charge}, in $d-u$ models neutrinos do not couple to the new force. As a result, the branching ratio to invisible states arises solely from decays to dark matter, thus enhancing the sensitivity of this model to the precise value of the coupling $\gx$. The $10+\bar{5}$ model also introduces rather weak couplings between the $\zp$ and neutrinos compared to the universal and $B-L$ models. This explains why the bounds on $\mzp$ are nearly constant as a function of $\gx$ in the universal model, but weaken in a much more pronounced manner for the $d-u$ model. However, in the small $\gx$ regime, all these models are excluded for $\zp$ masses below $\sim 2.6$ TeV and dark matter masses from 8 GeV to 5 TeV, except for the universal model where masses below $\sim 3$ TeV are excluded for the same dark matter masses. Our results for $\gx\sim0$ compare favorably to the ATLAS bounds on other $\zp$ models~\cite{Aad:2014cka}. For larger couplings, the bounds are softened, as discussed above, reaching $\sim 1.8$ TeV in the $d-u$ models with $\gx= 1$ and $\mchi=8$ GeV and $\sim 2.4$ TeV for the same couplings with $\mchi=1$ TeV. We point out that the limits from LEPII derived in \cite{Carena:2004xs} 
are not applicable here due to the existence of invisible state.

We also should comment that we are ignoring interference effects with the SM Z boson and photon. For the most part of the parameters spaces of the models under study, the $zp$ width is sufficiently narrow to justify that. Yet, even for larger widths, these effects are expected to be small for $\zp$ masses close to the bounds that we found. Moreover, ignoring the interferences gives us more conservative bounds, once the interference among the neutral vector states is constructive.

\begin{table}[t]
\centering 
{\color{blue}Collider Limits}\\
\begin{tabular}{|c||c|c|c|}
\hline
\hline 
Model & Mass  & Coupling & Bound\\ 
\hline
 & 8 GeV- 1 TeV & $\gzp=1$& $g_{\chi} \leq 0.1$, $\mzp \geq 2570$ GeV; $g_{\chi} \sim 1$, $\mzp \geq 2400$ GeV\\
$U(1)_{\rm universal}$ & 8 GeV- 1 TeV & $\gzp=0.5$& $g_{\chi} \leq 0.1$, $\mzp \geq 2196$ GeV; $g_{\chi} \sim 1$, $\mzp \geq 2100$ GeV\\
\hline
 & 8 GeV - 1 TeV & $\gzp=1$& $g_{\chi} \leq 0.1$, $\mzp \geq 2620$ GeV;$g_{\chi} \sim 1$, $\mzp \geq 2400$ GeV\\
$U(1)_{10+\bar{5}}$ & 8GeV - 1 TeV & $\gzp=0.5$& $g_{\chi} \leq 0.1$, $\mzp \geq 2480$ GeV;$g_{\chi} \sim 1$, $\mzp \geq 2040$ GeV\\
\hline 
 & 8 GeV - 1 TeV &  $\gzp=1$ & $g_{\chi} \leq 0.1$, $\mzp \geq 3000$ GeV;$g_{\chi} \sim 1$, $\mzp \geq 2930$ GeV\\
$U(1)_{\rm B-L}$ & 8 GeV - 1 TeV &  $\gzp=0.5$ & $g_{\chi} \leq 0.1$, $\mzp \geq 2570$ GeV;$g_{\chi} \sim 1$, $\mzp \geq 2280$ GeV\\
\hline 
 & 8 GeV-1 TeV &  $\gzp=1$ & $g_{\chi} \leq 0.1$, $\mzp \geq 2640$ GeV;$g_{\chi} \sim 1$, $\mzp \geq 2300$ GeV\\
$U(1)_{\rm d-u}$ & 8 GeV-1 TeV &  $\gzp=0.5$ & $g_{\chi} \leq 0.1$, $\mzp \geq 2430$ GeV;$g_{\chi} \sim 1$, $\mzp \geq 1800$ GeV\\
\hline 
\end{tabular}
\caption{Collider limits on the $\zp$ masses for the models in study. These limits are also reproduced in the plots of Figs.~4-27. We point out that the bounds for $g_{\chi} \sim 1$ are actually sensitive to the DM mass, we thus quote the most conservative ones.}
\label{colliderbounds}
\end{table}

\section{Results}\label{sec:results}

\begin{figure}[!t]
\centering
\includegraphics[scale=0.45]{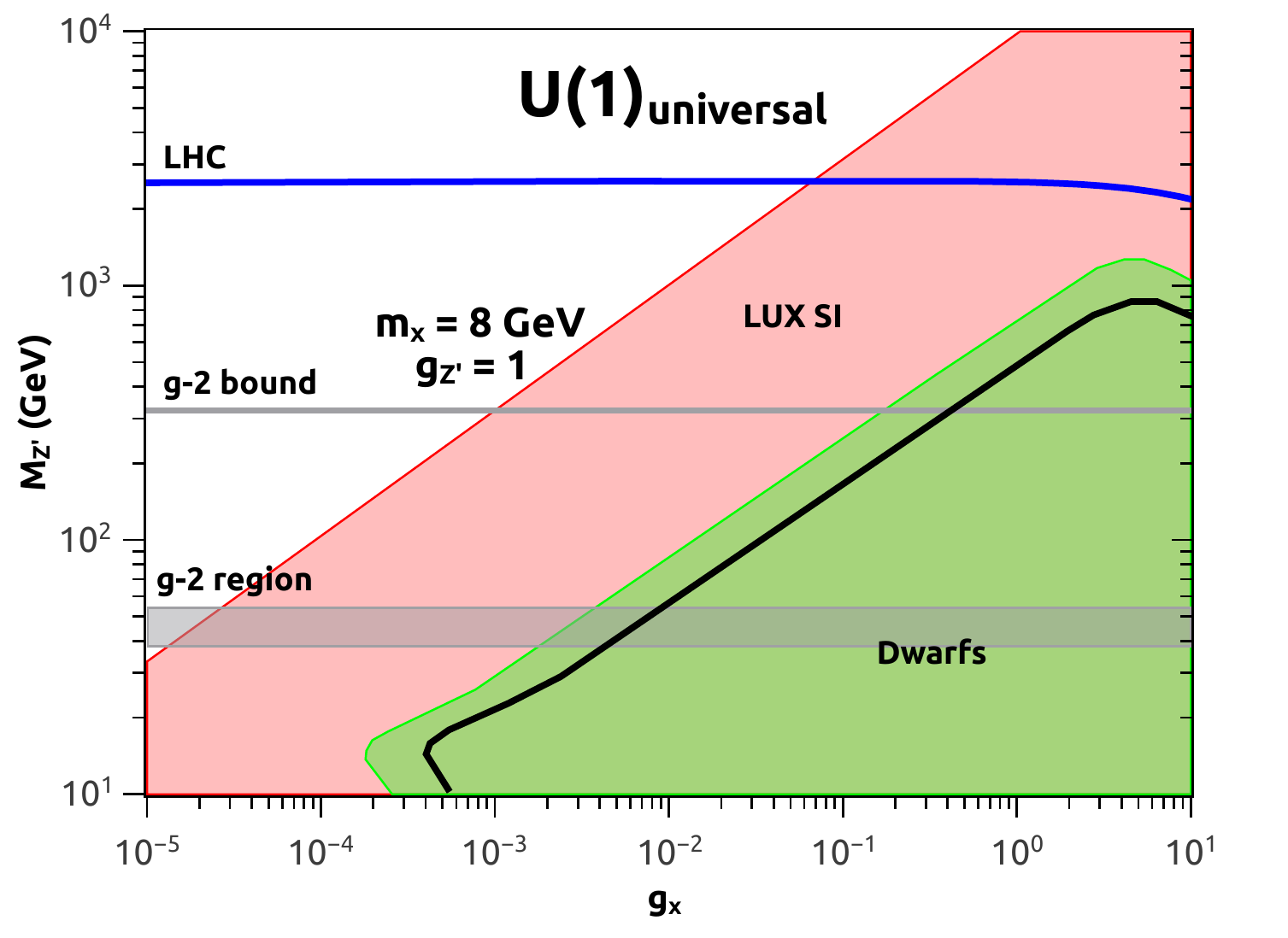}
\includegraphics[scale=0.45]{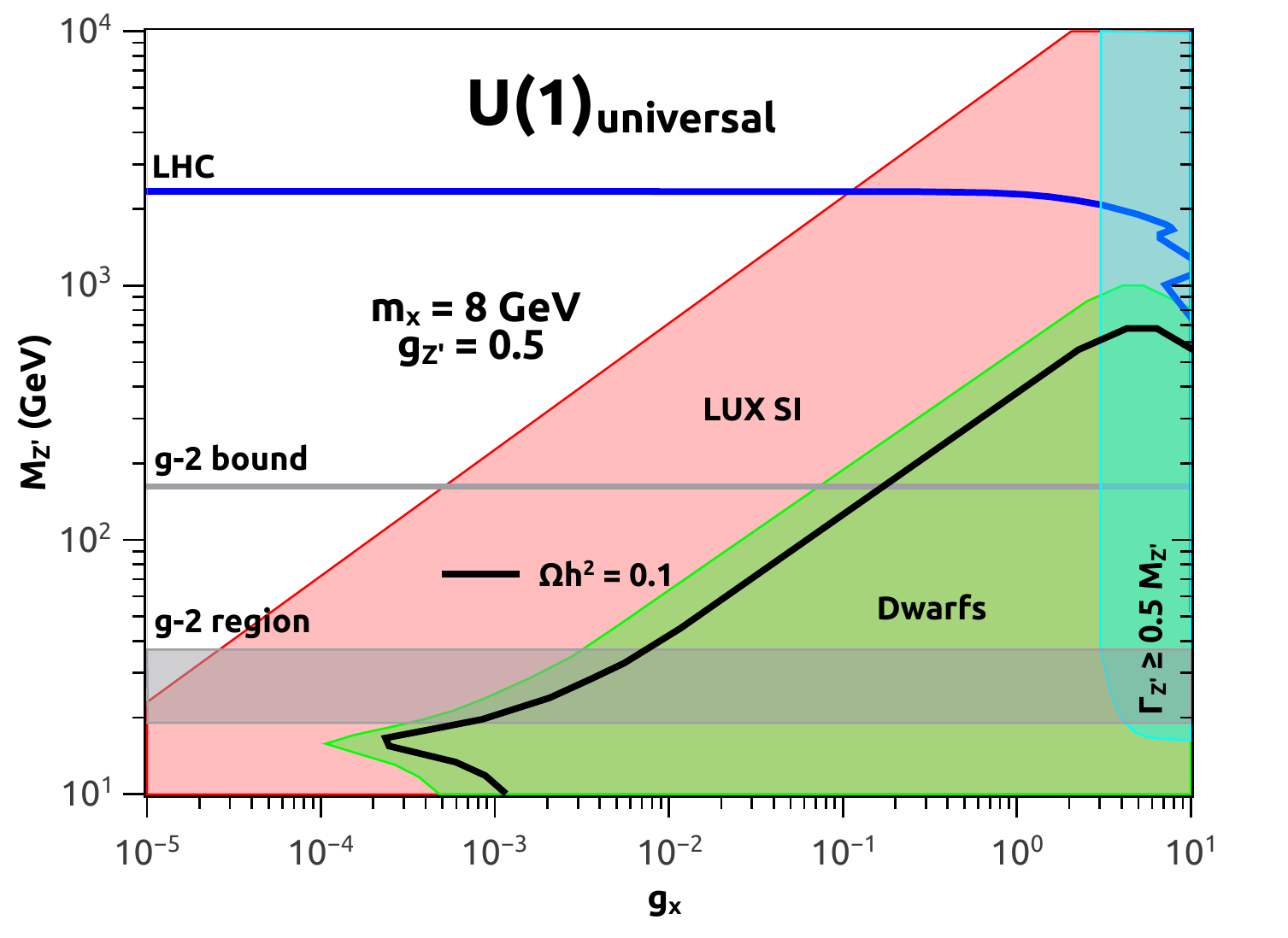}
\label{fig:results}
\caption{ Results for the $U(1)_{\rm universal}$ model. {\it Left:}  $m_{\chi}=8$~GeV, $g_{Z^{\prime}}=1$; {\it Right:}  $m_{\chi}=8$~GeV, $g_{Z^{\prime}}=0.5$. The blue horizontal line indicates the ATLAS LHC bound: everything below the curve is ruled out. The gray horizontal line is the $1\sigma$ bound from the muon magnetic moment by forcing the correction to lie within the error bar. The thick gray band delimits the $Z^{\prime}$ that accommodates the muon anomalous magnetic moment excess. The red (green) regions are ruled out by the LUX spin-independent direct detection results, while the green region is ruled out by Fermi-LAT dwarfs PASS8 limits. The black curve indicates the region of parameter space that features a DM relic density matching the universal DM abundance. The cyan shaded region corresponds to the violation of the perturbative limit, $\Gamma_{\zp} \gtrsim M_{\zp}/2$. In the left graph ($g_{\zp}=1$), the whole parameter space violates this pertubative limit, therefore the dilepton limit does not apply since it is based on the narrow width approach.}
\end{figure}
\begin{figure}[!t]
\centering
\includegraphics[scale=0.45]{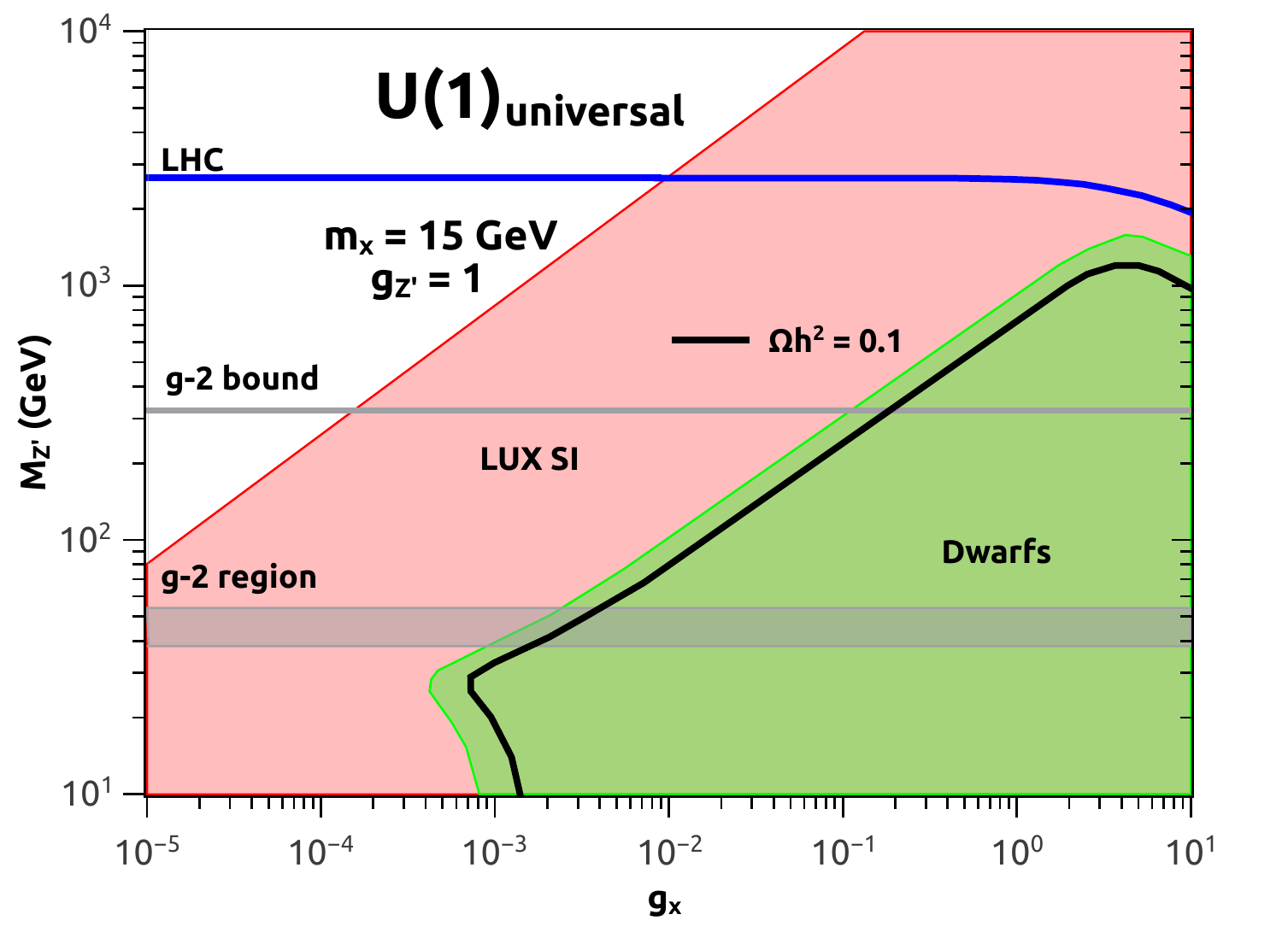}
\includegraphics[scale=0.45]{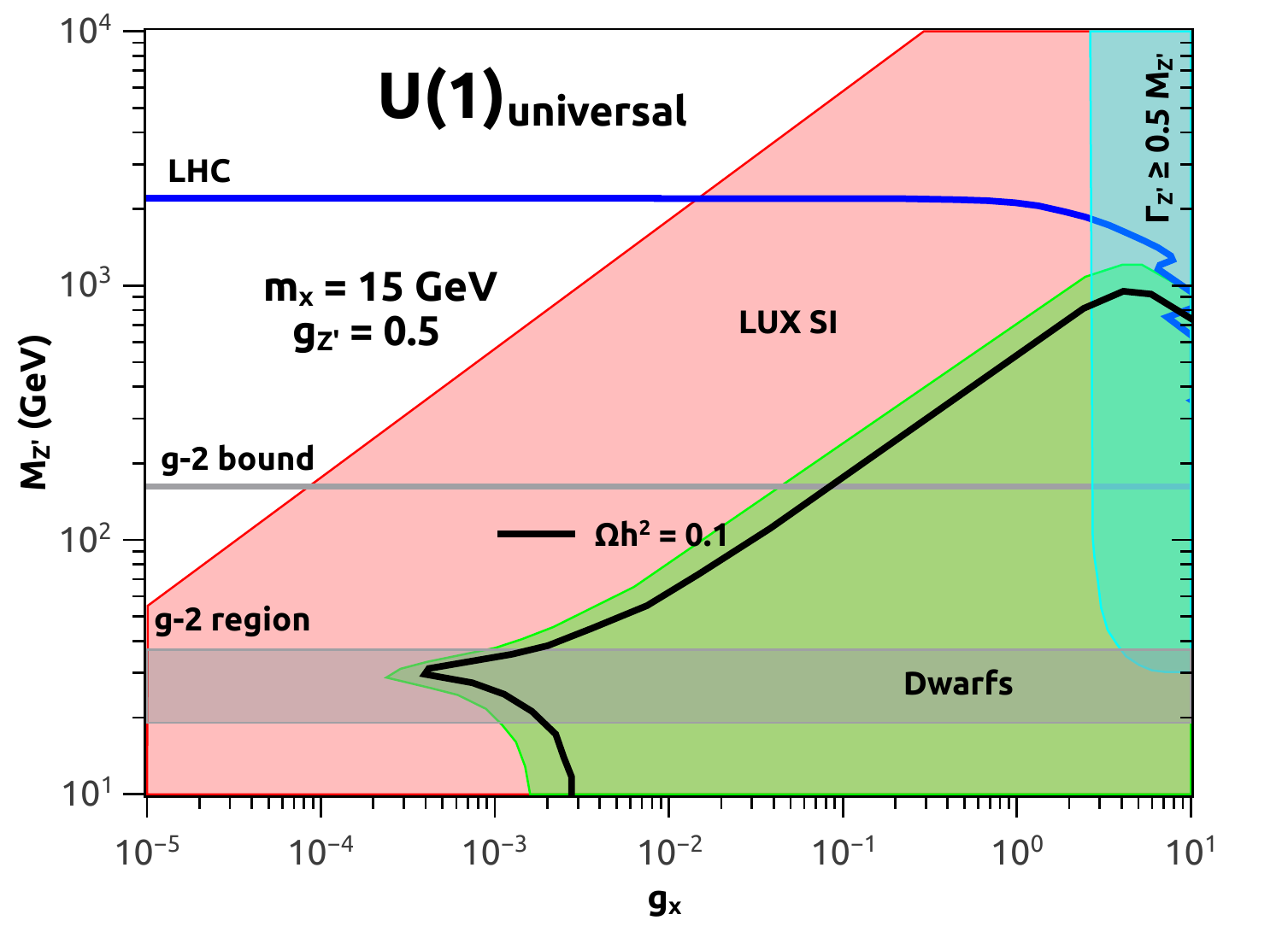}
\caption{Same as in Fig.~\ref{fig:results}, for the $U(1)_{\rm universal}$ model. {\it Left:}  $m_{\chi}=15$~GeV, $g_{Z^{\prime}}=1$; {\it Right:}  $m_{\chi}=15$~GeV, $g_{Z^{\prime}}=0.5$.}
\end{figure}
\begin{figure}[!t]
\includegraphics[scale=0.5]{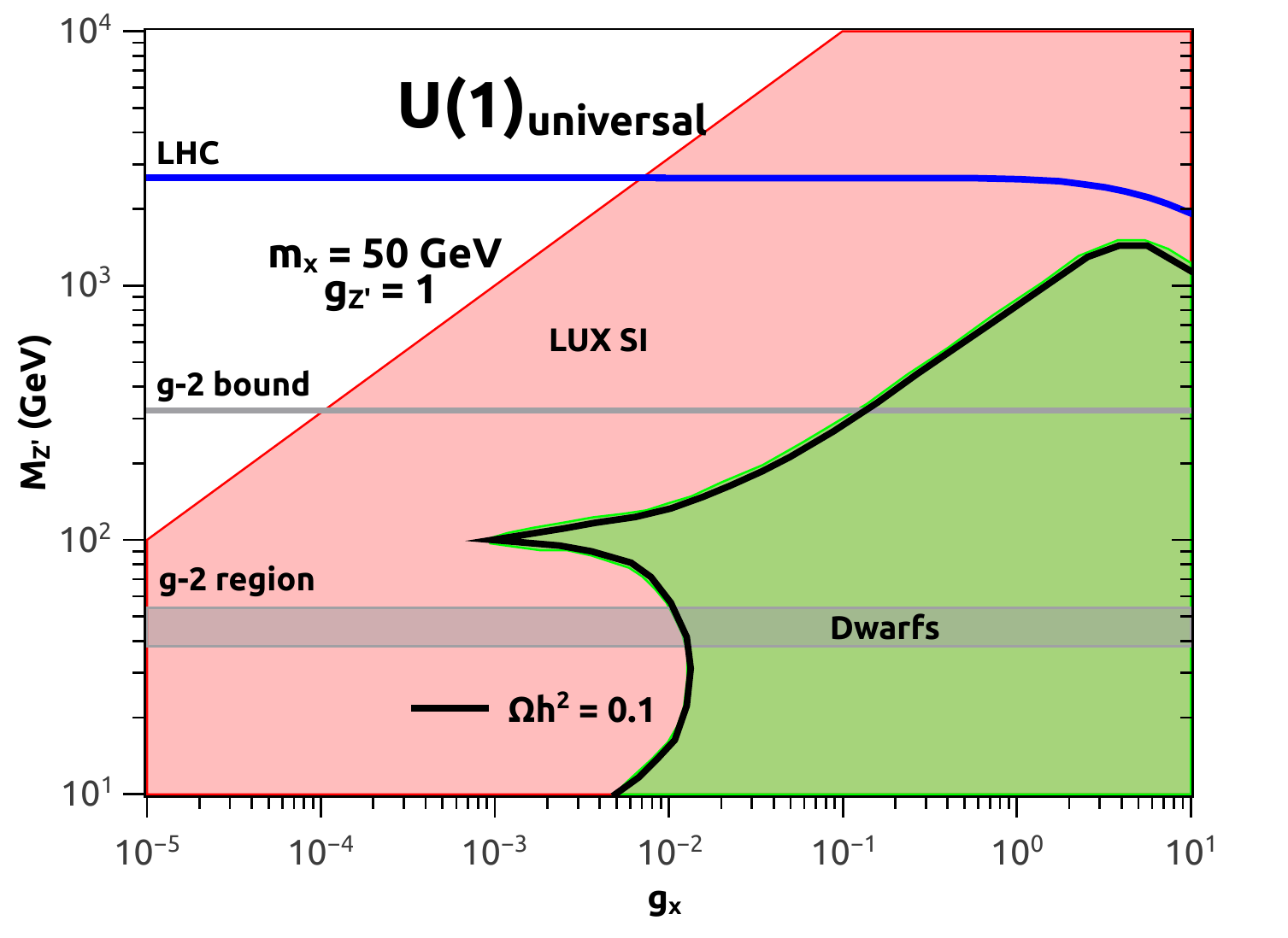}
\includegraphics[scale=0.5]{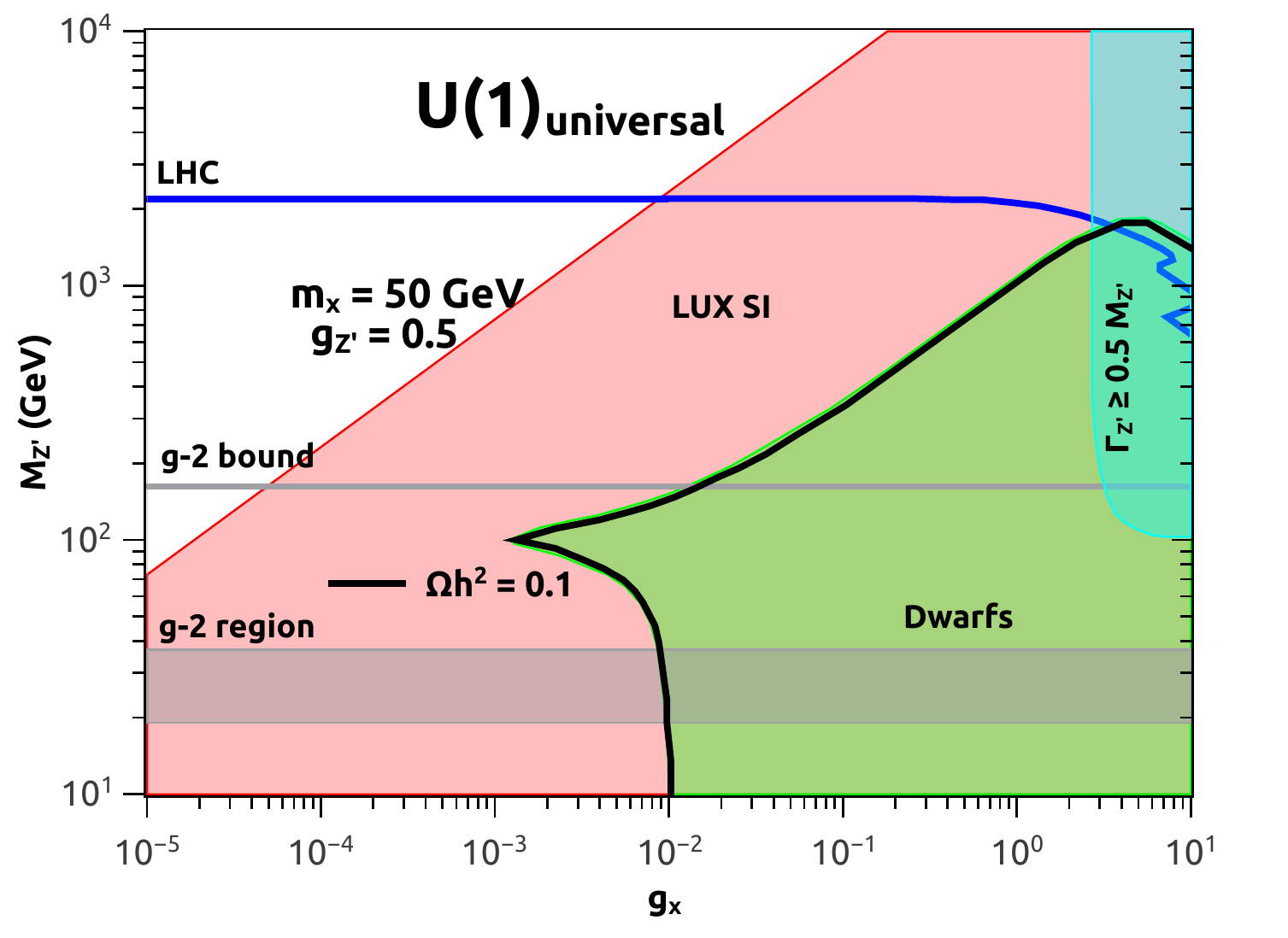}
\caption{Same as in Fig.~\ref{fig:results}, for the $U(1)_{\rm universal}$ model. {\it Left:}  $m_{\chi}=50$~GeV, $g_{Z^{\prime}}=1$; {\it Right:}  $m_{\chi}=50$~GeV, $g_{Z^{\prime}}=0.5$.}
\end{figure}
\begin{figure}[!t]
\includegraphics[scale=0.5]{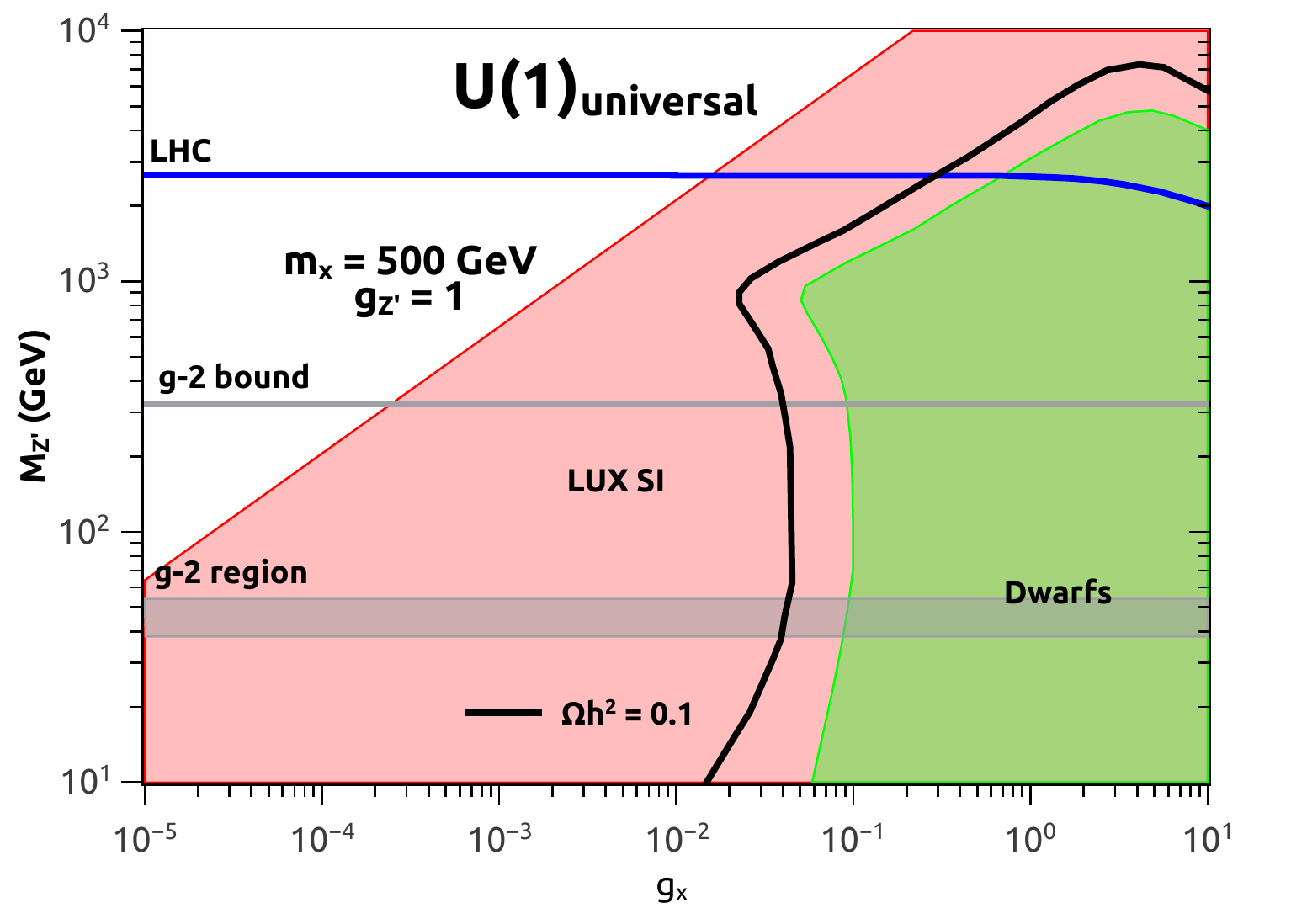}
\includegraphics[scale=0.5]{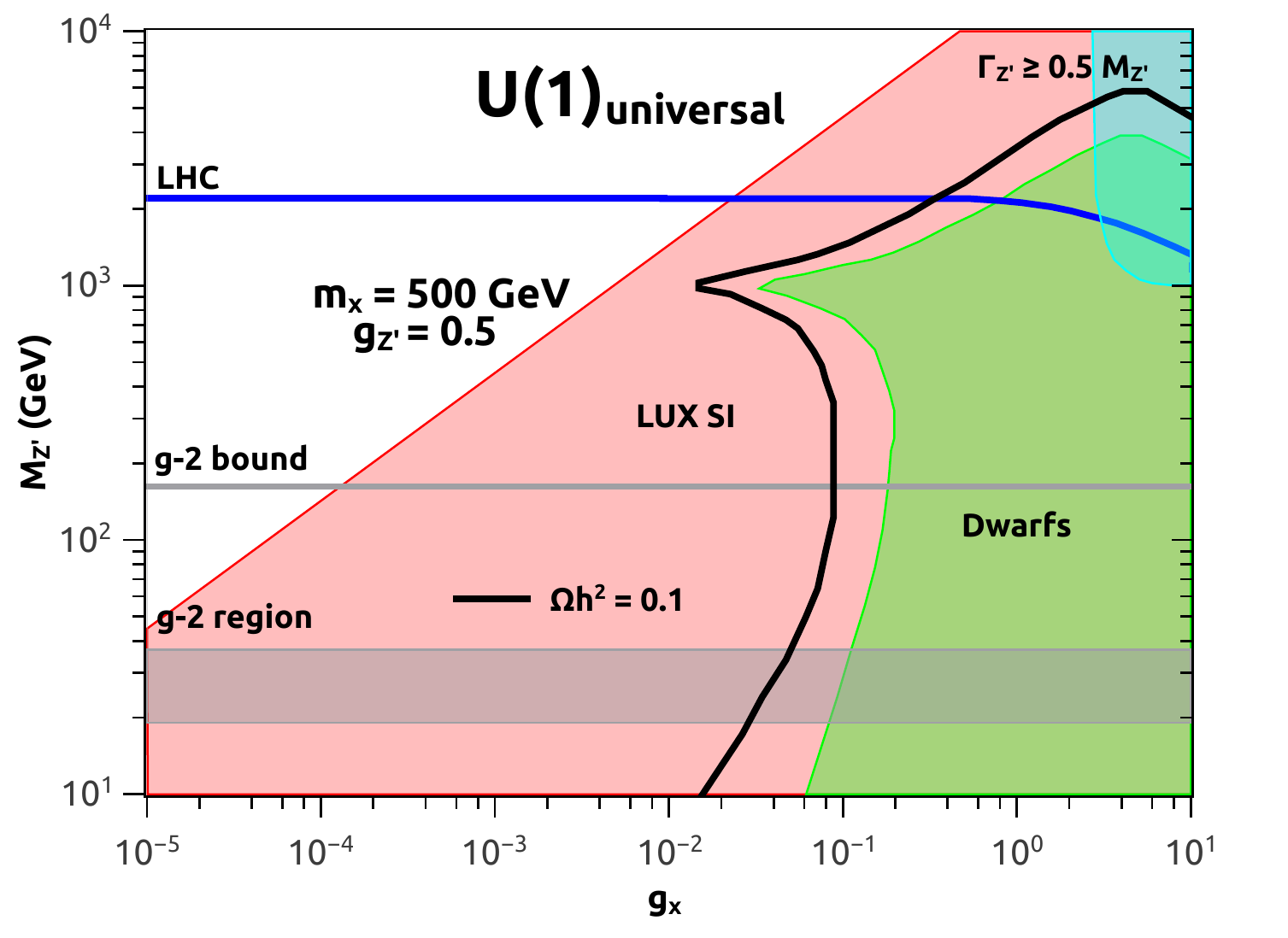}
\caption{Same as in Fig.~\ref{fig:results}, for the $U(1)_{\rm universal}$ model. {\it Left:}  $m_{\chi}=500$~GeV, $g_{Z^{\prime}}=1$; {\it Right:}  $m_{\chi}=500$~GeV, $g_{Z^{\prime}}=0.5$.}
\end{figure}
\begin{figure}[!t]
\includegraphics[scale=0.5]{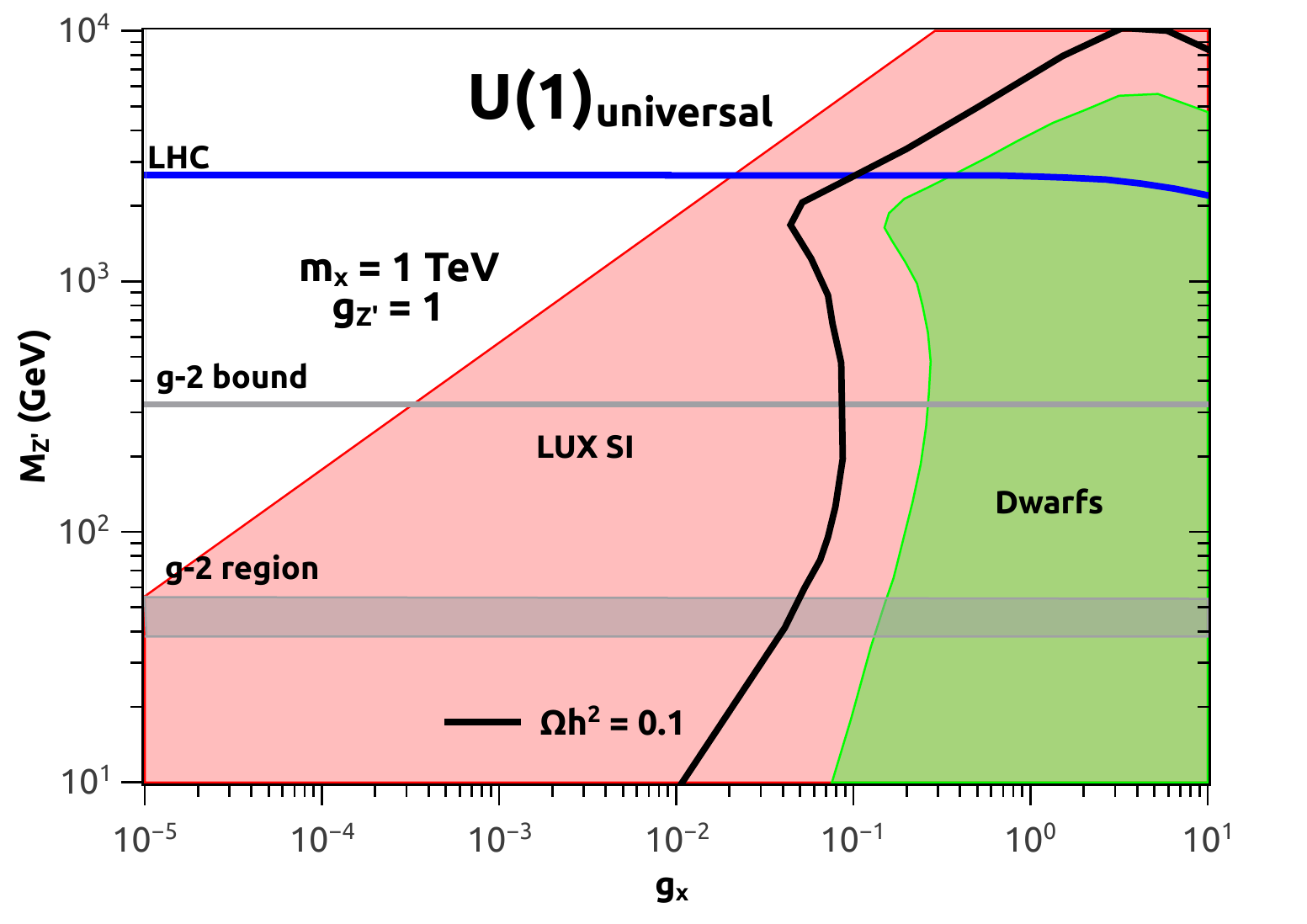}
\includegraphics[scale=0.5]{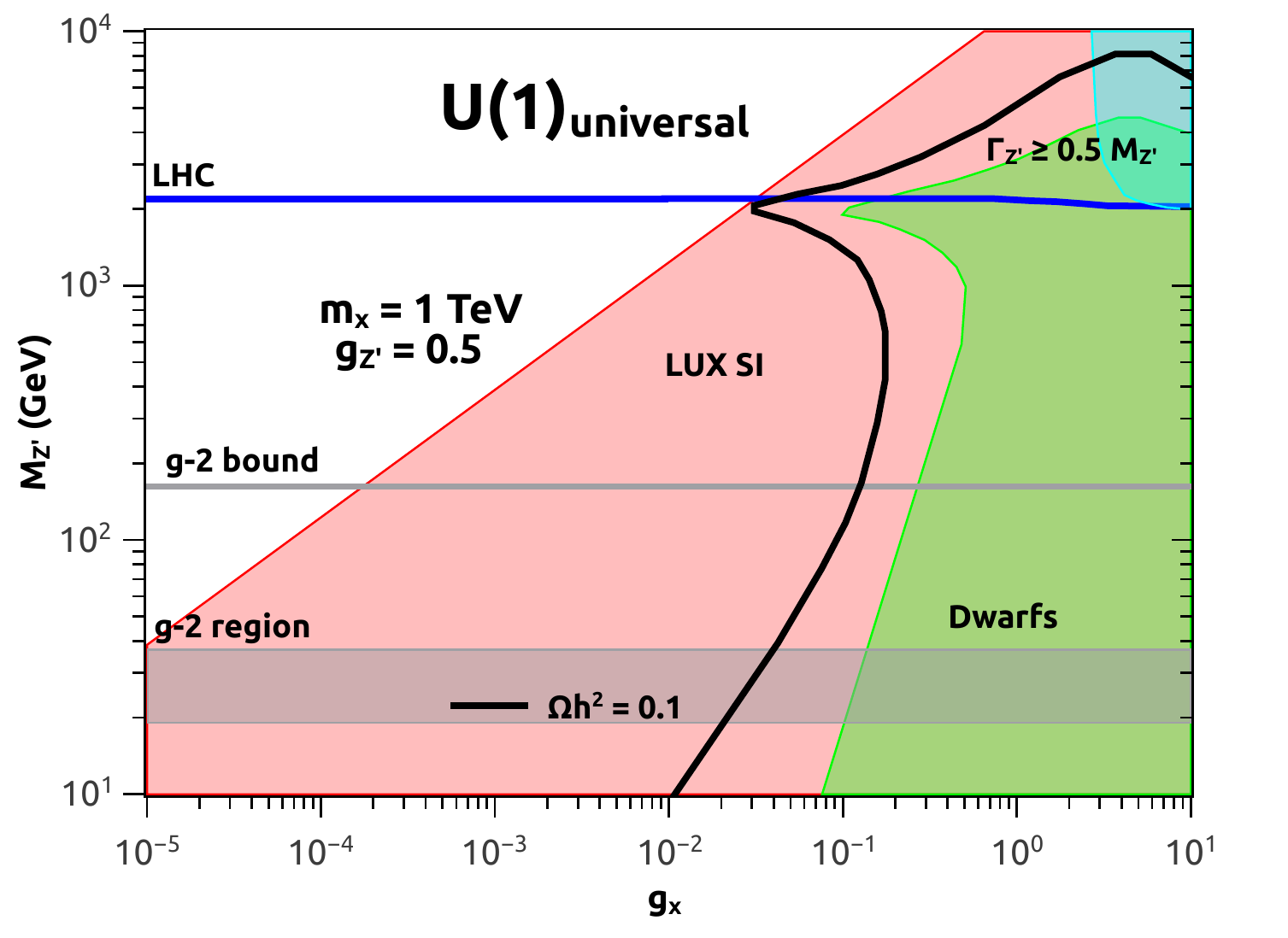}
\caption{Same as in Fig.~\ref{fig:results}, for the $U(1)_{\rm universal}$ model. {\it Left:} $m_{\chi}=1$~TeV, $g_{Z^{\prime}}=1$; {\it Right:}  $m_{\chi}=1$~TeV, $g_{Z^{\prime}}=0.5$.}
\end{figure}
\begin{figure}[!th]
\includegraphics[scale=0.5]{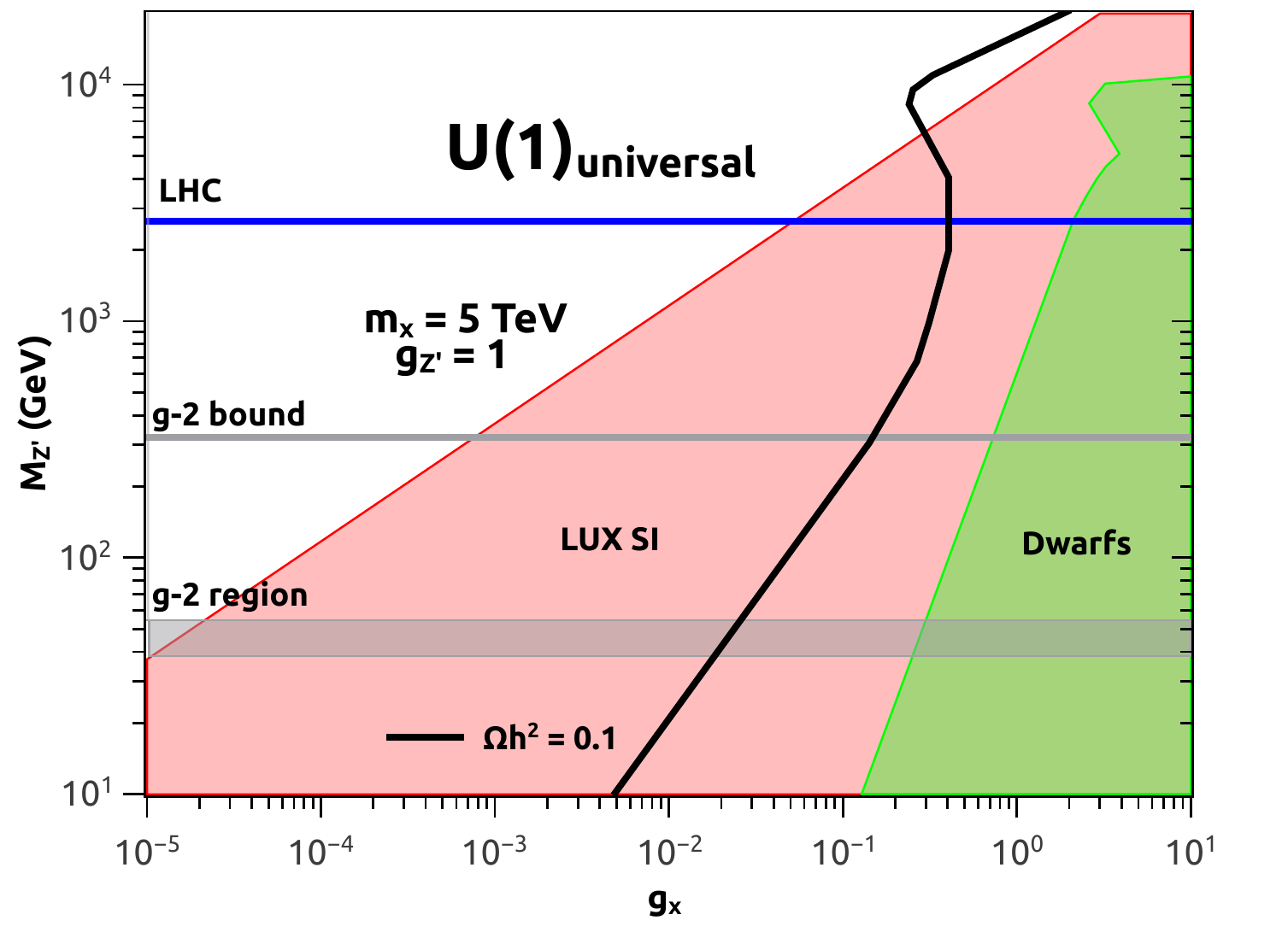}
\includegraphics[scale=0.5]{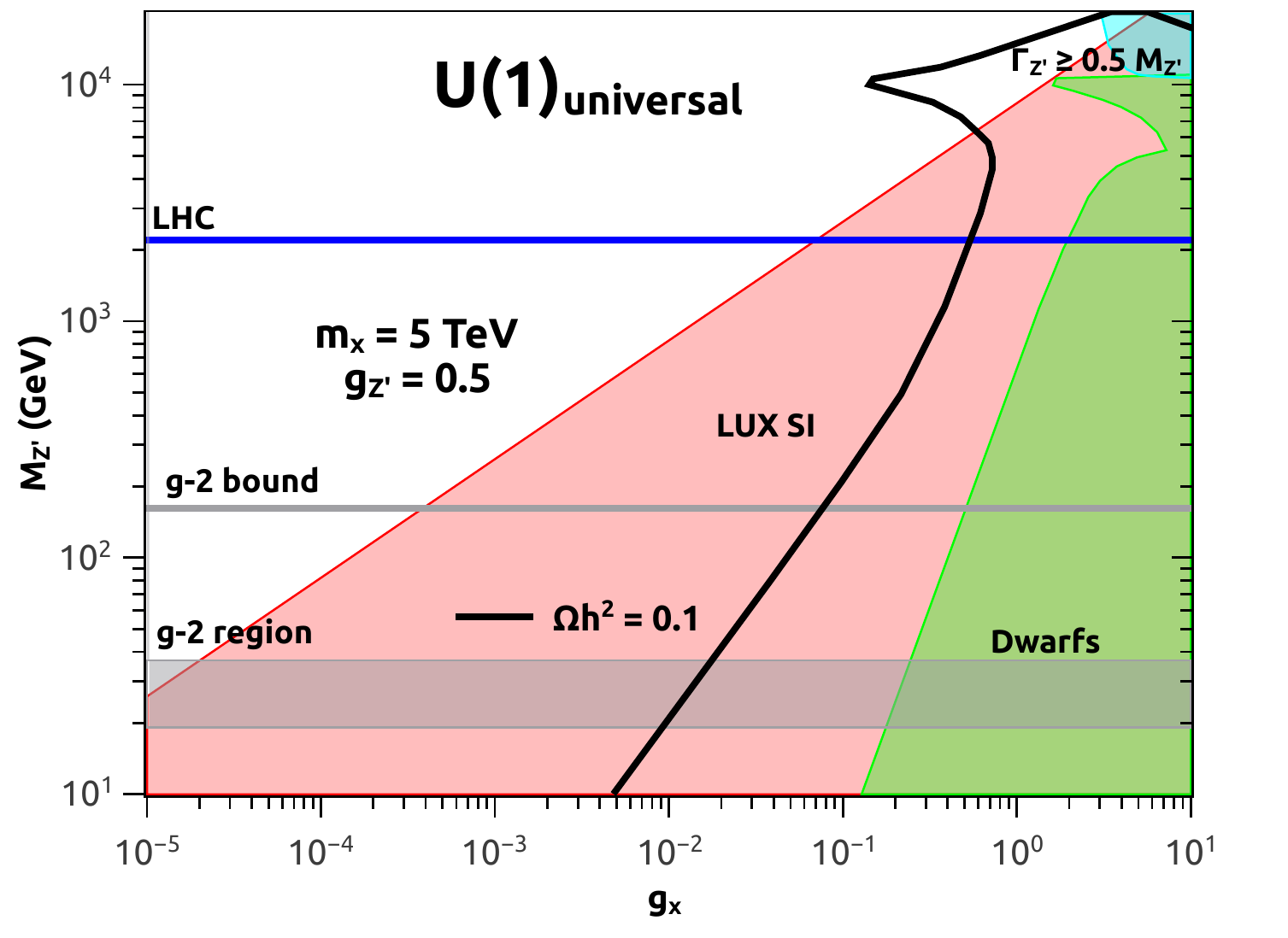}
\caption{Same as in Fig.~\ref{fig:results}, for the $U(1)_{\rm universal}$ model. {\it Left:} 5 TeV WIMP, $g_{Z^{\prime}}=1$; {\it Right:} 5 TeV WIMP, $g_{Z^{\prime}}=0.5$. Notice that a DM mass above a few TeV is required to be consistent with current limits and reproduce $\Omega h^2= 0.1$}
\end{figure}

We have outlined a comprehensive study of four different $U(1)_X$ realizations, in the context of new physics contributions to the muon magnetic moment, collider searches, and direct and indirect dark matter searches. In addition to considering different, theoretically well motivated fermion charge assignments, we focused on two values for the $g_{Z^{\prime}}$ coupling, 1 and 0.5.
The muon magnetic moment and collider limits are summarized, respectively, in  Tables~\ref{g2bounds} and \ref{colliderbounds}. The collider bounds were derived using dilepton search results from ATLAS, Ref.~\cite{Aad:2014cka}. We emphasize that the limits from LEPII derived in \cite{Carena:2004xs} are not applicable here due to the existence of dark matter.

As for the muon magnetic moment bounds, we utilized the public code described in  \cite{Queiroz:2014zfa}. For the direct and indirect detection limits we compared our analytical results with micromegas \cite{Belanger:2006is} concerning the thermal relic abundance calculation and utilized the {\tt PPPC4DMID} code \cite{Cirelli:2010xx} to calculate the gamma-ray spectra.
 
In this section we present a summary of our results, combining all constraints outlined above, for a variety of dark matter particle masses, and for all the charge assignments described, on the plane defined by the $\zp$ mass and the DM-$\zp$ coupling $g_\chi$. Each plot focuses on a particular value for $g_{\zp}=0.5,\ 1$ and $\mchi=8,\ 15,\ 50,\ 500,\ 1000$ and 5000 GeV, and a selected charge assignment structure. The broad range of DM masses allows us to highlight the importance and complementarity across different search strategies.

{\it General comment \# 1:} We employ the same color scheme for all figures: The red (purple) regions indicate the parameter space ruled out by LUX SI (XENON SD) direct dark matter searches (Sec.~\ref{section: directdetection}). The black curve indicates the region of the parameter space where the thermal relic density of $\chi$ matches the observed dark matter abundance, while the green shaded region is ruled out by gamma-ray observations of Dwarf satellites (Sec.~\ref{section: indirect}). The horizontal gray line represents the muon magnetic moment limit (Sec.~\ref{sec:gmu}), whereas the gray band delimits the region of parameter space that accommodates the g-2 excess. In all cases, the region compatible with a new physics interpretation of the anomalous magnetic dipole moment of the muon is ruled out by other other searches.

{\it General comment \# 2:} The blue curves indicate the minimal $\mzp$ compatible with null ATLAS searches in the dilepton channel (Sec.~\ref{sec:colliders}). The cyan region represent the parameter space where the total width of the $\zp$ exceeds the perturbative limit, $\Gamma_{\zp} > M_{\zp}/2$, invaliding the dilepton bounds which are based on the narrow width approximation. However, one should notice that the direct detection limits are most relevant in this regime leaving our general conclusions intact. In  particular, this perturbative bound on the total width of the $\zp$ is exceeded for all cases where we consider the $U(1)_{\rm universal}$ model with $g_{\zp}=1$. However, for completeness, we choose to still display these results.

{\it General comment \# 3:} Concerning the collider bounds, one can see from Table 1, that some gauge structures induce stronger $\zp$-leptons couplings, and consequently larger branching ratios to charged leptons, such as the $B-L$ model. Moreover, gauge structures such as $U(1)_{\rm universal}$ lead to sizable $\zp$-quarks couplings and thus large production rates. The balance between production rate and branching ratio to charged leptons sets the number of dilepton events at each invariant mass bin, and therefore our limits. 

{\it General comment \# 4:} We notice that in the absence of vector couplings to quarks, RG running from the $Z^\prime$ mass scale down to nucleon energy scales effectively induces spin-independent interactions, as pointed out and investigated in Ref.~\cite{D'Eramo:2014aba}. The resulting limits, which we do not compute here, might be comparable or even stronger than those from spin-dependent searches.

We now summarize the results presented in Figs.4-27:

\begin{itemize}

\item $U(1)_{\rm universal}$: In Figs.~4-9 we show the exclusion region for several dark matter masses and conclude that only for DM masses larger than $\sim1$ TeV can one obtain models with the right relic abundance which evade both direct detection and collider constraints. For $g_{\zp}=1 (0.5)$, collider bounds roughly require $\mzp \gtrsim 2570 (2200)$~GeV. This excludes this setup from providing an explanation to the ATLAS diboson excess \cite{Aad:2015owa}, since a large $g_{\zp}$ is needed to produce a large enough number of events \cite{Hisano:2015gna}. Note that these limits are marginally sensitive to the magnitude of the $\zp$-dark matter coupling ($g_{\chi}$). When $g_{\chi}$ becomes sufficiently large the limits weaken because the DM branching fraction increases and the charged lepton one decreases. Keeping $g_{\zp}=1$, in particular we find
$\mzp \gtrsim 2400$~GeV, $g_{\chi} \sim 1$. A summary of our limits is provided in Table~\ref{colliderbounds}. 

In this model the correction to g-2 has the right sign, however the region of parameter space that accommodated the g-2 excess is ruled out as one can see. For $g_{\chi} \sim 10^{-2}-10^{-1}$ the direct detection limits the most constraining ones, whereas the indirect detection bounds are always weaker. In summary, DM masses above few TeV are required to match the universal DM abundance and obey the existing bounds.

\item $U(1)_{\rm B-L}$: The combination of g-2, collider, direct a indirect dark matter detection limits are depicted in Figs.~10-15. The collider limits result in $\mzp  \gtrsim 3$~TeV for $g_{\zp}=1$ and $\mzp \gtrsim 2570$~GeV for $g_{\zp}=0.5$. Again, this is incompatible with the ATLAS diboson excess. See Table~\ref{colliderbounds} for different coupling choices. Similarly to the $U(1)_{\rm universal}$ model, in principle it could accommodate the g-2 excess, but the region of parameter space is excluded by collider data, and DM masses larger than few TeV are needed to reproduce $\Omega h^2=0.1$ while obeying the current limits.

\item $U(1)_{10+\bar{5}}$: We present our findings for this gauge model in Figs.~16-21. 
Keeping $g_{\chi} \sim 0.1$, the collider bounds are in the ballpark of $\mzp \gtrsim 2.6$~TeV for $g_{\zp}=1$ and $\mzp \gtrsim 2480$~GeV for $g_{\zp}=0.5$. For $g_{\chi} \sim 1$ those limits drop to $\mzp \gtrsim 2400$~GeV and $\mzp \gtrsim 2040$~GeV respectively. In this model the usual leading contribution to the SI cross section vanishes at tree level, since the vector couplings to valence quarks vanishes. However, spin-dependent limits from XENON100 are still relevant, and are complementary to the Fermi Dwarf limits for certain DM masses, although generically weaker. As mentioned in comment \# 4 above, loop effects might produce sizable spin-independent scattering as well. It is clear from Fig.19 that one can have a Dirac dark matter WIMP with mass of 500 GeV or larger in this model.

\item $U(1)_{d-u}$: Our results are summarized in Figs.~22-27 and Table 3. The importance of taking a dark matter complementarity approach is clear in this model, since for $g_{\chi} < 1$, the dilepton limits are the strongest, whereas for $g_{\chi} > 1$, the direct and indirect detection bounds are the leading ones. The dilepton limits are again very stringent, excluding $\zp$ masses below $2640 (2430)$~GeV for $g_{\zp} = 1 (0.5)$.  For $g_{\zp}$ of order of unit those limits drop to $2430 (1800)$ ~GeV respectively. In this model, similarly to the previous case, for masses of 500 GeV one can accommodate a Dirac fermion as DM \footnote{We point out our results only apply for Dirac fermions which were thermally produced in the early universe. For recent discussion concerning non-thermal DM production see Ref.~\cite{Hooper:2011aj,Allahverdi:2012wb,Allahverdi:2014bva,Allahverdi:2013noa,
Anandakrishnan:2013tqa,Baer:2014eja,Kelso:2013nwa,Kelso:2013paa,Dev:2013yza}.}.

\section{Diboson Excess}

We show in this section that it is possible to accommodate the recent ATLAS  diboson excesses \cite{Aad:2015owa} within the $U(1)_{d-u}$ model discussed above. The diboson excesses in the WZ,WW and ZZ channels reported by the ATLAS Collaboration are well fitted by resonances whose peaks are in the ballpark of 2 TeV, potentially implying the detection of a new  particle. Since the tagging selections for each mode used in the analysis are relatively poor at this stage ($\sim 20\%$) it is hard to conclude that one resonance is responsible for the excesses in all channels. Indeed, it is possible that a single 2-TeV particle such as a $\zp$ might account for only one part of the channels, and that the peaks in the other channels are contaminations due to  incomplete tagging selections, see e.g. \cite{Hisano:2015gna}. 

In the context of $U(1)_{d-u}$ models, we find that we can accommodate the excess while reproducing the right thermal relic abundance for the Dirac fermion dark matter candidate we discuss here, avoiding at the same time dilepton, direct and indirect detection bounds. We incorporated a $\zp W\,W$ interaction in the $U(1)_{d-u}$ setup, proportional to $g_{\zp}\theta$, where $\theta$ is the $Z-\zp$ mixing angle, bounded to be smaller than $10^{-3}$ \cite{Carena:2004xs} due to constraints from SM $Z$ properties. We then computed the relic abundance, dilepton bounds and the $pp \rightarrow \zp \rightarrow W\,W$ production cross section, finding that for  $M_{\zp} \sim 2$~TeV, $g_{\zp}=1$; $g_{\chi} \sim 1$, $m_{\chi} =500$~GeV; $\theta=10^{-3}$, we can reproduce the right abundance ($\Omega h^2 \sim 0.1$) and obtain $\sigma (pp \rightarrow \zp \rightarrow W\,W) \sim 30$~fb, which accommodates the excess. 

We note that we employed $m_{\chi}=500$~GeV because for lighter masses we cannot simultaneously get the right abundance and avoid dileptons exclusion limits. Furthermore, for $m_{\chi} > M_{\zp}$ the branching ratio into $WW$ is sizable, inducing a $WW$ production cross section too large to reproduce the excess. However, we find that it is possible to accommodate heavier dark matter, with masses up to $\sim 1$ TeV by adjusting the production cross section of a 2 TeV $\zp$ in order to explain the diboson excess, and at the same time we respect all the other bounds considered in this work, since the inclusion of a $WW$ decay channel weakens the dilepton collider bounds, thus opening a broader region of the allowed parameters space which could explain the ATLAS excess.


\end{itemize}


\section{Conclusions}\label{sec:conclusions}

In this work we studied several anomaly-free $U(1)$ gauge structures that arise in several, well-motivated extensions of the standard model, namely  $U(1)_{B-L}, U(1)_{d-u}, U(1)_{universal}$, and $U(1)_{10+\bar{5}}$. By postulating a Dirac fermion $\chi$ coupled to the $\zp$ as the dark matter particle candidate, we computed the thermal relic abundance, annihilation cross section in the low velocity limit, and spin-dependent and -independent scattering cross sections off of nuclei. We then used the current bounds from dwarfs observations using Fermi-LAT data and from null results reported by LUX and XENON to set constraints on the $\mzp$ {\it versus} $g_{\chi}$ parameter space, where $g_{\chi}$ is the $\zp$-DM coupling. 

We derived dilepton limits from null searches with ATLAS, which provide the strongest constraints for all those models for $g_{\chi} < 10^{-2}$, with direct detection limits leading for larger couplings. Dilepton constraints rule out $\zp$ masses up to $\sim 3$~TeV. In general, the collider limits depend quite sensitively on the assumed gauge charge structure, but not as much on the $g_{\chi}$ coupling. Albeit, the collider limits for the $U(1)_{d-u}$ gauge group do weaken significantly for $g_{\chi} \gtrsim 1$. 

In no case we find large enough contributions to the anomalous magnetic moment of the muon large enough to explain the observed anomaly compatible with collider constraints. typically  the combination of collider and direct detection limits forces the mass of the dark matter particle to values larger than a TeV, assuming thermal production in the early universe, unless very suppressed couplings to leptons are in place (as is the case for the $U(1)_{10+\bar{5}}$ model, for which the limits is around 0.5 TeV). Relaxing the thermal production requirement, but enforcing a thermal abundance at most as large as the 2$\sigma$ upper limit on the inferred universal DM density does not change this conclusion. Underabundant models generically prefer small values for $\mzp$ and large values for $g_\chi$, but are severely constrained by indirect detection constraints.

Finally, we addressed the question of whether the models under investigation could accommodate the recently reported ATLAS diboson excess; we found that for large enough $\zp$ couplings a $\zp$ mass on the order of 2 TeV is generically excluded for all but the $U(1)_{d-u}$ model by a combination of LHC and direct detection constraints. For the $U(1)_{d-u}$ model a 2 TeV $\zp$ and a thermal relic Dirac fermion dark matter particle with a mass of 500 GeV appears to be a possibility, although further investigation is needed to probe whether the required number of events could be produced in this scenario.

\section*{Acknowledgements}

A. Alves thanks Funda\c{c}\~ao de Amparo \`a Pesquisa de Estado de S\~ao Paulo (FAPESP), grant 2013/22079-8, and Conselho Nacional de Desenvolvimento Cientifico e Tecnologico (CNPq), grant 307098/2014-1. SP is partly supported by the US Department of Energy, Contract DE-SC0010107-001. AB is supported by the Kavli Institute for Cosmological Physics at the University of Chicago through grant NSF PHY-1125897. FSQ thanks FERMILAB, ARGONNE and North Western University for the hospitality, where this projected was initially carried out. The authors are greatly indebted to Dan Hooper for crucial discussions and comments.  The authors also thank Jim Cline, Francesco D'Eramo, Ian Low, Yann Mambrini, Kuver Sinha and Carlos Wagner for important discussions.

\begin{figure}[!t]
\includegraphics[scale=0.5]{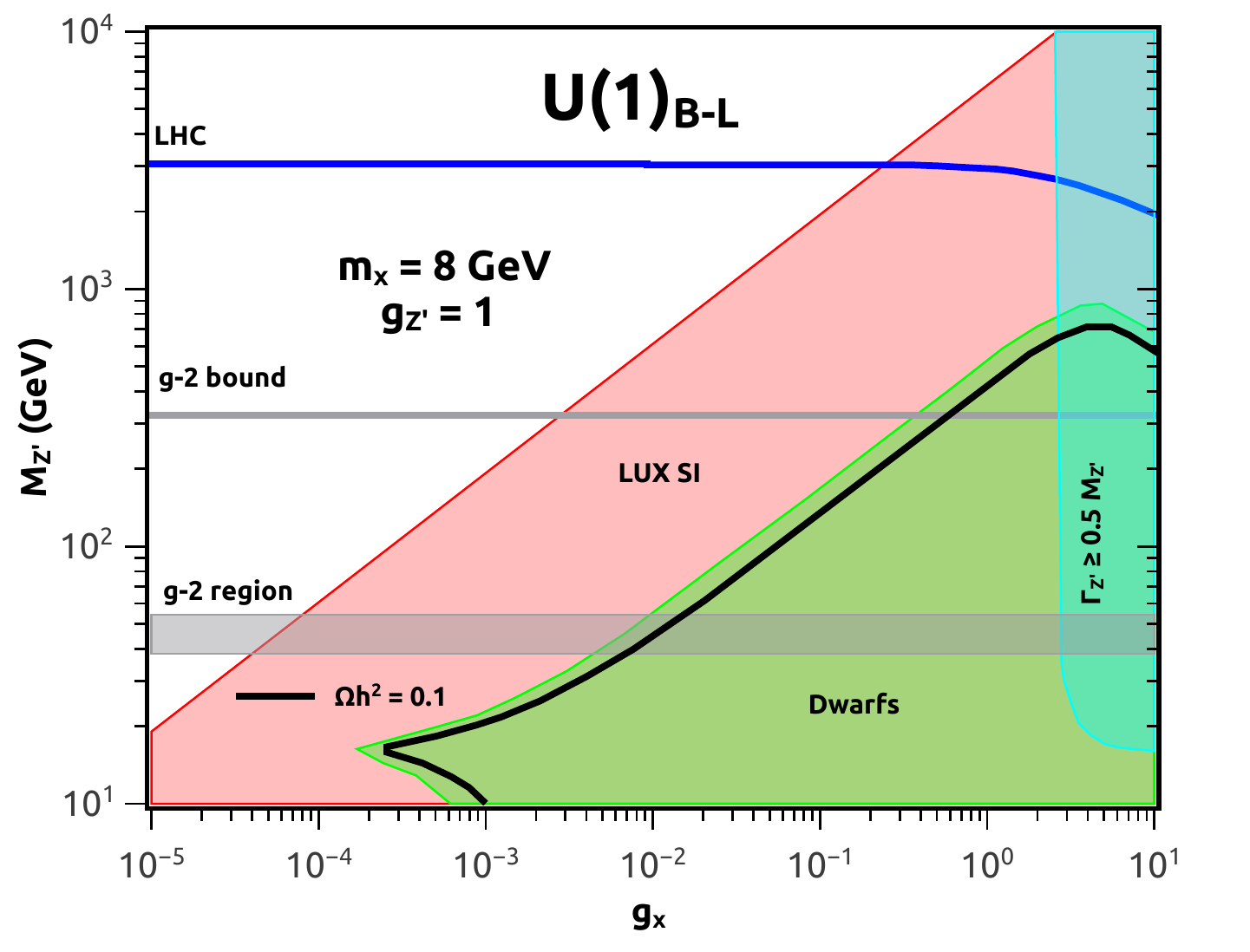}
\includegraphics[scale=0.5]{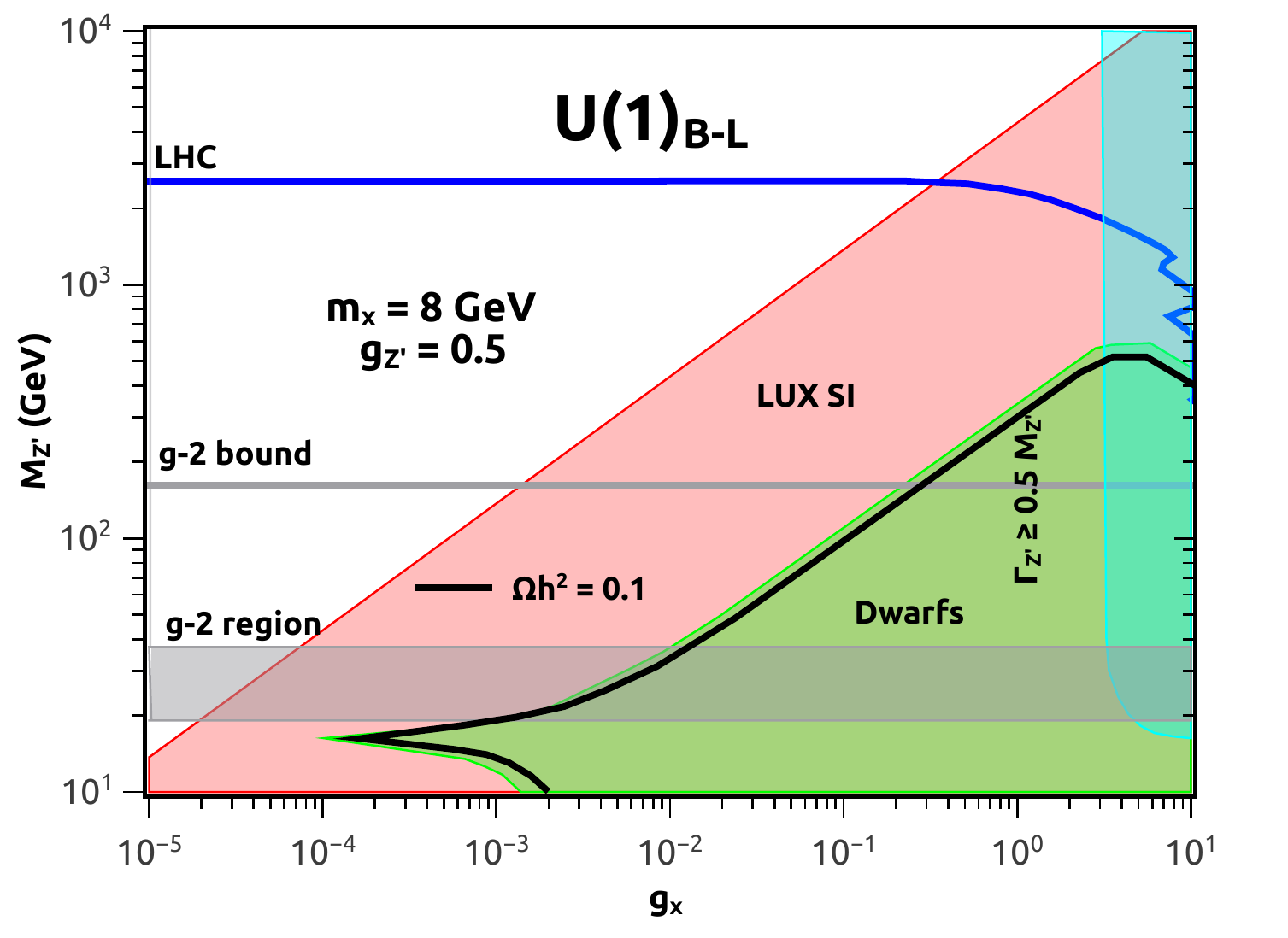}
\caption{Same as in Fig.~\ref{fig:results}, for the $U(1)_{\rm B-L}$ model. {\it Left:}  $m_{\chi}=8$~GeV, $g_{Z^{\prime}}=1$; {\it Right:}  $m_{\chi}=8$~GeV, $g_{Z^{\prime}}=0.5$. Blue horizontal line is LHC bound. Everything below the curve is ruled out. Gray horizontal line is the $1\sigma$ bound from the muon magnetic moment. In red (green) we exhibit the LUX spin-independent (Fermi Galactic Center) limit. The black curve sets region of parameter space that reproduced the right abundance. The cyan shaded region corresponds to the violation of the perturbative limit, $\Gamma_{\zp} \gtrsim M_{\zp}/2$.}
\end{figure}
\begin{figure}[!th]
\includegraphics[scale=0.5]{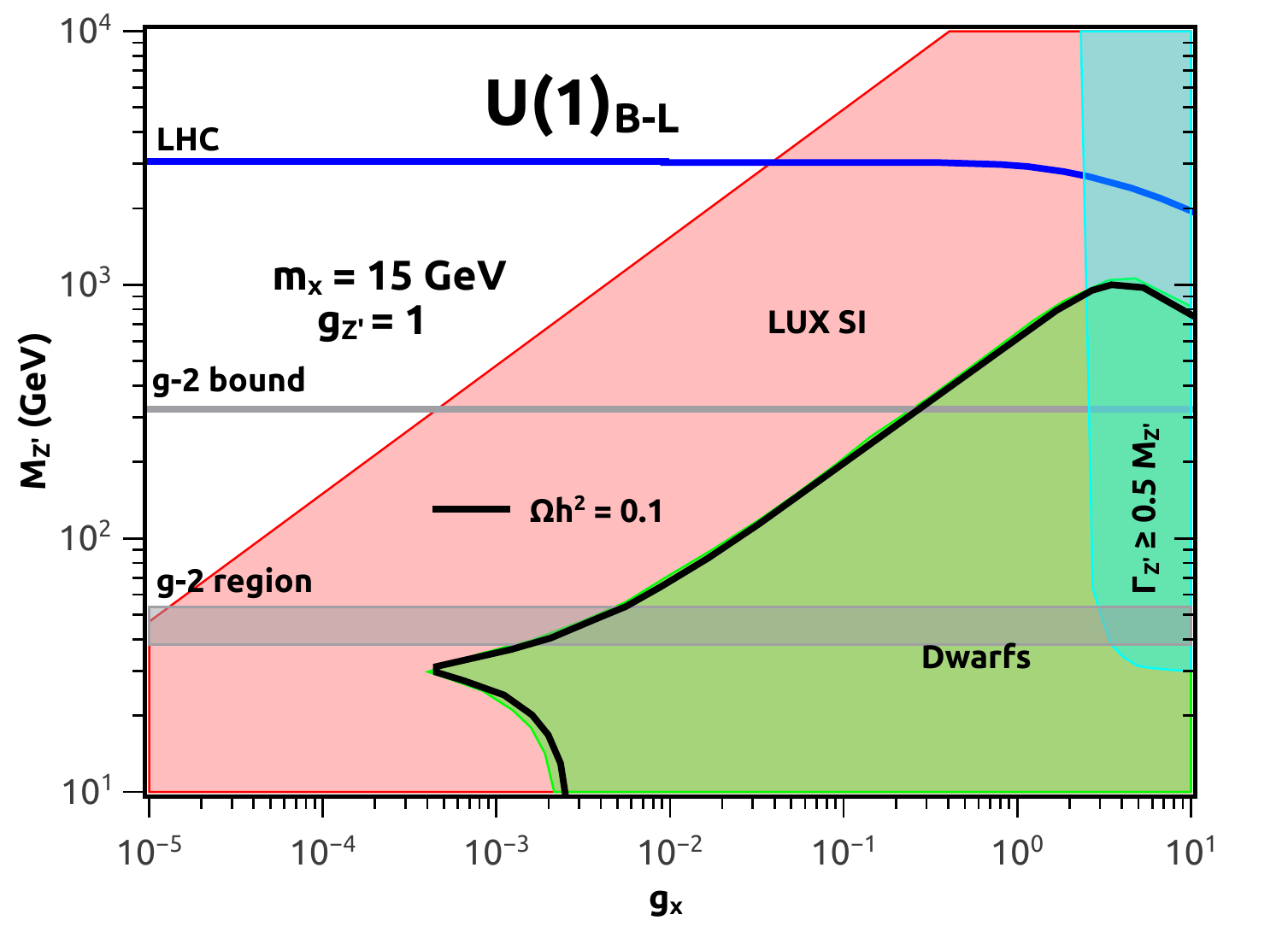}
\includegraphics[scale=0.5]{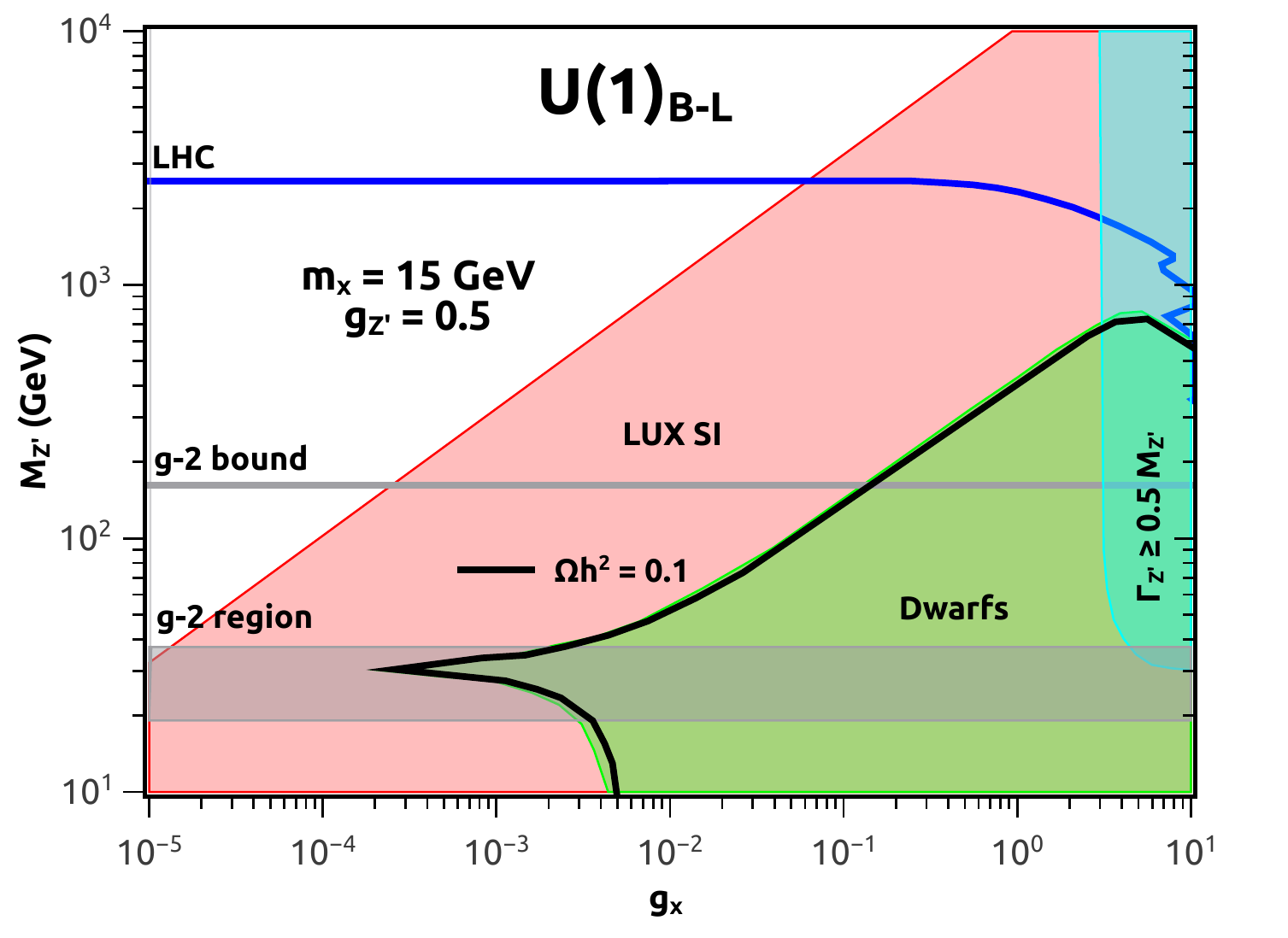}
\caption{Same as in Fig.~\ref{fig:results}, for the $U(1)_{\rm B-L}$ model. {\it Left:}  $m_{\chi}=15$~GeV, $g_{Z^{\prime}}=1$; {\it Right:}  $m_{\chi}=15$~GeV, $g_{Z^{\prime}}=0.5$.}
\end{figure}

\begin{figure}[!th]
\includegraphics[scale=0.5]{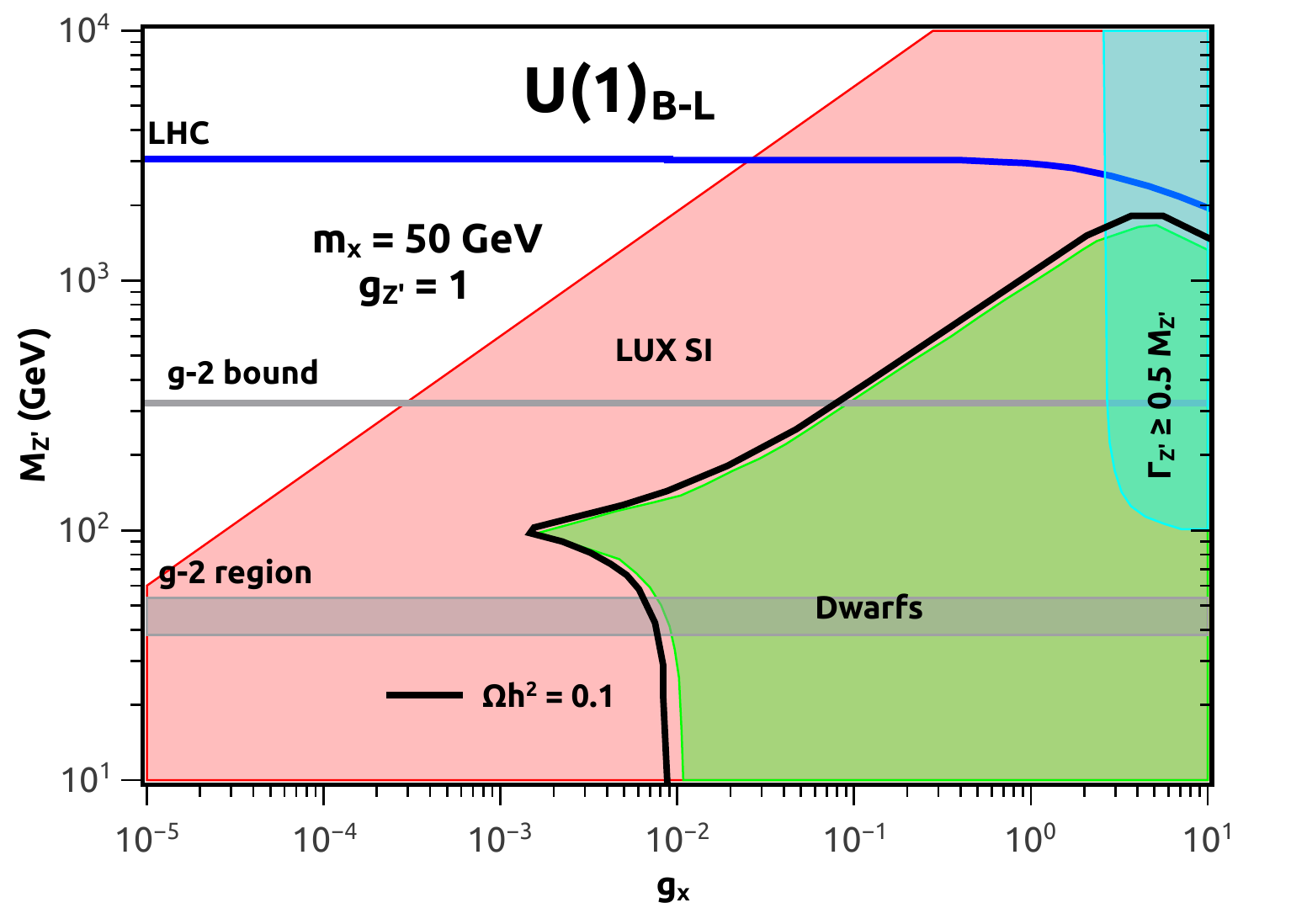}
\includegraphics[scale=0.5]{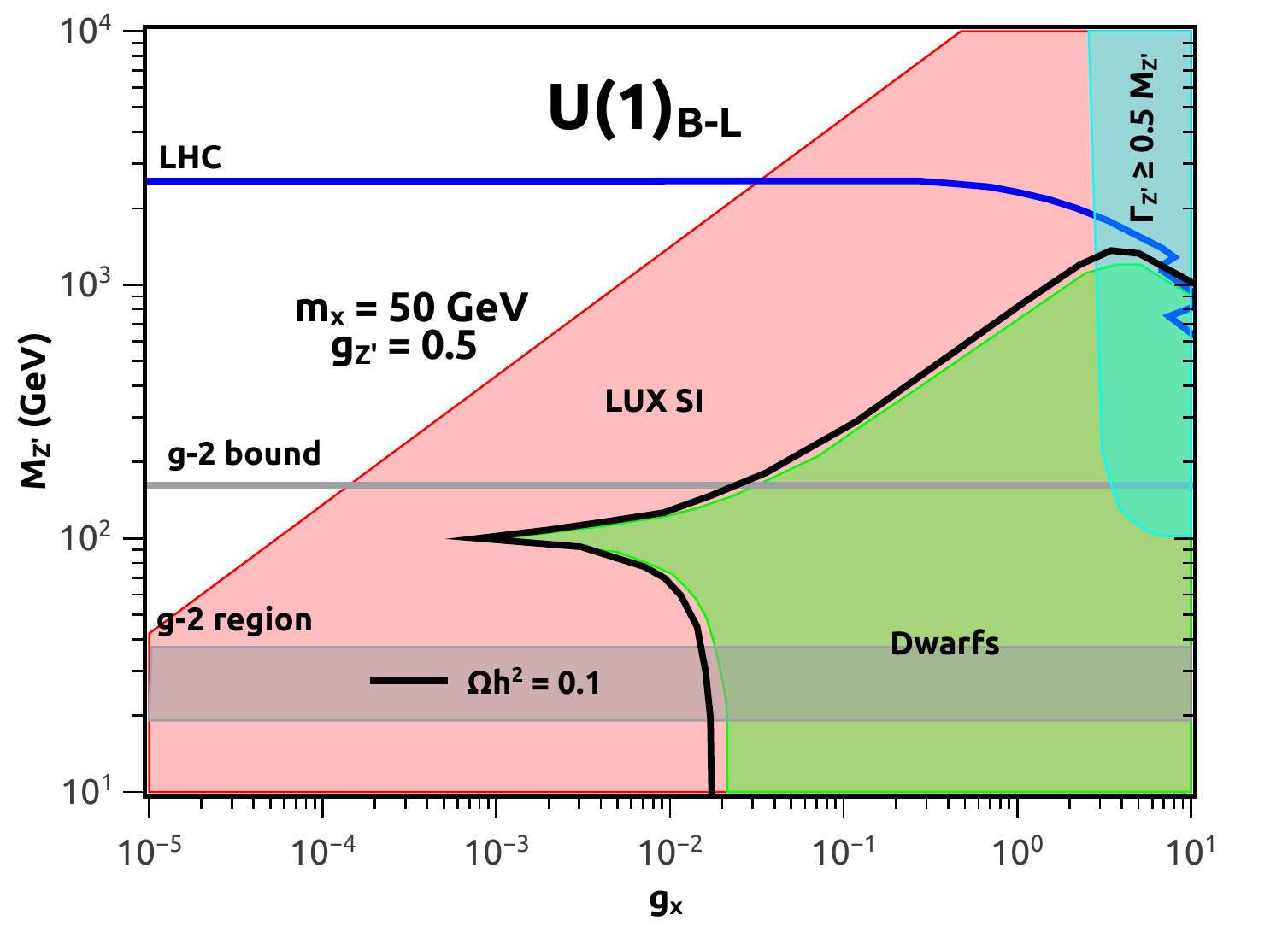}
\caption{Same as in Fig.~\ref{fig:results}, for the $U(1)_{\rm B-L}$ model. {\it Left:}  $m_{\chi}=50$~GeV, $g_{Z^{\prime}}=1$; {\it Right:}  $m_{\chi}=50$~GeV, $g_{Z^{\prime}}=0.5$.}
\end{figure}

\begin{figure}[!th]
\includegraphics[scale=0.5]{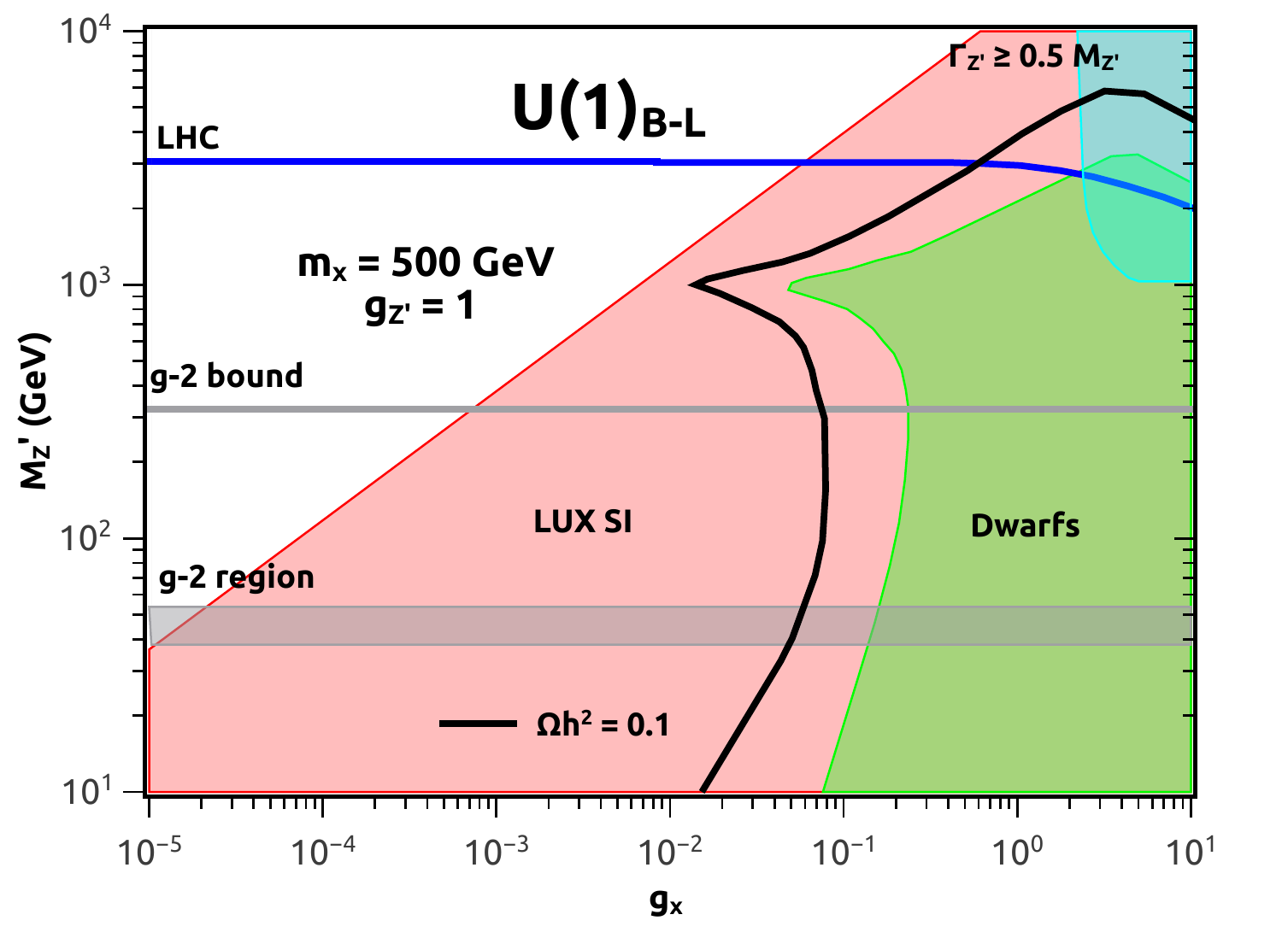}
\includegraphics[scale=0.5]{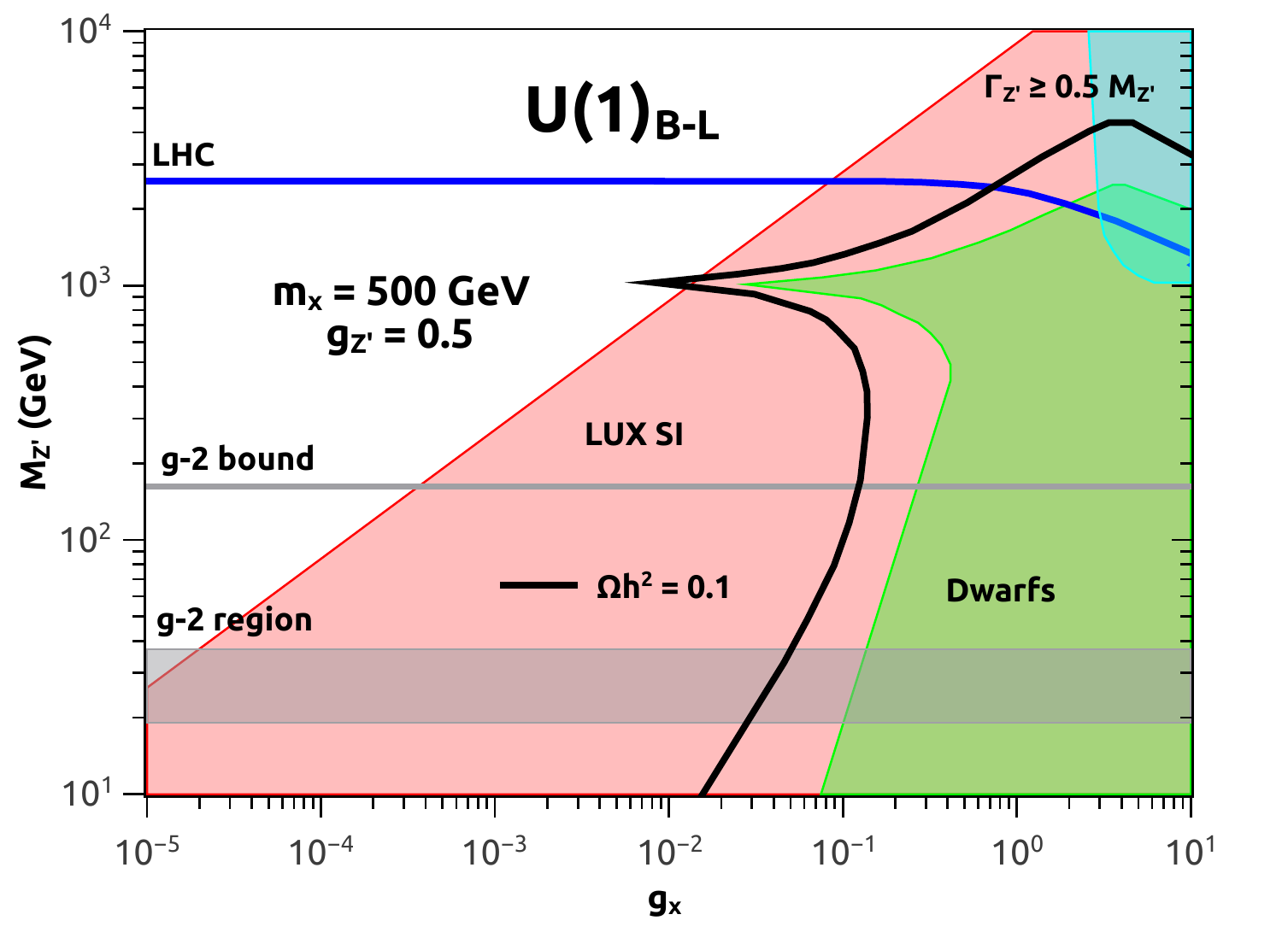}
\caption{Same as in Fig.~\ref{fig:results}, for the $U(1)_{\rm B-L}$ model. {\it Left:}  $m_{\chi}=500$~GeV, $g_{Z^{\prime}}=1$; {\it Right:}  $m_{\chi}=500$~GeV, $g_{Z^{\prime}}=0.5$.}
\end{figure}

\begin{figure}[!th]
\includegraphics[scale=0.5]{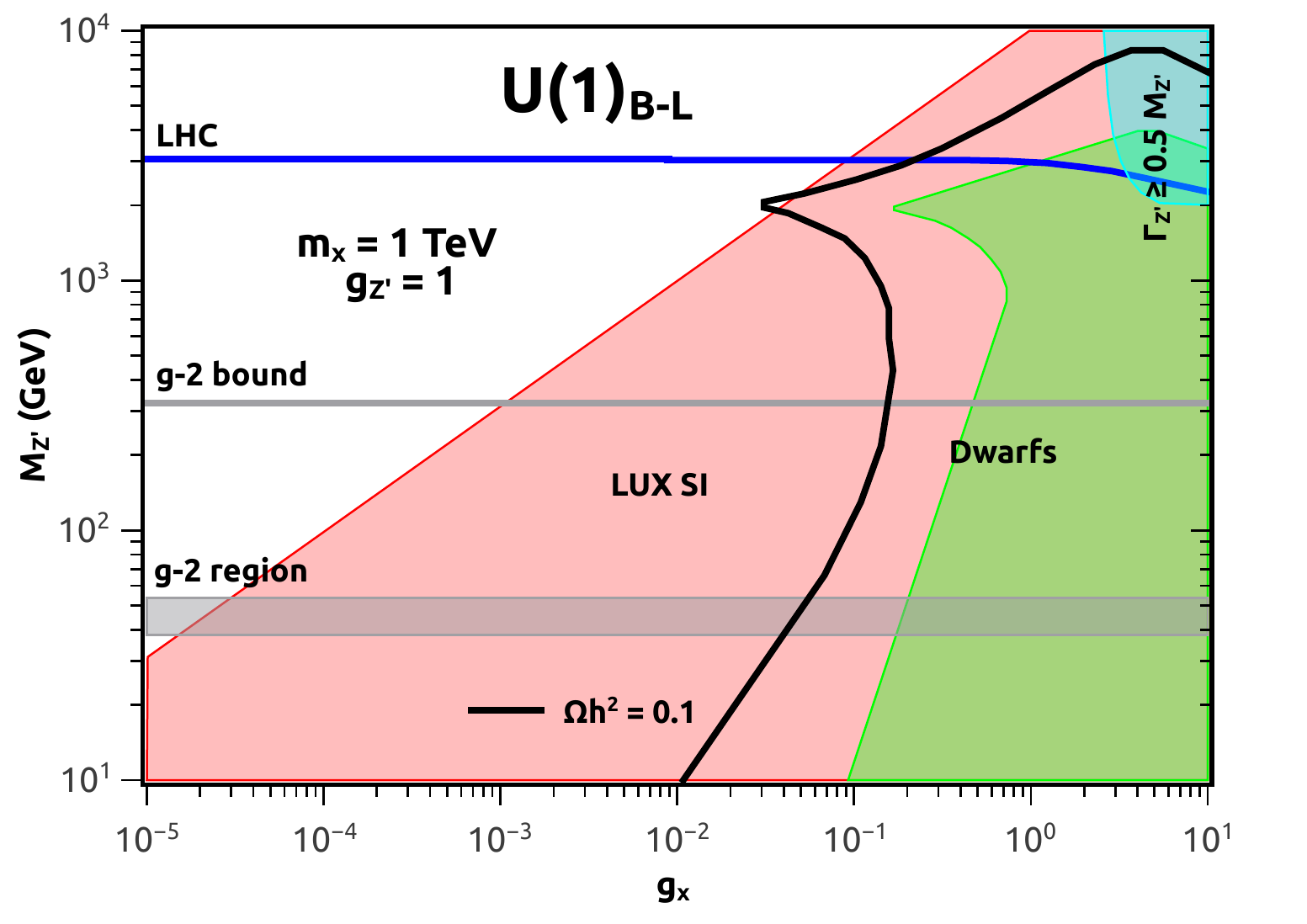}
\includegraphics[scale=0.5]{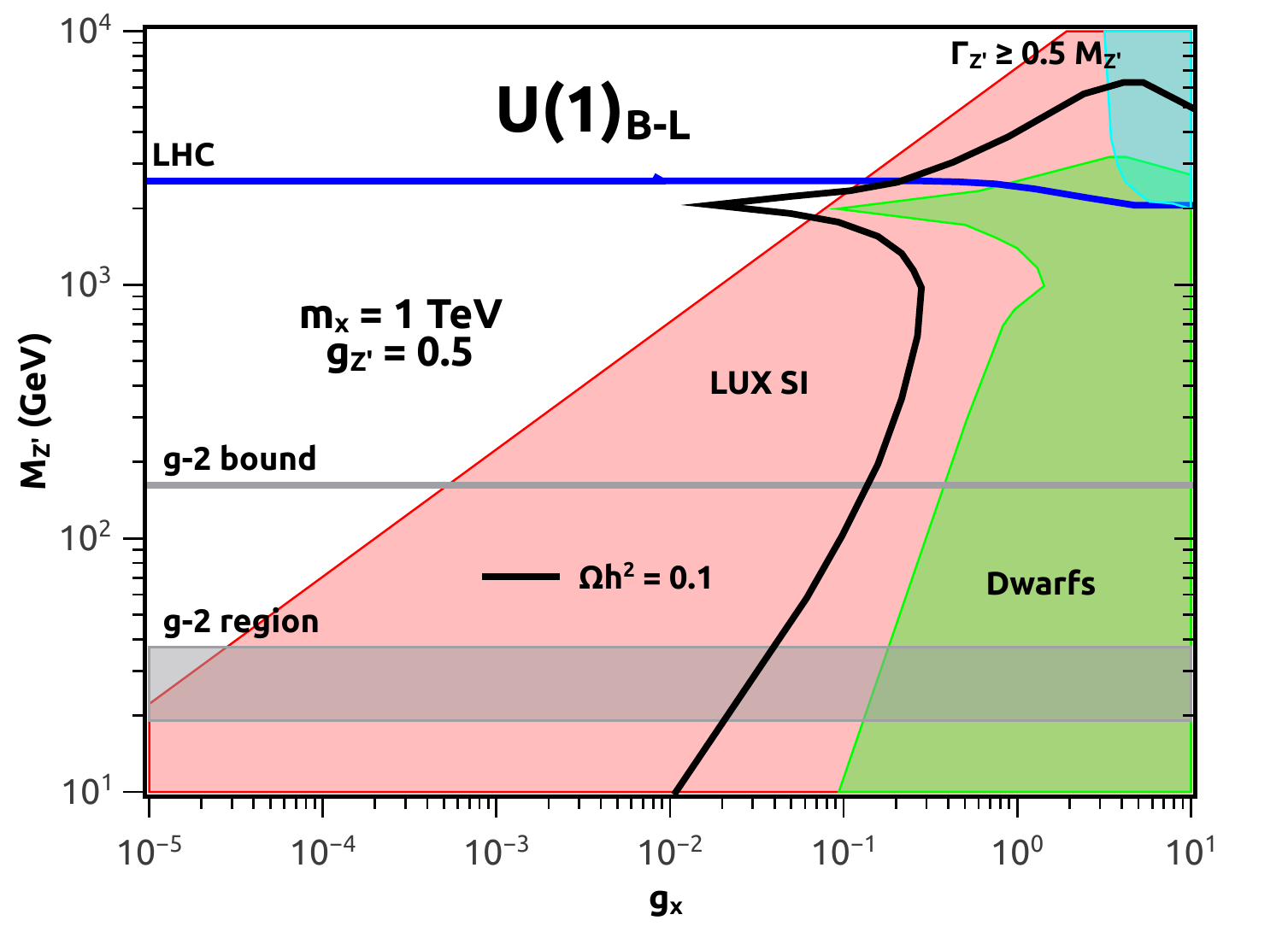}
\caption{Same as in Fig.~\ref{fig:results}, for the $U(1)_{\rm B-L}$ model. {\it Left:}  $m_{\chi}=1$~TeV, $g_{Z^{\prime}}=1$; {\it Right:}  $m_{\chi}=1$~TeV, $g_{Z^{\prime}}=0.5$.}
\end{figure}
\begin{figure}[!th]
\includegraphics[scale=0.5]{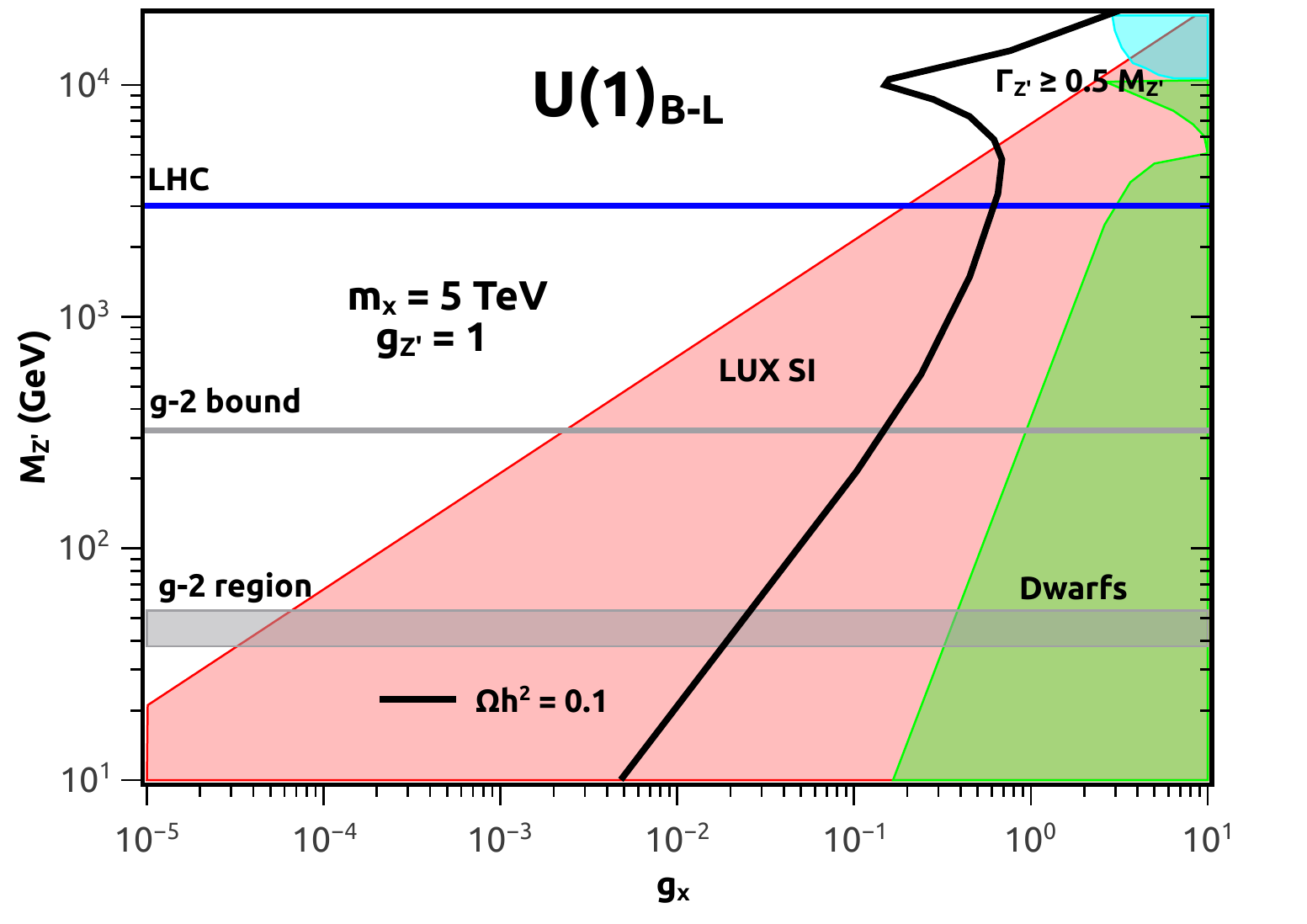}
\includegraphics[scale=0.5]{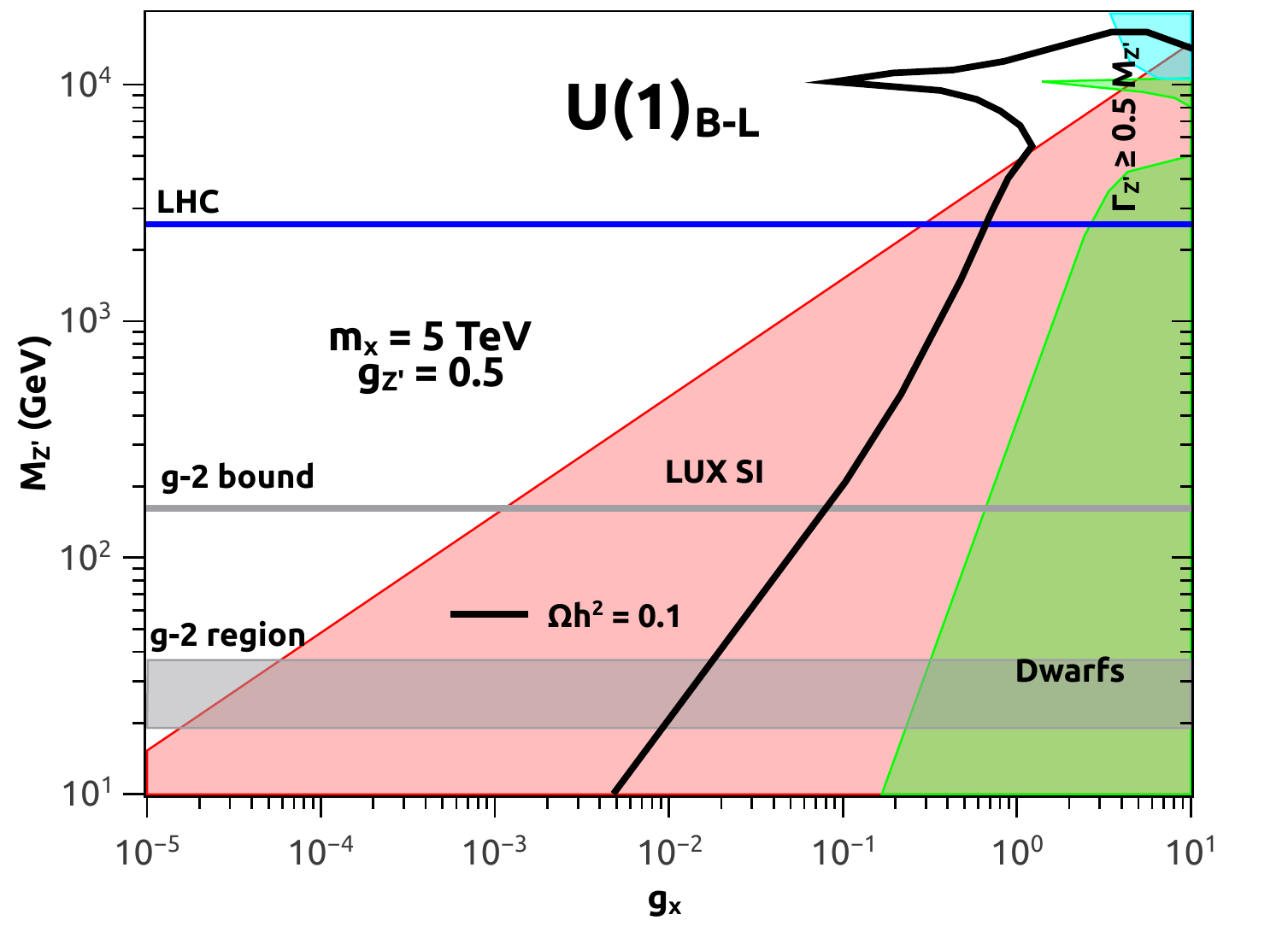}
\caption{Same as in Fig.~\ref{fig:results}, for the $U(1)_{\rm B-L}$ model. {\it Left:}  $m_{\chi}=5$~TeV, $g_{Z^{\prime}}=1$; {\it Right:}  $m_{\chi}=5$~TeV, $g_{Z^{\prime}}=0.5$.}
\end{figure}

\begin{figure}[!t]
\includegraphics[scale=0.5]{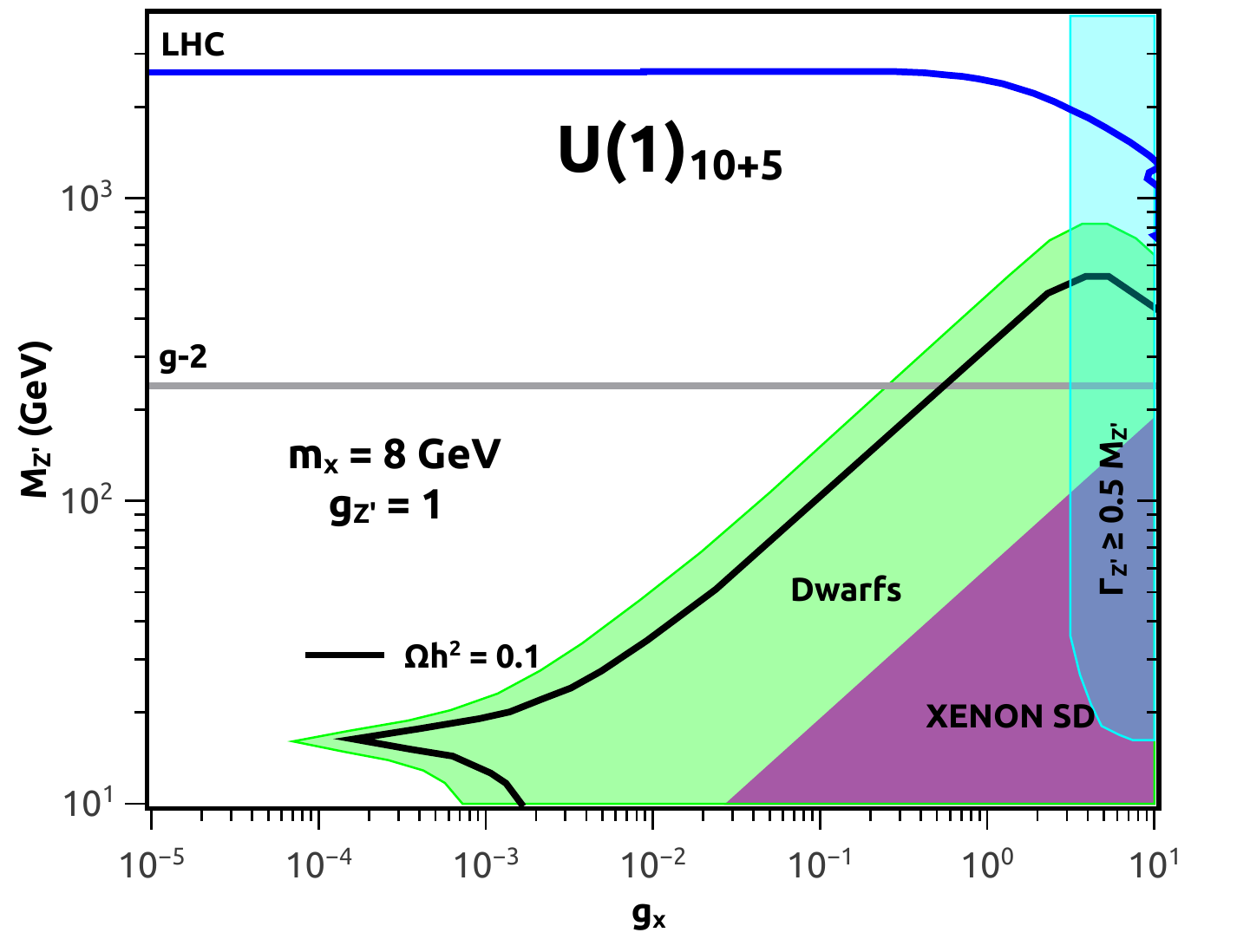}
\includegraphics[scale=0.5]{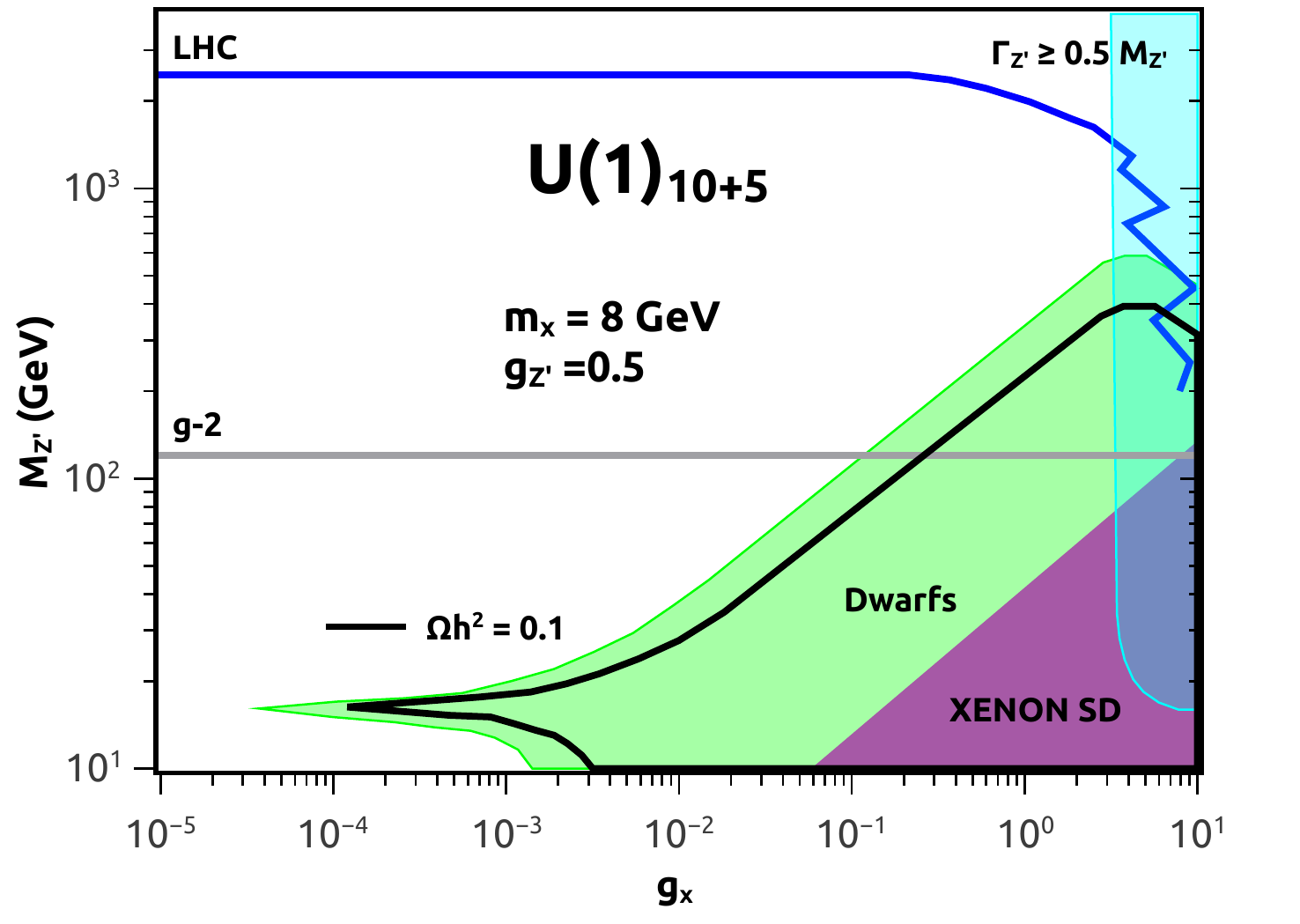}
\caption{Same as in Fig.~\ref{fig:results}, for the $U(1)_{10+\bar{5}}$ model. {\it Left:}  $m_{\chi}=8$~GeV, $g_{Z^{\prime}}=1$; {\it Right:}  $m_{\chi}=8$~GeV, $g_{Z^{\prime}}=0.5$. Blue horizontal line is LHC bound. Everything below the curve is ruled out. Gray horizontal line is the $1\sigma$ bound from the muon magnetic moment. In purple (green) we exhibit the XENON spin-dependent (Fermi Galactic Center) limit. The black curve sets region of parameter space that reproduced the right abundance. The cyan shaded region corresponds to the violation of the perturbative limit, $\Gamma_{\zp} \gtrsim M_{\zp}/2$.}
\end{figure}
\begin{figure}[!th]
\includegraphics[scale=0.5]{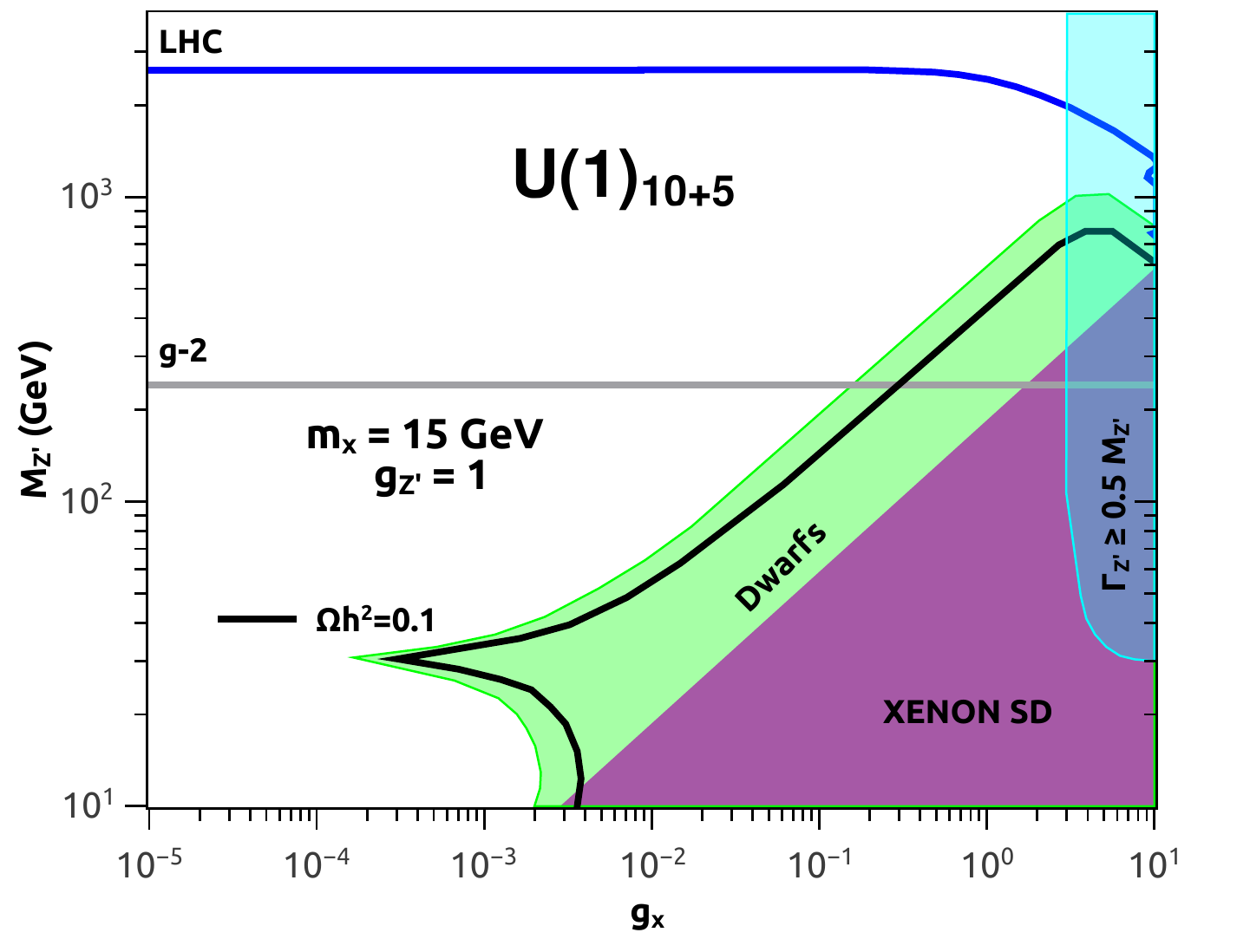}
\includegraphics[scale=0.5]{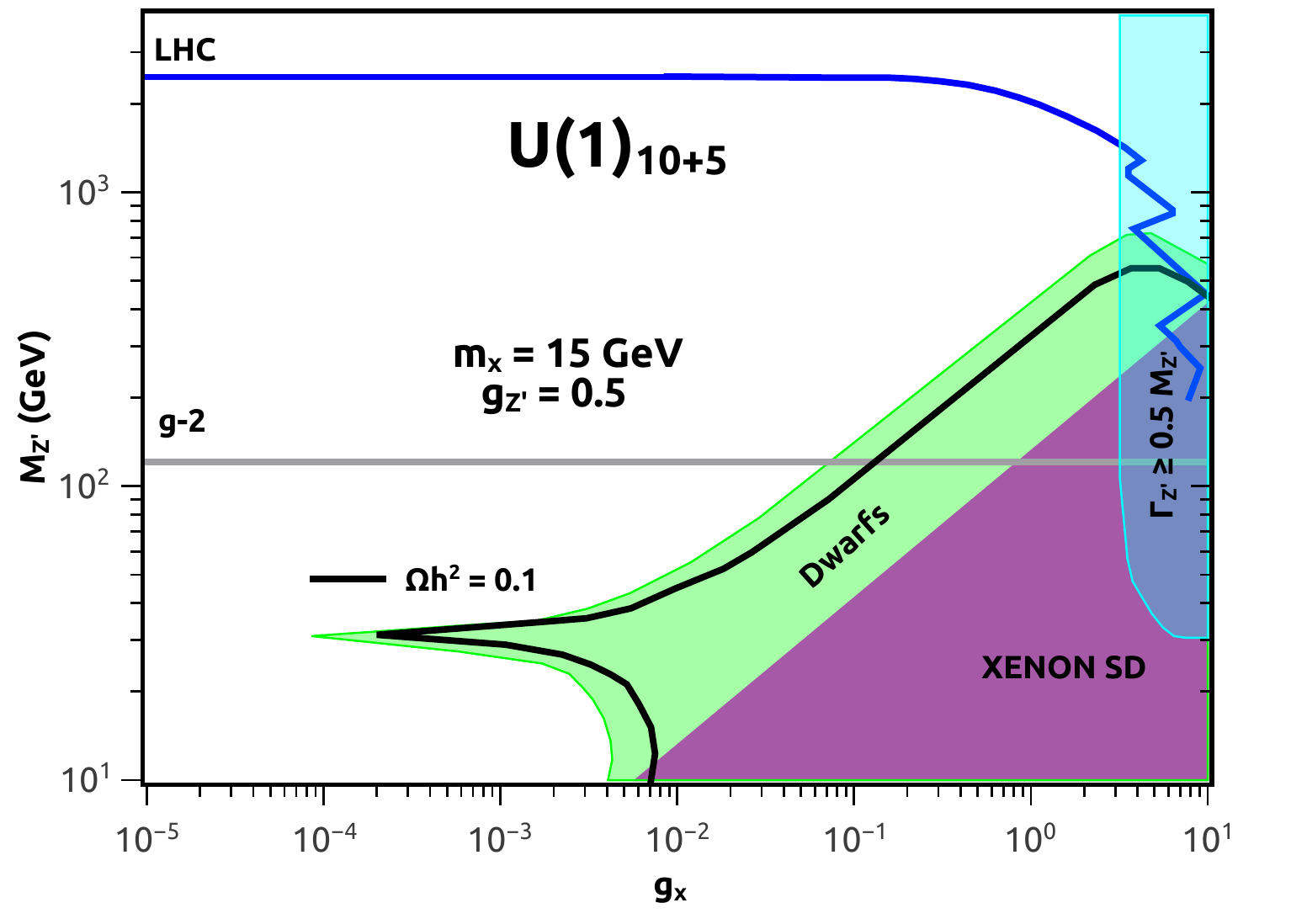}
\caption{Same as in Fig.~\ref{fig:results}, for the $U(1)_{10+\bar{5}}$ model. {\it Left:}  $m_{\chi}=15$~GeV, $g_{Z^{\prime}}=1$; {\it Right:}  $m_{\chi}=15$~GeV, $g_{Z^{\prime}}=0.5$.}
\end{figure}

\begin{figure}[!th]
\includegraphics[scale=0.5]{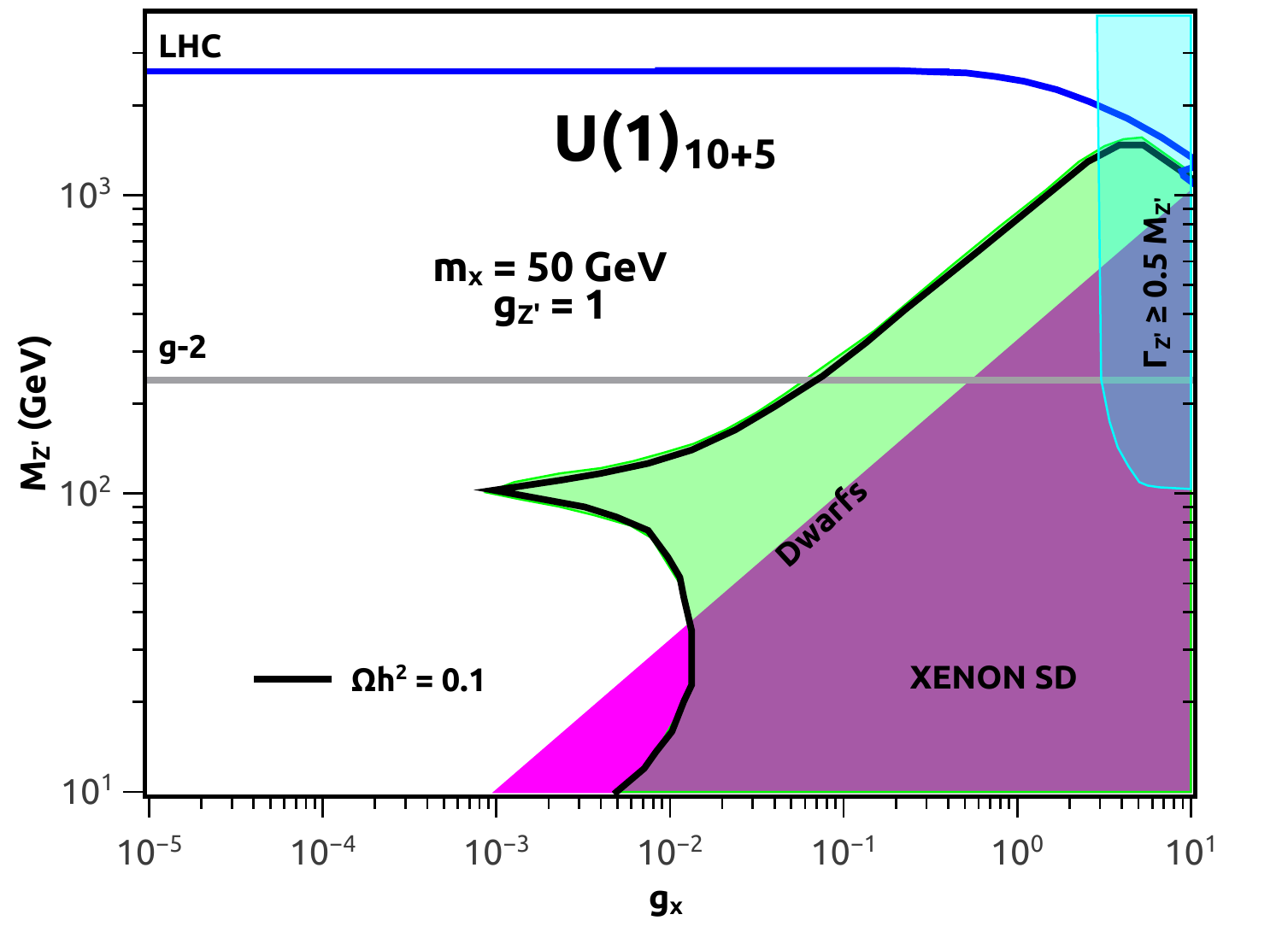}
\includegraphics[scale=0.5]{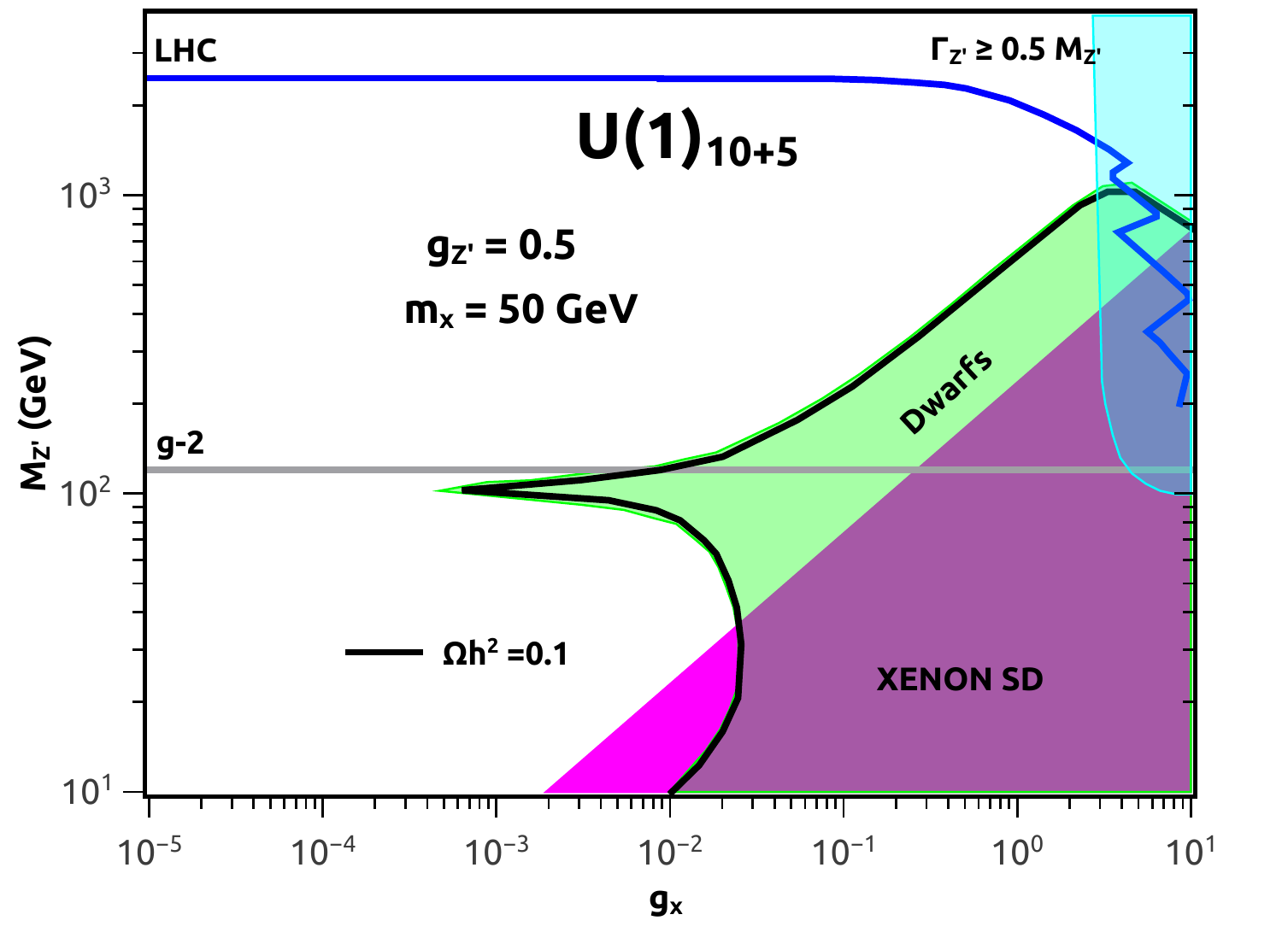}
\caption{Same as in Fig.~\ref{fig:results}, for the $U(1)_{10+\bar{5}}$ model. {\it Left:}  $m_{\chi}=50$~GeV, $g_{Z^{\prime}}=1$; {\it Right:}  $m_{\chi}=50$~GeV, $g_{Z^{\prime}}=0.5$.}
\end{figure}

\begin{figure}[!th]
\includegraphics[scale=0.5]{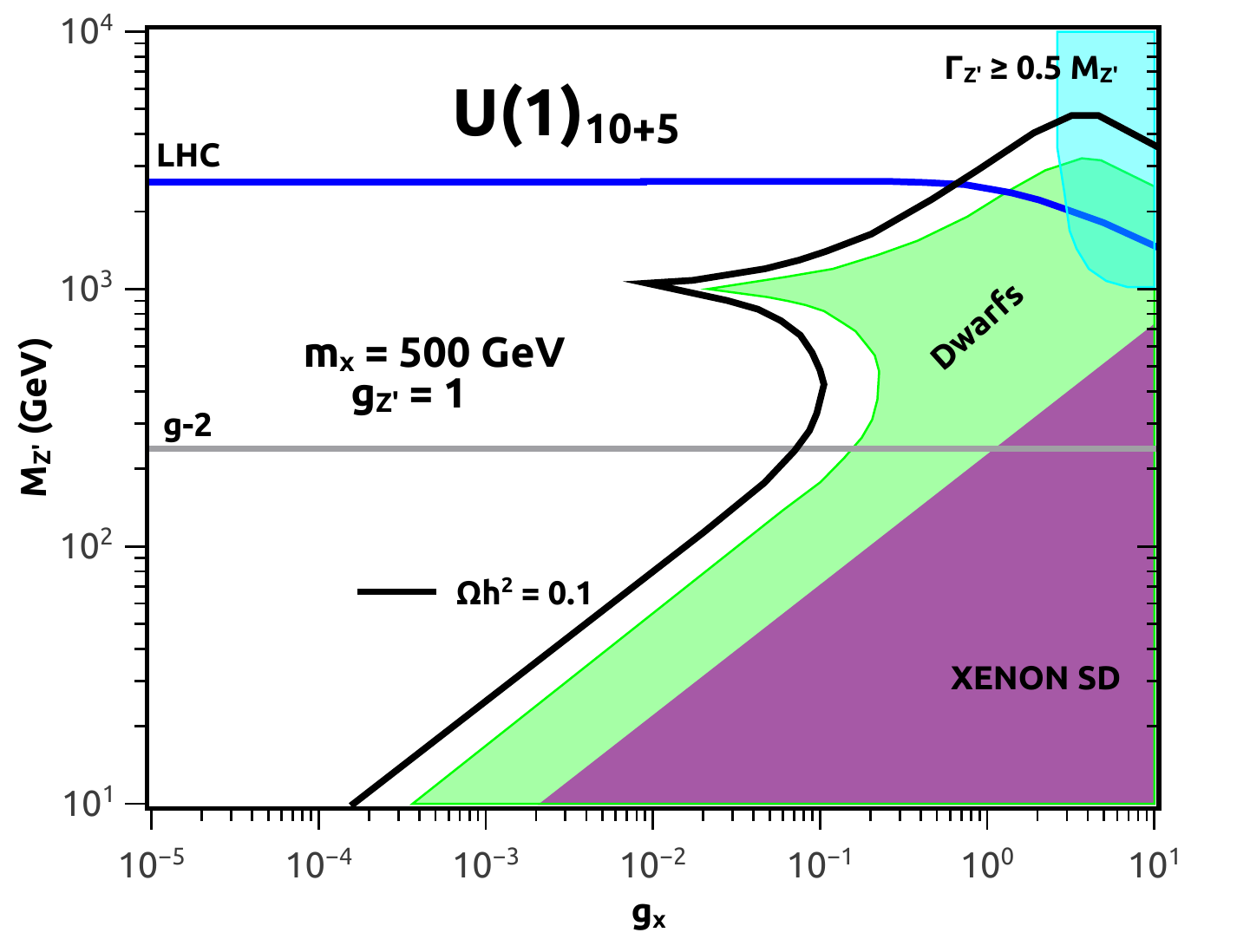}
\includegraphics[scale=0.5]{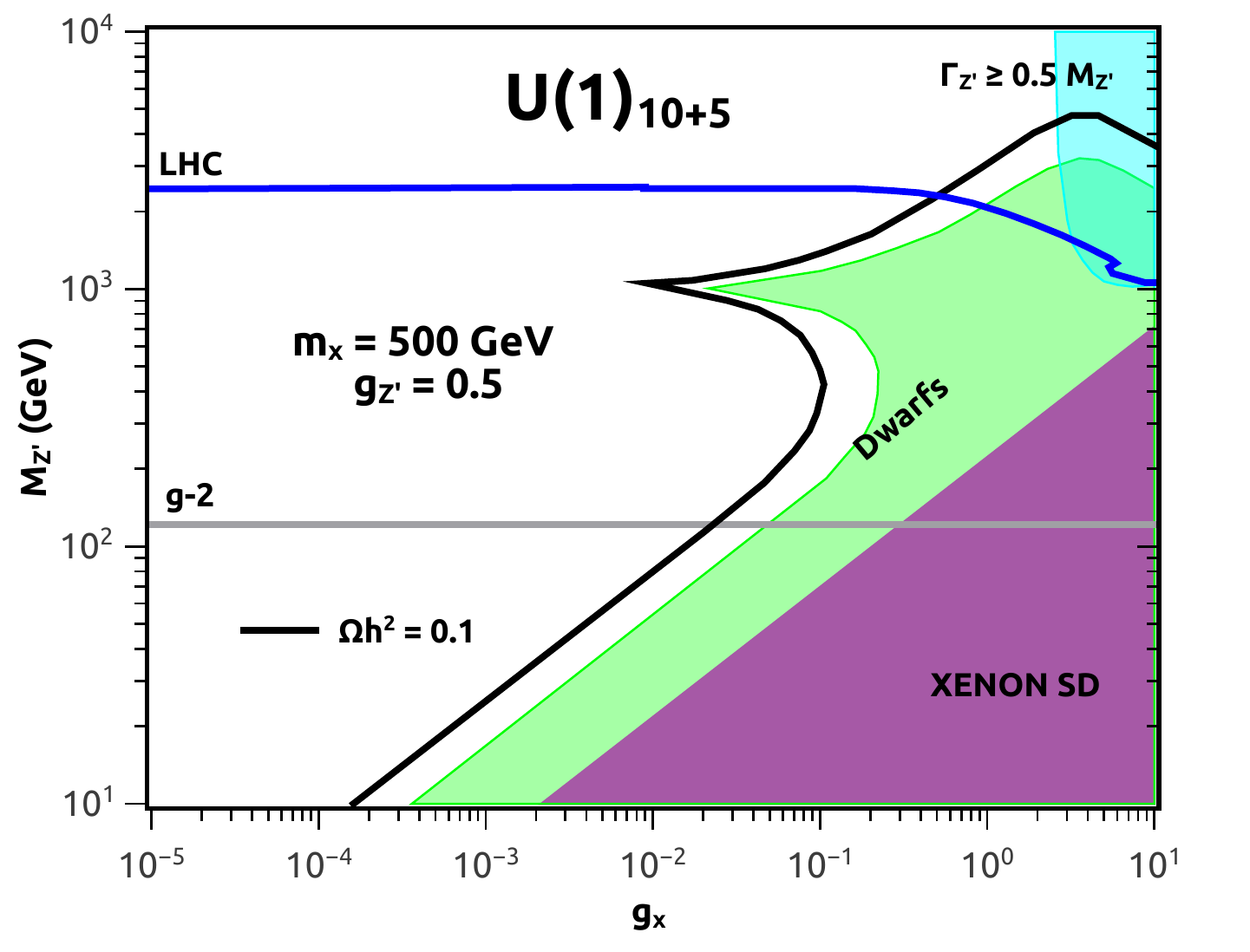}
\caption{Same as in Fig.~\ref{fig:results}, for the $U(1)_{10+\bar{5}}$ model. {\it Left:}  $m_{\chi}=500$~GeV, $g_{Z^{\prime}}=1$; {\it Right:}  $m_{\chi}=500$~GeV, $g_{Z^{\prime}}=0.5$.}
\end{figure}

\begin{figure}[!th]
\includegraphics[scale=0.5]{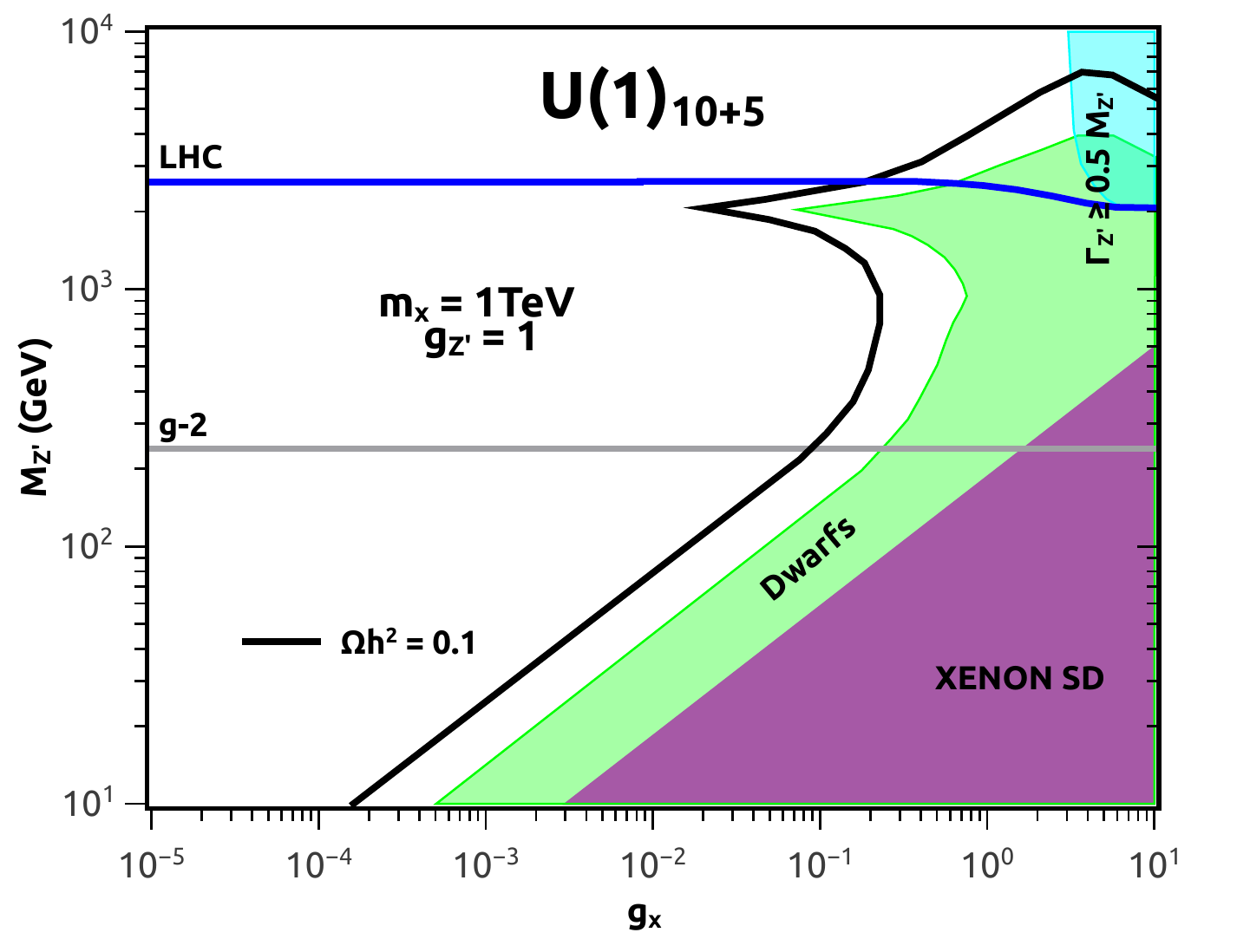}
\includegraphics[scale=0.5]{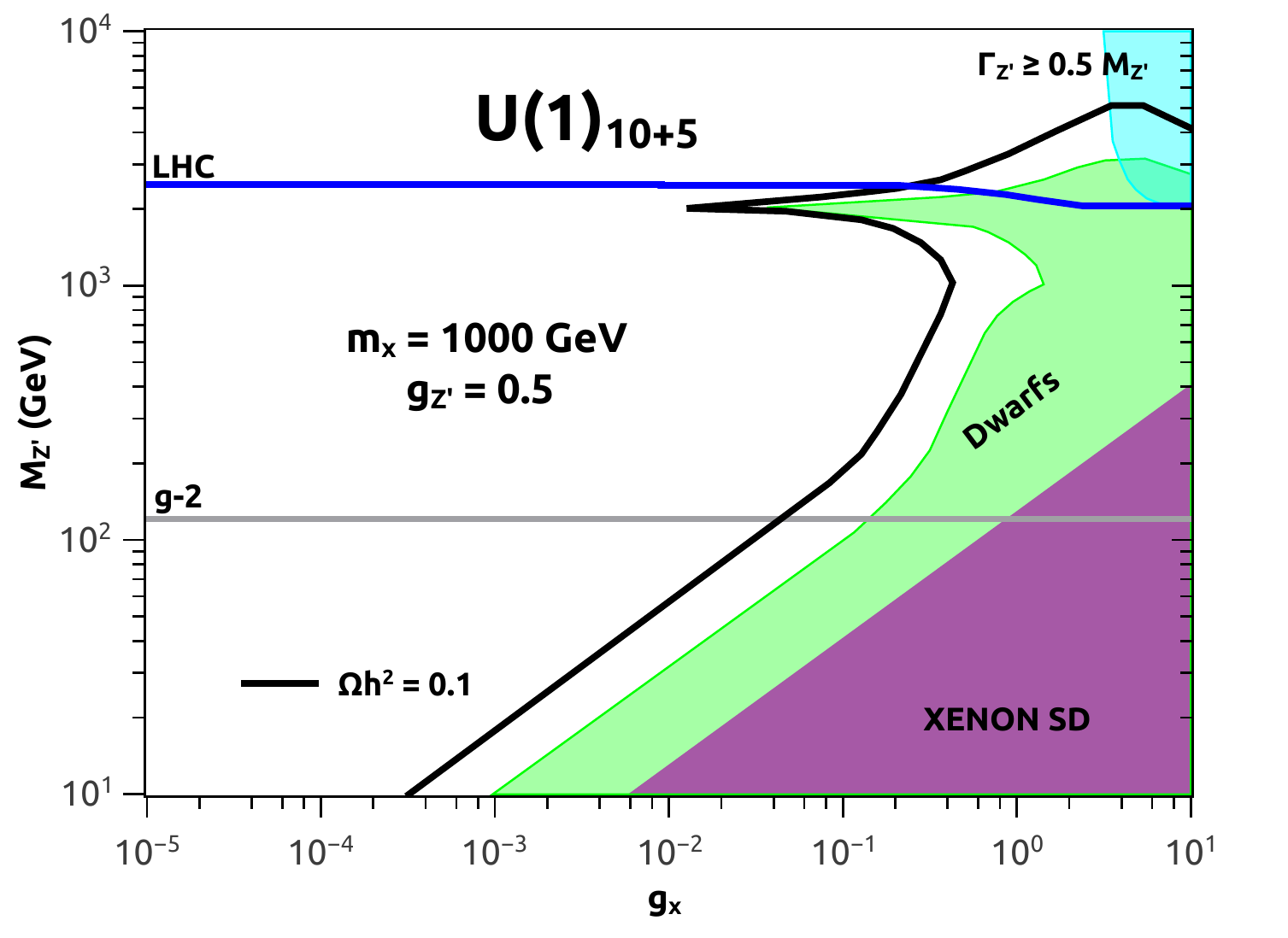}
\caption{Same as in Fig.~\ref{fig:results}, for the $U(1)_{10+\bar{5}}$ model. {\it Left:}  $m_{\chi}=1$~TeV, $g_{Z^{\prime}}=1$; {\it Right:}  $m_{\chi}=1$~TeV, $g_{Z^{\prime}}=0.5$.}
\end{figure}

\begin{figure}[!th]
\includegraphics[scale=0.5]{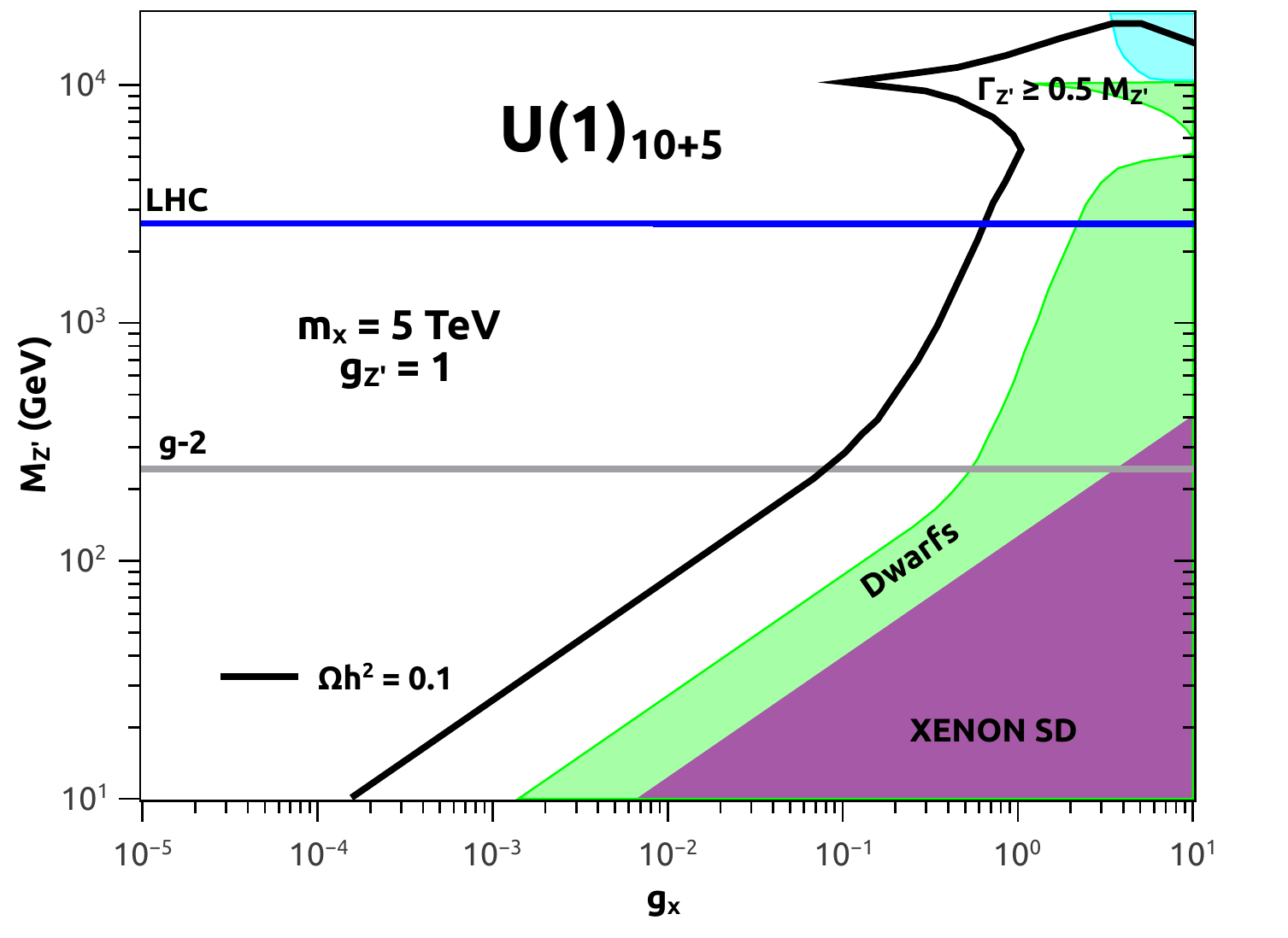}
\includegraphics[scale=0.5]{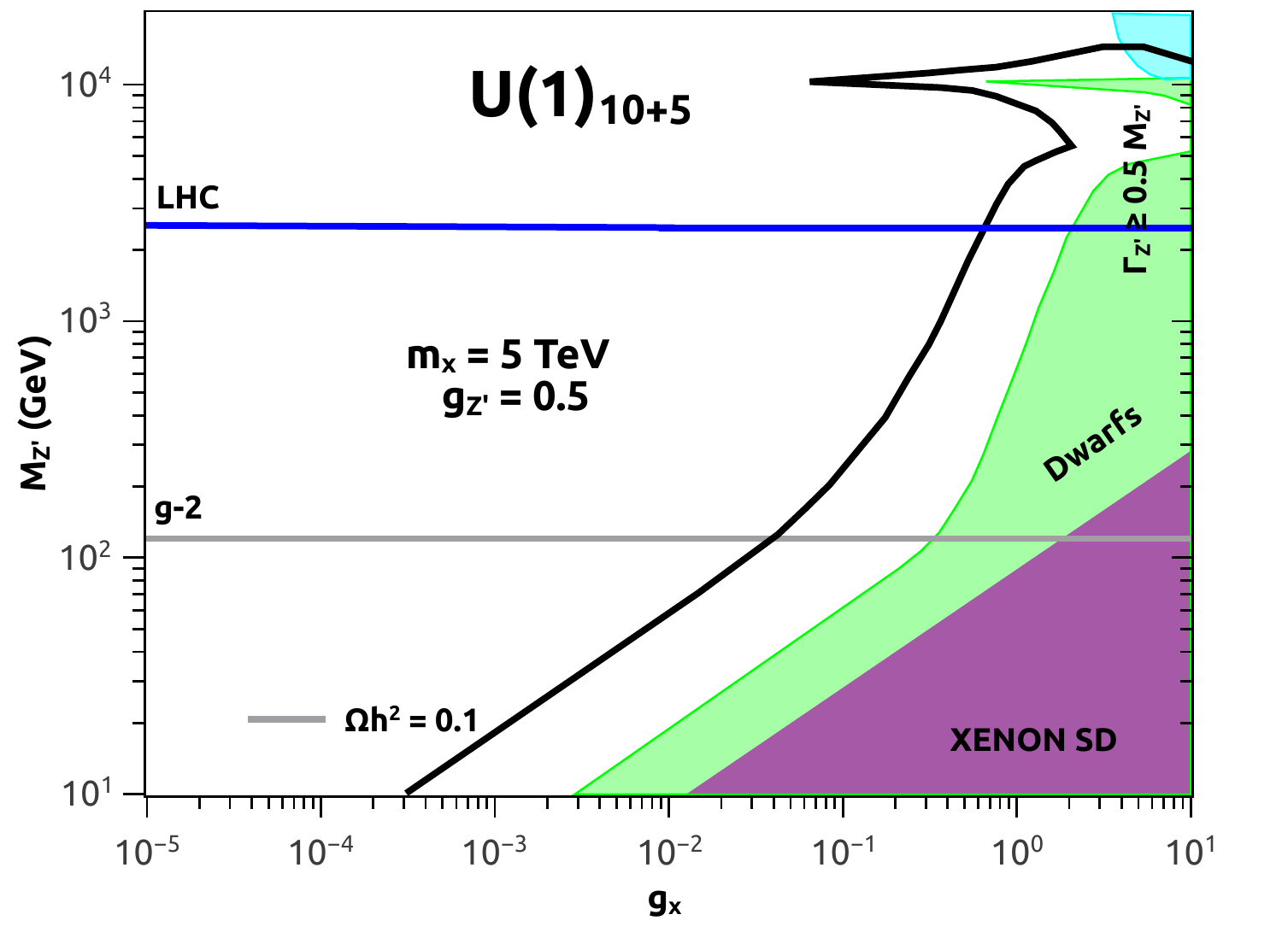}
\caption{Same as in Fig.~\ref{fig:results}, for the $U(1)_{10+\bar{5}}$ model. {\it Left:}  $m_{\chi}=5$~TeV, $g_{Z^{\prime}}=1$; {\it Right:}  $m_{\chi}=5$~TeV, $g_{Z^{\prime}}=0.5$.}
\end{figure}

\begin{figure}[!th]
\centering
\includegraphics[scale=0.45]{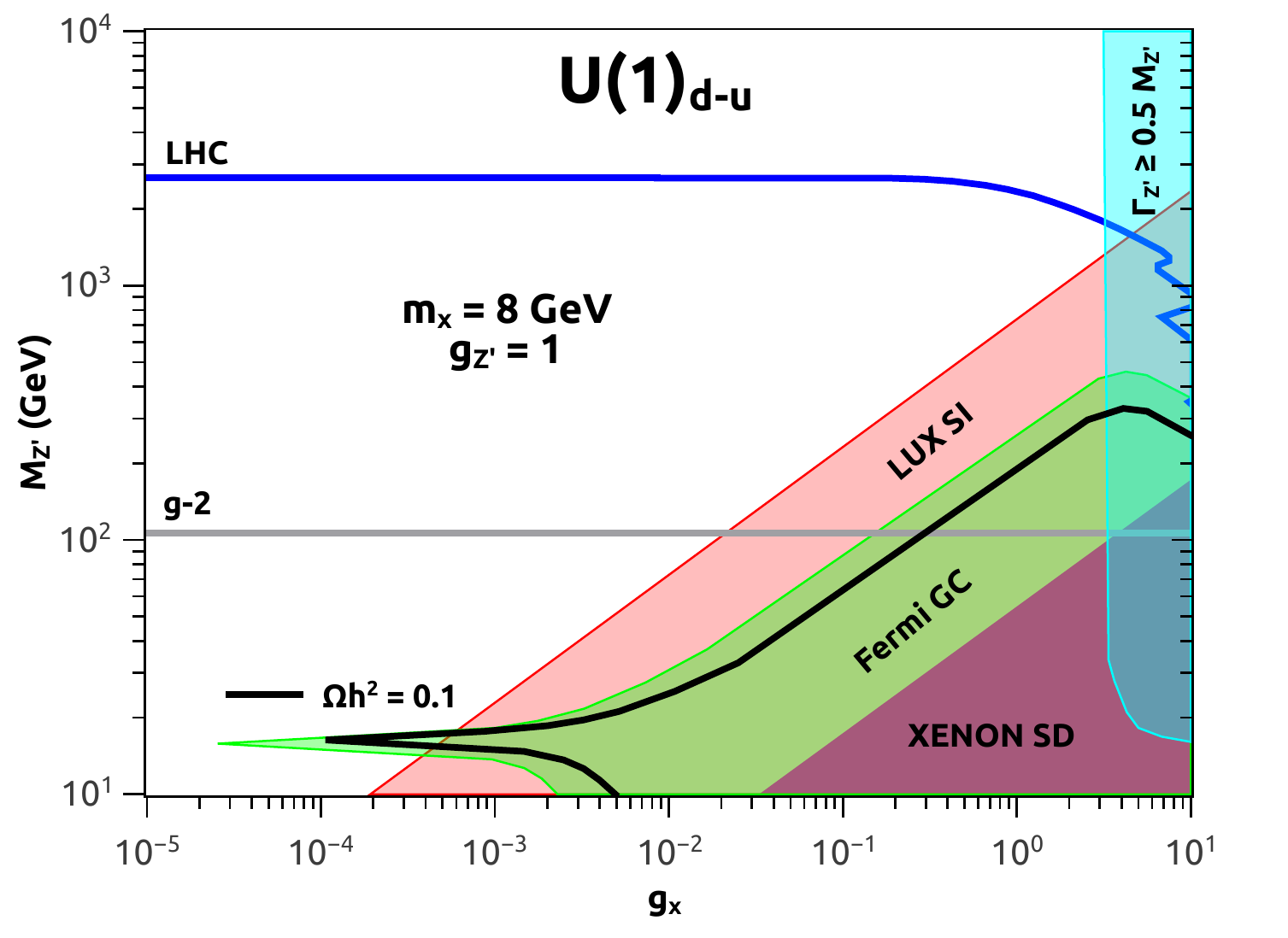}
\includegraphics[scale=0.45]{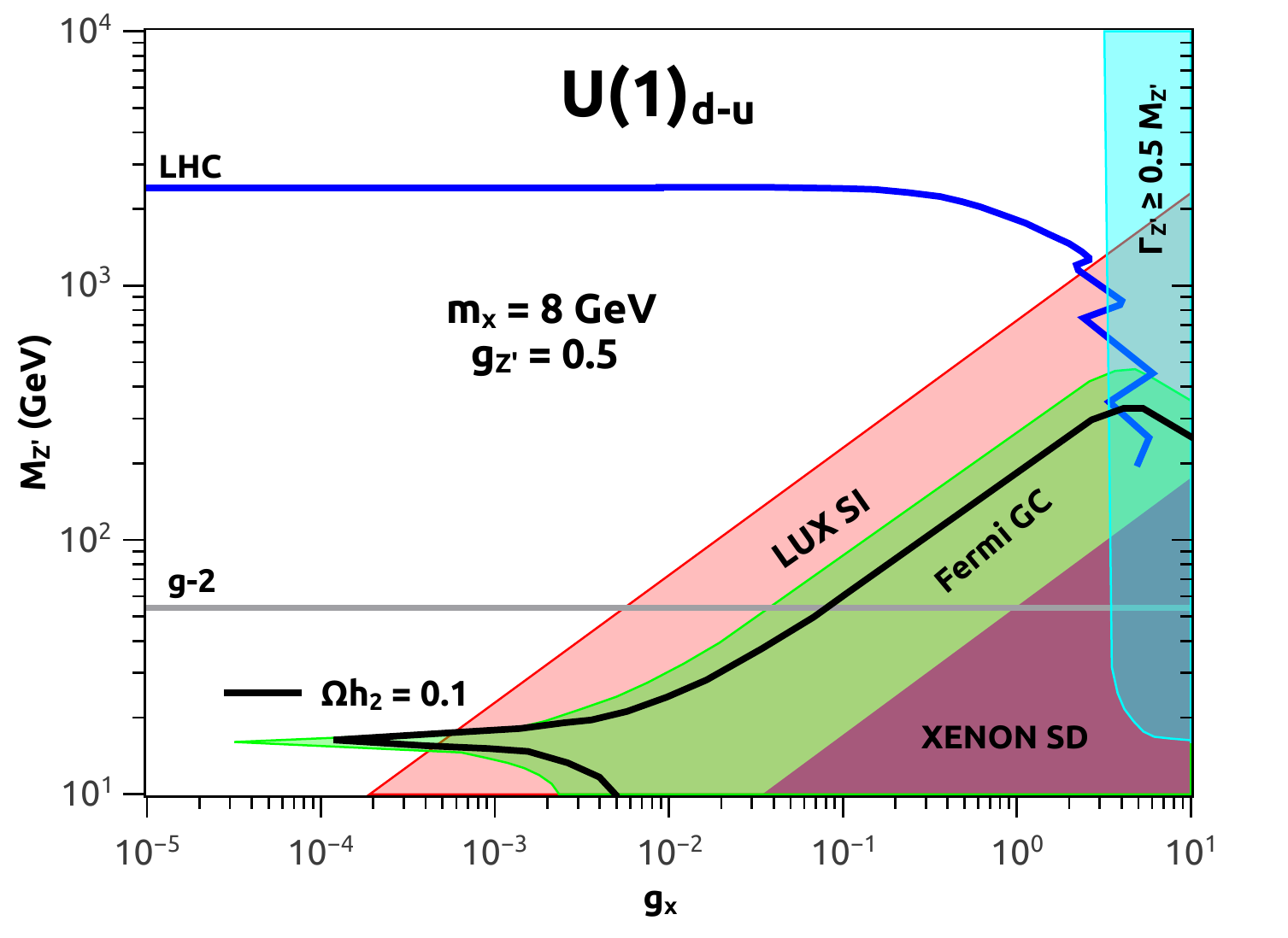}
\caption{Same as in Fig.~\ref{fig:results}, for the $U(1)_{d-u}$ model. {\it Left:}  $m_{\chi}=8$~GeV, $g_{Z^{\prime}}=1$; {\it Right:}  $m_{\chi}=8$~GeV, $g_{Z^{\prime}}=0.5$. Blue horizontal line is LHC bound. Everything below the curve is ruled out. Gray horizontal line is the $1\sigma$ bound from the muon magnetic moment. In red (purple) we exhibit the LUX (XENON) spin-independent (spin-dependent) bounds. In green we show the Fermi Galactic Center limit. The black curve sets region of parameter space that reproduced the right abundance. The cyan shaded region corresponds to the violation of the perturbative limit, $\Gamma_{\zp} \gtrsim M_{\zp}/2$.}
\end{figure}
\begin{figure}[!th]
\centering
\includegraphics[scale=0.45]{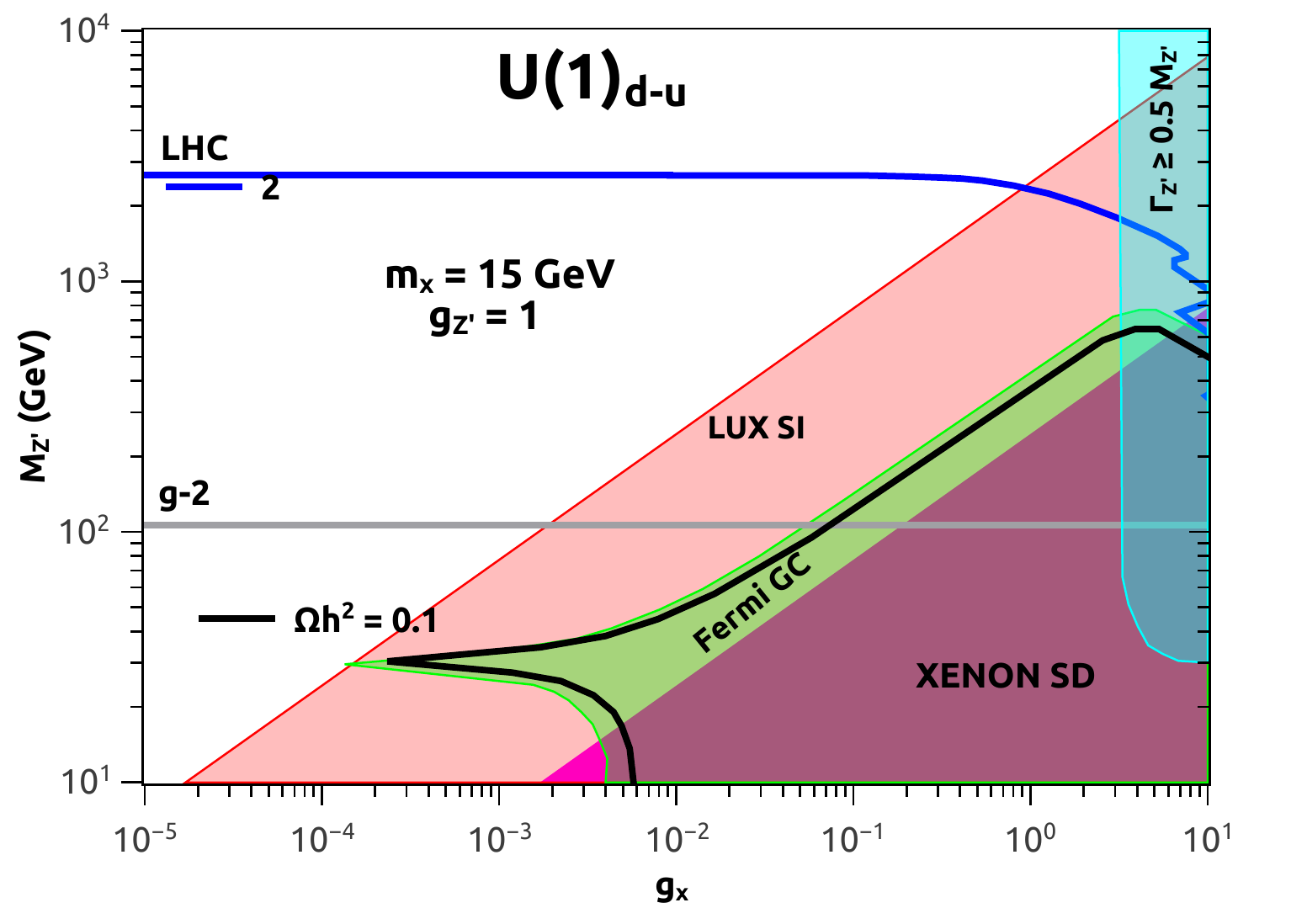}
\includegraphics[scale=0.45]{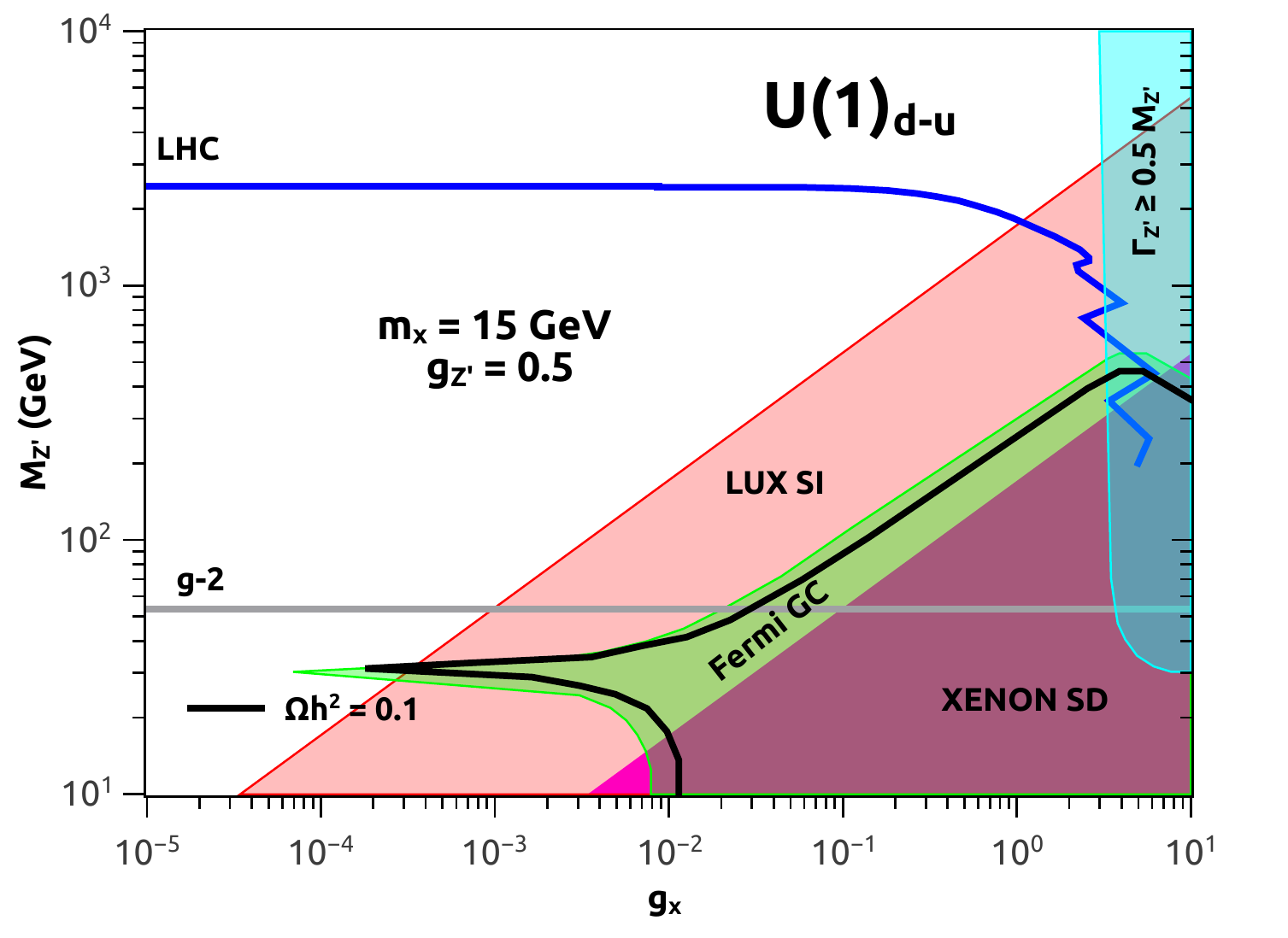}
\caption{Same as in Fig.~\ref{fig:results}, for the $U(1)_{d-u}$ model. {\it Left:}  $m_{\chi}=15$~GeV, $g_{Z^{\prime}}=1$; {\it Right:}  $m_{\chi}=15$~GeV, $g_{Z^{\prime}}=0.5$.}
\end{figure}
\begin{figure}[!t]
\includegraphics[scale=0.5]{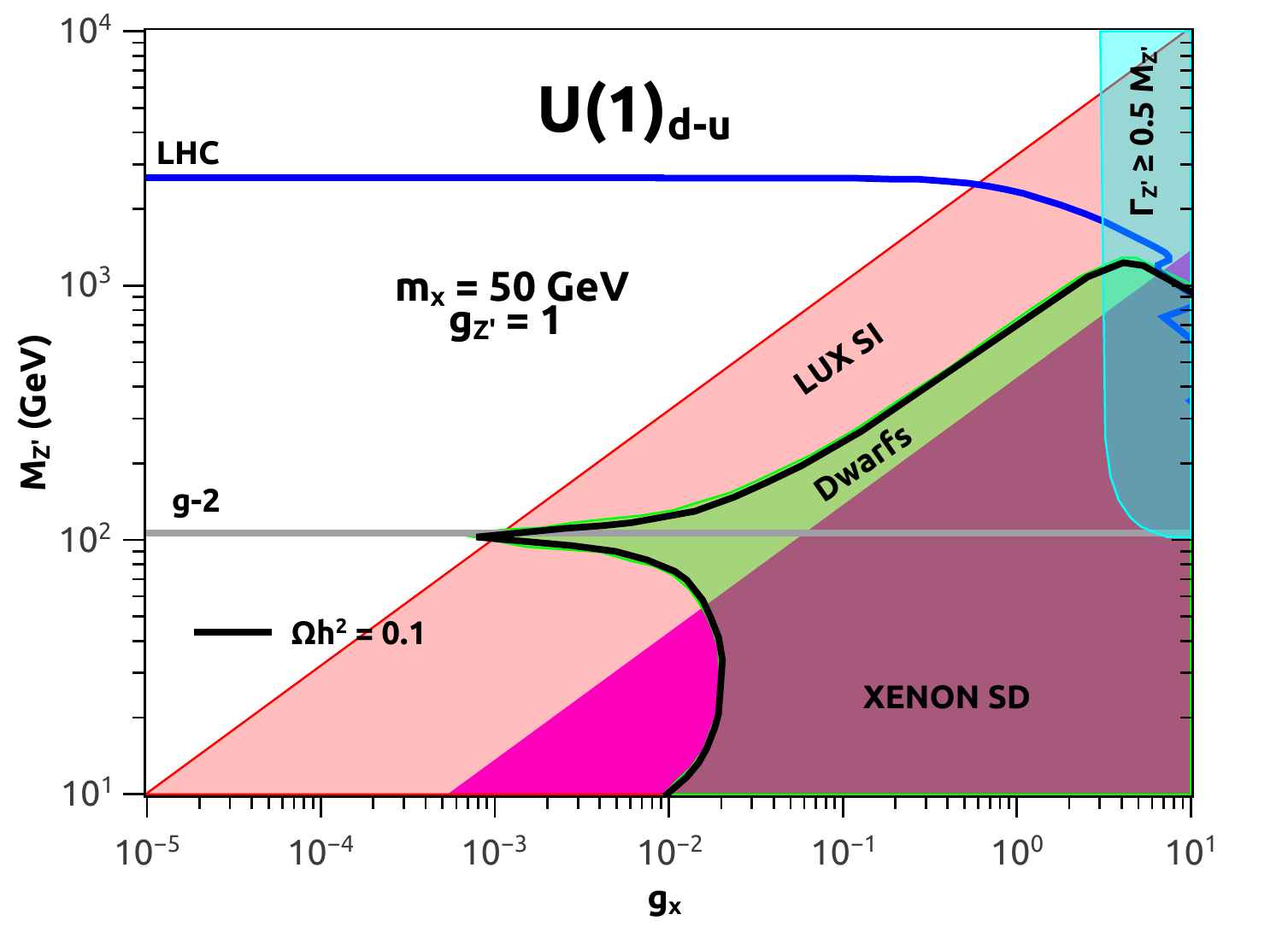}
\includegraphics[scale=0.5]{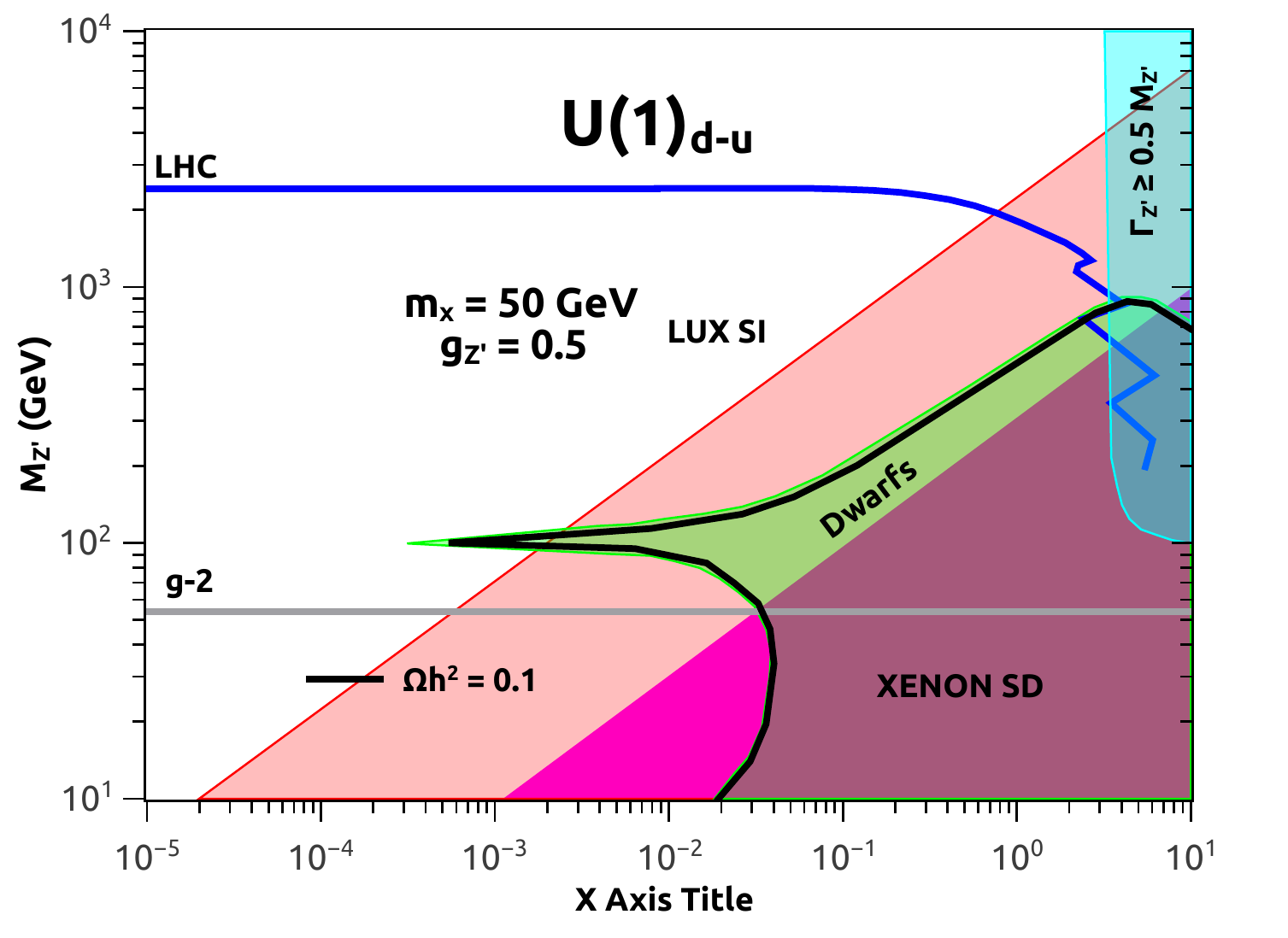}
\caption{Same as in Fig.~\ref{fig:results}, for the $U(1)_{d-u}$ model. {\it Left:}  $m_{\chi}=50$~GeV, $g_{Z^{\prime}}=1$; {\it Right:}  $m_{\chi}=50$~GeV, $g_{Z^{\prime}}=0.5$.}
\end{figure}
\begin{figure}[!t]
\includegraphics[scale=0.5]{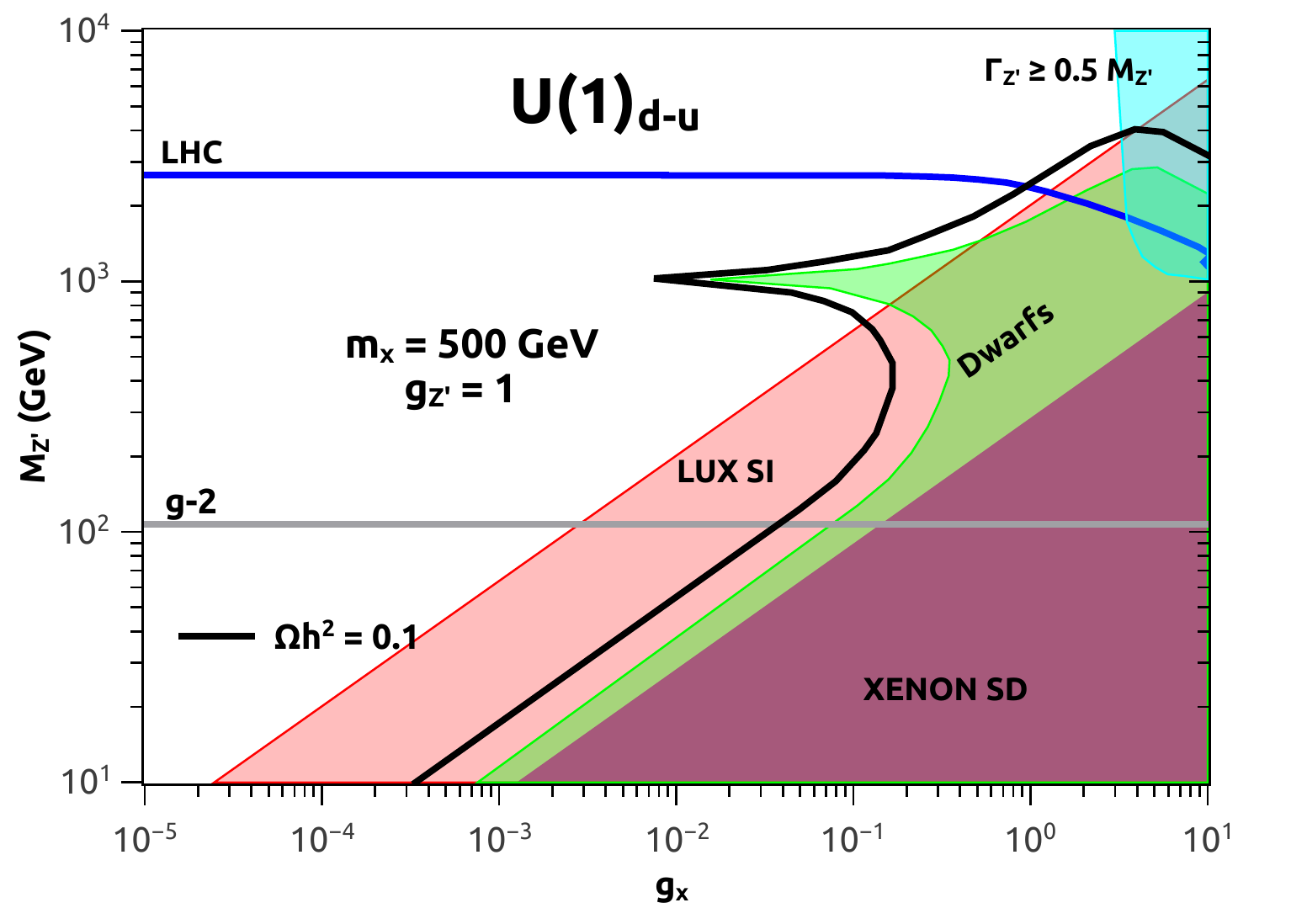}
\includegraphics[scale=0.5]{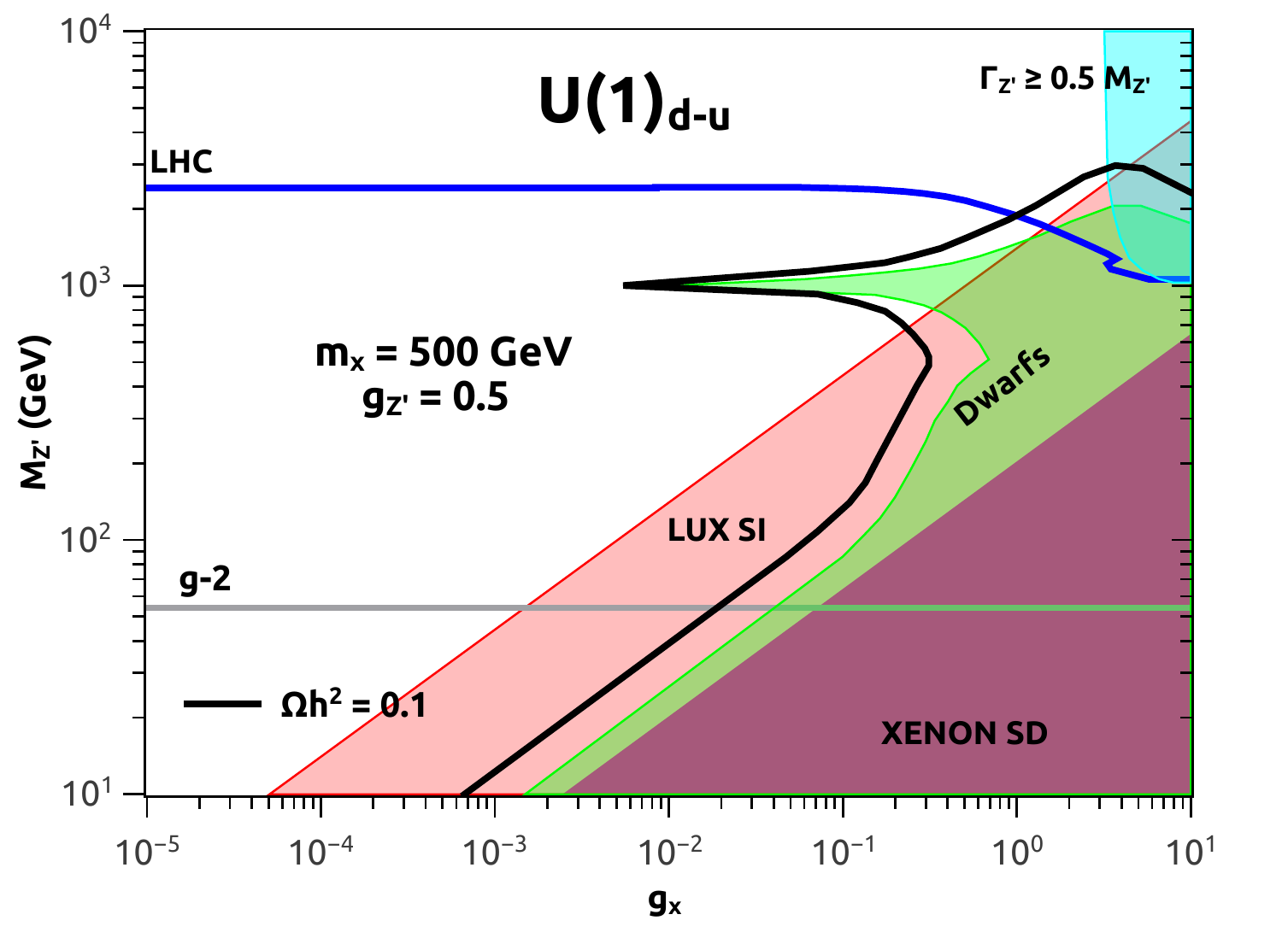}
\caption{Same as in Fig.~\ref{fig:results}, for the $U(1)_{d-u}$ model. {\it Left:}  $m_{\chi}=500$~GeV, $g_{Z^{\prime}}=1$; {\it Right:}  $m_{\chi}=500$~GeV, $g_{Z^{\prime}}=0.5$.}
\end{figure}
\begin{figure}[!t]
\includegraphics[scale=0.5]{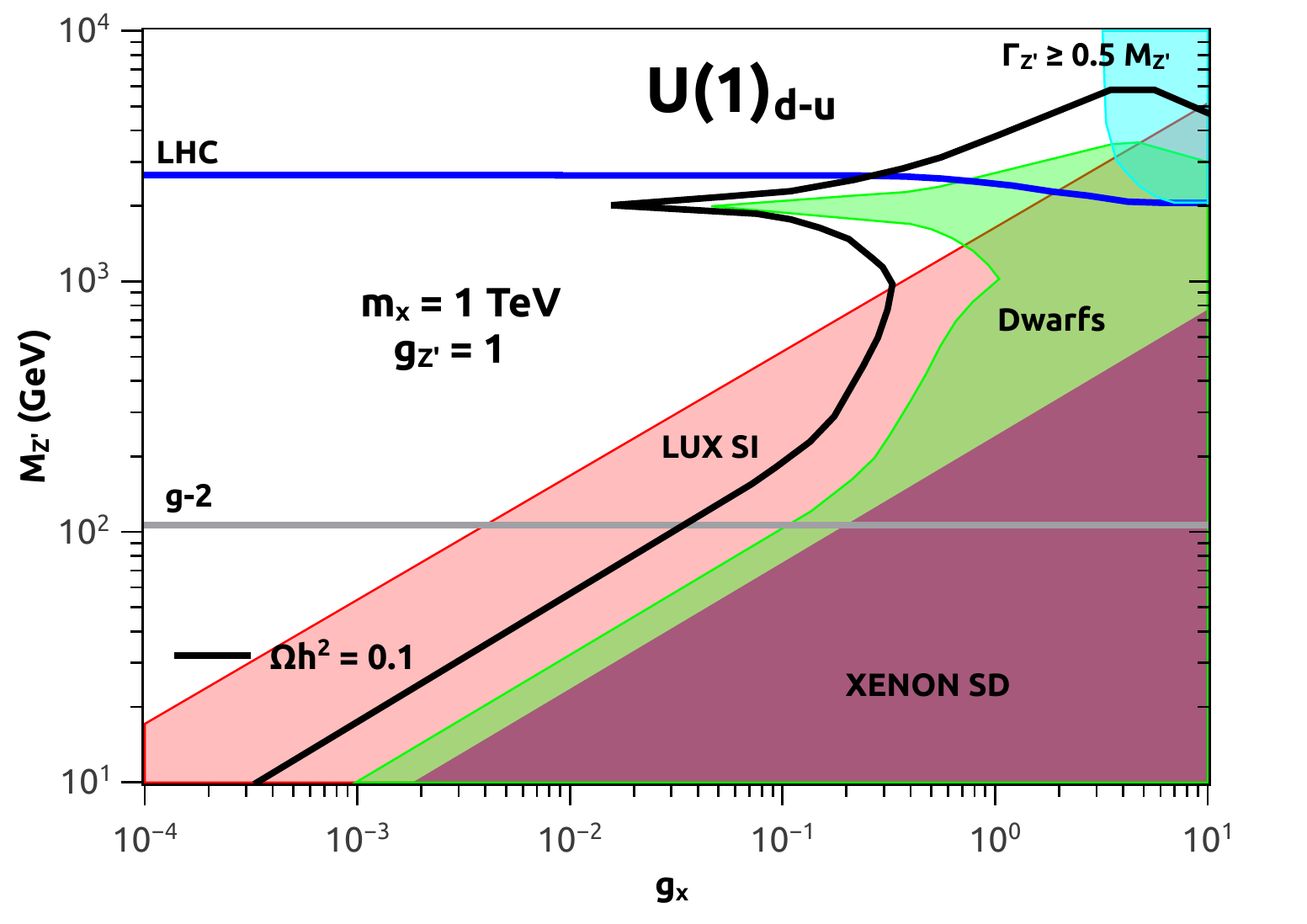}
\includegraphics[scale=0.5]{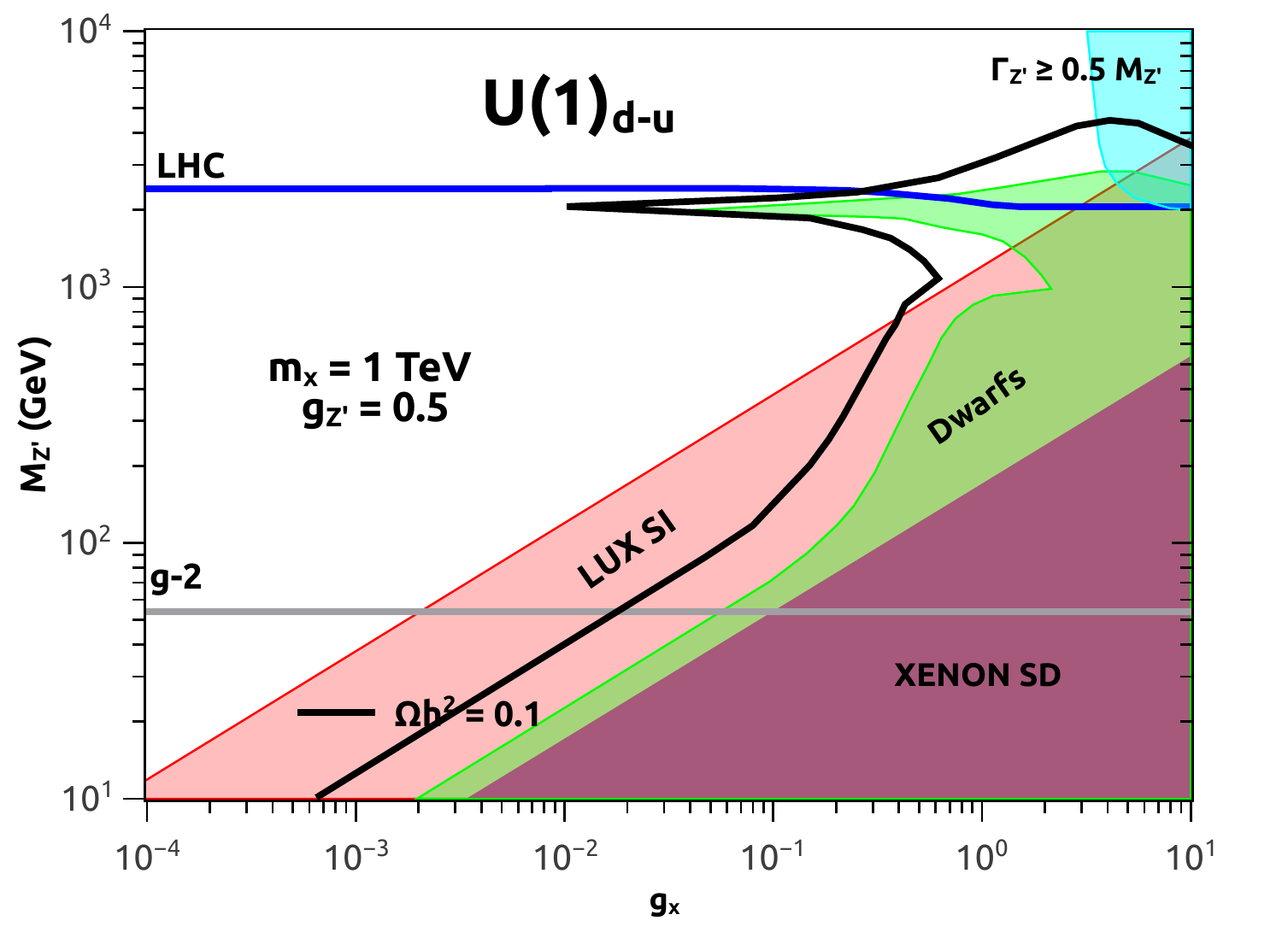}
\caption{Same as in Fig.~\ref{fig:results}, for the $U(1)_{d-u}$ model. {\it Left:}  $m_{\chi}=1$~TeV, $g_{Z^{\prime}}=1$; {\it Right:}  $m_{\chi}=1$~TeV, $g_{Z^{\prime}}=0.5$.}
\end{figure}

\begin{figure}[!t]
\includegraphics[scale=0.5]{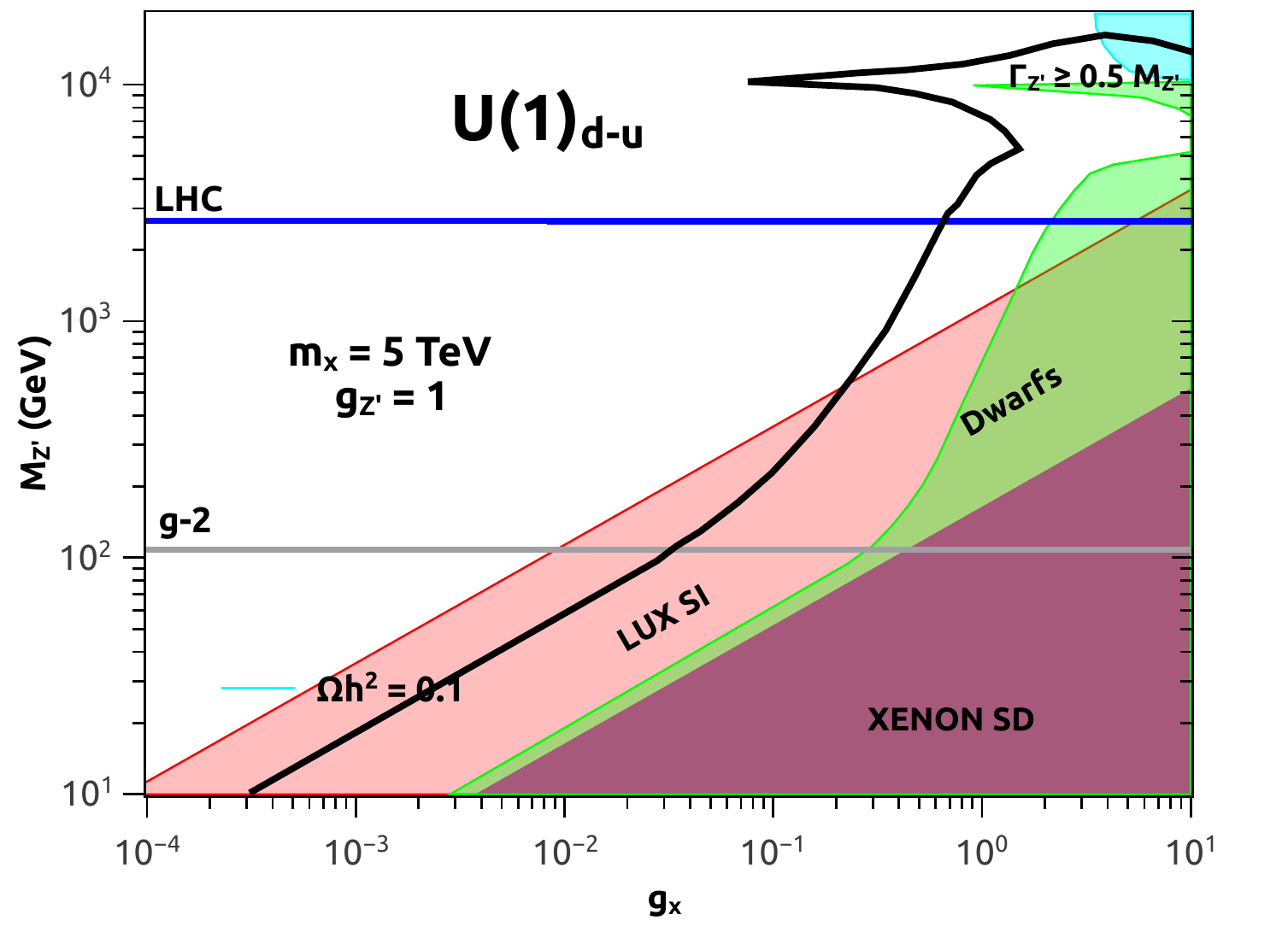}
\includegraphics[scale=0.5]{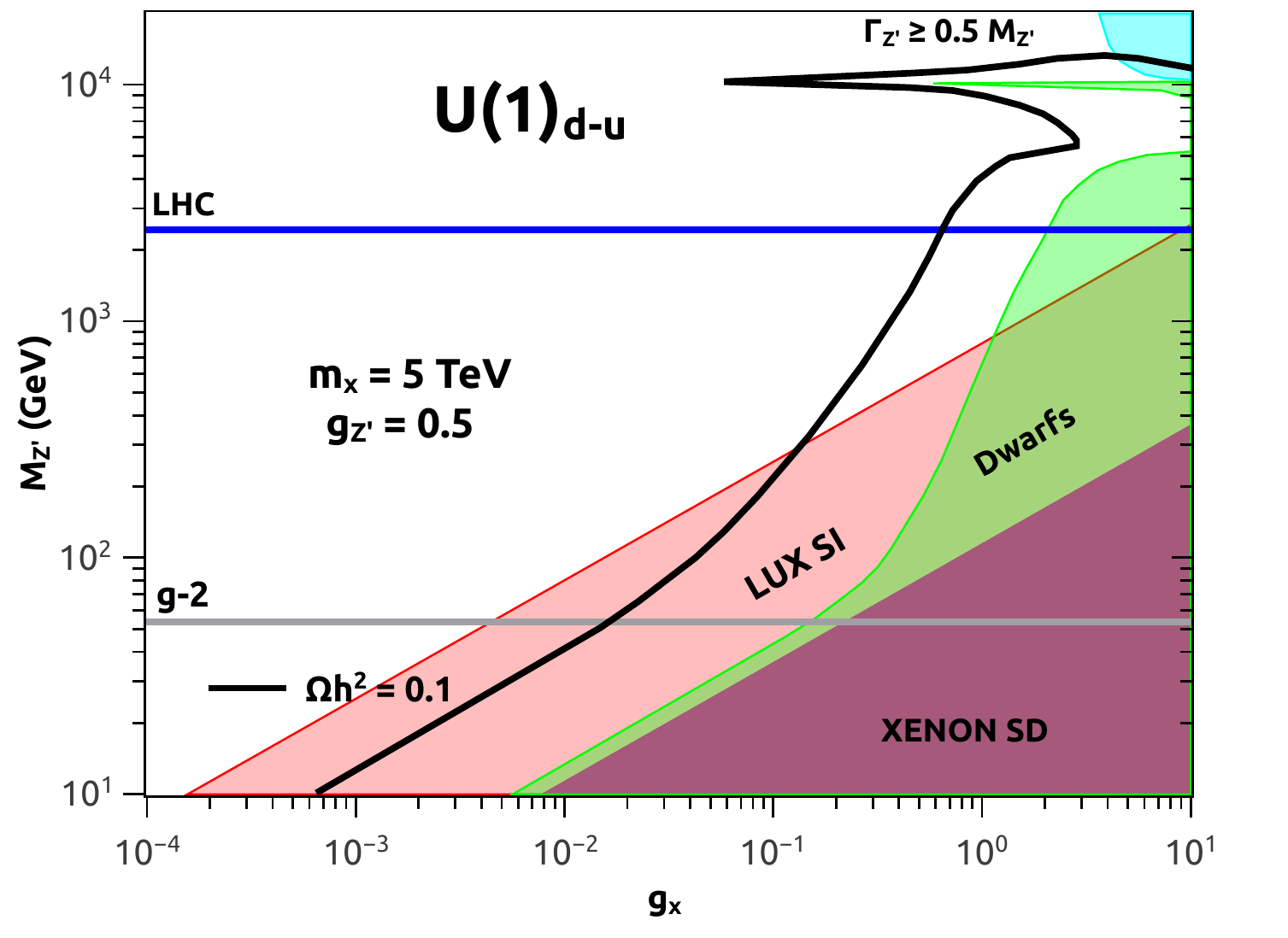}
\caption{Same as in Fig.~\ref{fig:results}, for the $U(1)_{d-u}$ model. {\it Left:}  $m_{\chi}=5$~TeV, $g_{Z^{\prime}}=1$; {\it Right:}  $m_{\chi}=5$~TeV, $g_{Z^{\prime}}=0.5$.}
\end{figure}

\end{document}